\documentclass[
superscriptaddress,
 amsmath,amssymb,
 aps,
pra,
twocolumn,
]{revtex4-2}

\usepackage{xcolor}
\usepackage{graphicx} 
\usepackage{dcolumn} 
\usepackage{bm,graphics}
\usepackage{dsfont}
\usepackage{braket} 
\usepackage{amsthm}
\usepackage{mathtools}

\usepackage{times}
\usepackage[normalem]{ulem}
\usepackage{multirow}
 \usepackage{ragged2e}

\usepackage[colorlinks=true,linkcolor=teal, citecolor=teal
]{hyperref}

\usepackage{booktabs}

\usepackage[most]{tcolorbox}
\tcbuselibrary{breakable}
\newtcolorbox[auto counter, number within=section]{mybox}[2][]{
  float, floatplacement=tp,
  colback=blue!5!white, colframe=blue!75!black,
  fonttitle=\bfseries, center title,
  title=Box~\thetcbcounter: #2, #1
}

\newcounter{mybox}
\renewcommand{\themybox}{\arabic{mybox}}

\usepackage[ruled]{algorithm2e}

\newcommand{\bx}{\bm{x}}
\newcommand{\by}{\bm{y}}
\newcommand{\bs}{\bm{s}}
\newcommand{\bhaty}{\hat{\bm{y}}}
\newcommand{\ham}{\mathsf{H}}

\newcommand{\Tr}{\text{Tr}}
\newcommand{\TML}{\mathcal{T}_{\text{ML}}}
\newcommand{\TDL}{\mathcal{T}_{\text{DL}}}
\newcommand{\TLM}{\mathcal{T}_{\text{LM}}}
\newcommand{\TLMP}{\mathcal{T}_{\text{LM,P}}}
\newcommand{\TLMF}{\mathcal{T}_{\text{LM,F}}}
\newcommand{\bxi}{\bm{x}^{(i)}}
\newcommand{\bzi}{\bm{z}^{(i)}}
\newcommand{\bsi}{\bm{s}^{(i)}}

\newcommand{\btheta}{\bm{\theta}}
\newcommand{\bomega}{\bm{\omega}}

\newcommand{\bw}{\bm{w}}
\newcommand{\bhatyi}{\hat{\bm{y}}^{(i)}}

\newcommand{\tilderho}{\tilde{\rho}}

\newcommand{\PP}{\mathbb{P}}
\newcommand{\QQ}{\mathbb{Q}}

\newcommand{\je}[1]{\textcolor{cyan}{#1}}

\begin{document}

\title{Artificial intelligence for representing and characterizing quantum systems}

\author{Yuxuan Du\href{https://https://orcid.org/0000-0002-1193-9756}{\includegraphics[scale=0.05]{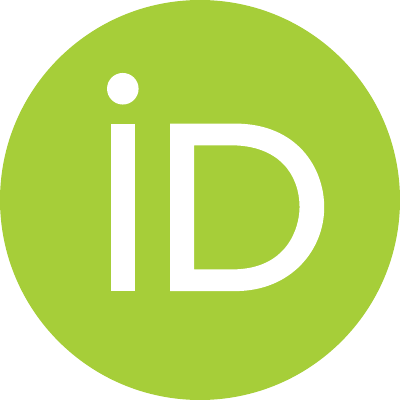}}
}

\affiliation{College of Computing and Data Science, Nanyang Technological University, Singapore 639798, Singapore}
 
\author{Yan Zhu\href{https://orcid.org/0000-0002-4963-160X}
{\includegraphics[scale=0.05]{orcidid.pdf}}
}
\affiliation{QICI Quantum Information and Computation Initiative, Department of Computer Science,
The University of Hong Kong, Pokfulam Road, Hong Kong}

\author{Yuan-Hang Zhang\href{https://orcid.org/0000-0002-5415-3307}{\includegraphics[scale=0.05]{orcidid.pdf}}
}
\affiliation{Department of Physics, University of California, San Diego, La Jolla, CA 92093, USA}

\author{Min-Hsiu Hsieh\href{https://orcid.org/0000-0002-3396-8427}{\includegraphics[scale=0.05]{orcidid.pdf}}
}
\affiliation{Hon Hai (Foxconn) Research Institute, Taipei, Taiwan}

\author{Patrick Rebentrost\href{https://orcid.org/0000-0002-6728-8163}{\includegraphics[scale=0.05]{orcidid.pdf}}
}
\affiliation{Centre for Quantum Technologies,
National University of Singapore}
\affiliation{Department of Computer Science, National University of Singapore}

\author{Weibo Gao\href{https://orcid.org/0000-0003-3971-621X}{\includegraphics[scale=0.05]{orcidid.pdf}}
}
\affiliation{School of Electrical \& Electronic Engineering, Nanyang Technological University, Singapore 639798, Singapore}
\affiliation{Centre for Quantum Technologies,
National University of Singapore}

\author{Ya-Dong Wu\href{https://orcid.org/0000-0002-9940-6128}{\includegraphics[scale=0.05]{orcidid.pdf}}
}
\email{wuyadong301@sjtu.edu.cn}
\affiliation{John Hopcroft Center for Computer Science, Shanghai Jiao Tong University, Shanghai 200240, China}

\author{Jens Eisert\href{https://orcid.org/0000-0003-3033-1292}{\includegraphics[scale=0.05]{orcidid.pdf}}
}

\affiliation{
Dahlem Center for Complex Quantum Systems,
Freie Universitat Berlin, 14195 Berlin, Germany}
\affiliation{Helmholtz-Zentrum Berlin f{\"u}r Materialien und Energie, 14109 Berlin, Germany}

\author{Giulio Chiribella\href{https://orcid.org/0000-0002-1339-0656}{\includegraphics[scale=0.05]{orcidid.pdf}}
}
\affiliation{QICI Quantum Information and Computation Initiative, Department of Computer Science,
The University of Hong Kong, Pokfulam Road, Hong Kong}
\affiliation{Department of Computer Science, Parks Road, Oxford, OX1 3QD, United Kingdom}
\affiliation{Perimeter Institute for Theoretical Physics, Waterloo, Ontario N2L 2Y5, Canada}

\author{Dacheng Tao\href{https://orcid.org/0000-0001-7225-5449}{\includegraphics[scale=0.05]{orcidid.pdf}}
}
\affiliation{College of Computing and Data Science, Nanyang Technological University, Singapore 639798, Singapore}

\author{Barry C. Sanders\href{https://orcid.org/0000-0002-8326-8912}{\includegraphics[scale=0.05]{orcidid.pdf}}
}
\affiliation{Institute for Quantum Science and Technology, University of Calgary, Alberta T2N 1N4, Canada}

\begin{abstract}
Efficient characterization of large-scale quantum systems, especially those produced by quantum analog simulators and megaquop quantum computers, poses a central challenge in quantum science due to the exponential scaling of the Hilbert space with respect to system size. Recent advances in artificial intelligence (AI), with its aptitude for high-dimensional pattern recognition and function approximation, have emerged as a powerful tool to address this challenge. A growing body of research has leveraged AI to represent and characterize scalable quantum systems, spanning from theoretical foundations to experimental realizations. Depending on how prior knowledge and learning architectures are incorporated, the integration of AI into quantum system characterization can be categorized into three synergistic paradigms: machine learning, and, in particular, deep learning and language models. This review discusses how each of these AI paradigms contributes to two core tasks in quantum systems characterization: quantum property prediction and the construction of surrogates for quantum states. These tasks underlie diverse applications, from quantum certification and benchmarking to the enhancement of quantum algorithms and the understanding of strongly correlated phases of matter. Key challenges and open questions are also discussed, together with future prospects at the interface of AI and quantum science.
\end{abstract}

\maketitle

\tableofcontents

\section*{Key Points}

\begin{itemize}
  \item AI models can be leveraged to represent and characterize scalable quantum systems in a data-driven manner for the tasks of quantum property prediction and 
  implicit and approximate quantum state reconstruction. 
\item Provably efficient machine learning models have been designed to characterize linear properties of scalable quantum systems and to classify quantum phases. 
\item Deep learning models offer powerful tools for predicting a wide range of quantum properties through representation learning, as well as for implicitly reconstructing quantum states using generative modeling approaches.
\item Language models, building on the GPT architecture, provide a flexible framework for auto-regressively representing large families of quantum states, paving the way toward foundation models for quantum systems and enabling new directions for research and application.
\end{itemize}

\section{INTRODUCTION}
Recent advances in quantum engineering have made it increasingly routine to fabricate and control highly complex quantum devices~\cite{van2024advanced,shaw2024benchmarking,manetsch2024tweezer,king2025beyond,Bourgund2025Formation}. These developments enable the exploration of quantum many-body systems using scalable quantum simulators~\cite{xu2024non,manovitz2025quantum,andersen2025thermalization} and the construction of quantum computers~\cite{decross2024computational,bluvstein2024logical,Quantum2025Google,gao2025establishing,preskill2025beyond}, taking steps towards the megaquop era ({\em e.g.},  handle $\sim$100 logical qubits at depths $\sim$10000~\cite{preskill2025beyond}). However, the increasing size of quantum systems accessible in the laboratory poses new challenges. Due to the exponential growth of the state space with the number of qubits, describing and characterizing quantum systems produced by modern quantum simulators and quantum computers becomes extremely difficult. Classical simulators, such as tensor networks~\cite{orus2019tensor} capture important classes of states but fail to model the behavior of highly entangled states. While approaches tailored for Clifford circuits enable efficient simulation, extensions to circuits containing non-Clifford gates typically incur a computational cost that grows exponentially with the number of non-Clifford operations, usually captured in terms of what is called 
``magic''~\cite{aaronson2004improved,bravyi2016improved,aharonov2023polynomial}. 
These challenges necessitate novel approaches that allow us to learn properties of quantum systems that cannot be exactly stored with the  
computational capabilities of classical simulators. 
 
Over the past decade, methods from AI have emerged as a promising candidate for addressing the challenges in characterizing quantum systems, owing to their ability to identify patterns and relationships in large datasets.  
A review over early applications of AI techniques, such as 
Bayesian inference and neural networks with shallow 
architectures and basic training algorithms, to the study 
of quantum systems is provided in Ref.~\cite{gebhart2023learning}. Recently, breakthroughs in AI, such as the development of \emph{generative pre-trained transformers} (GPT)~\cite{radford2018improving}, and advancements in quantum learning theory~\cite{anshu2024survey}, have spurred the creation of sophisticated methods for representing and characterizing quantum systems at scale. These developments have led to notable progress in designing AI models for understanding quantum many-body physics and quantum computing~\cite{huang2022provably,zhu2022,wang2022,PhysRevB.107.075147,tangtowards,wu2023,wang2022quest,xiao2022,liao2024machine,qian2023multimodal,lewis2024improved,che2024exponentially,cho2024machine,wu2024learning,PhysRevApplied.21.014037,du2025efficient,kim2024attention,yao2024shadowgpt,tang2024ssl4q,mohseni2024deep,dehghani2023neural,vsmid2025efficient,huang2023learning,zhu2023predictive,fitzek2024rydberggpt,onorati2023provably,rouze2024efficient,zhao2025Rethink,moss2023enhancing,lange2025transformer,huang2025direct}. In light of these rapid developments, it is essential to consolidate central findings and identify open questions to guide future research.
 
 \begin{figure*}[htp]
	\centering 
	\includegraphics[width=0.83\textwidth]{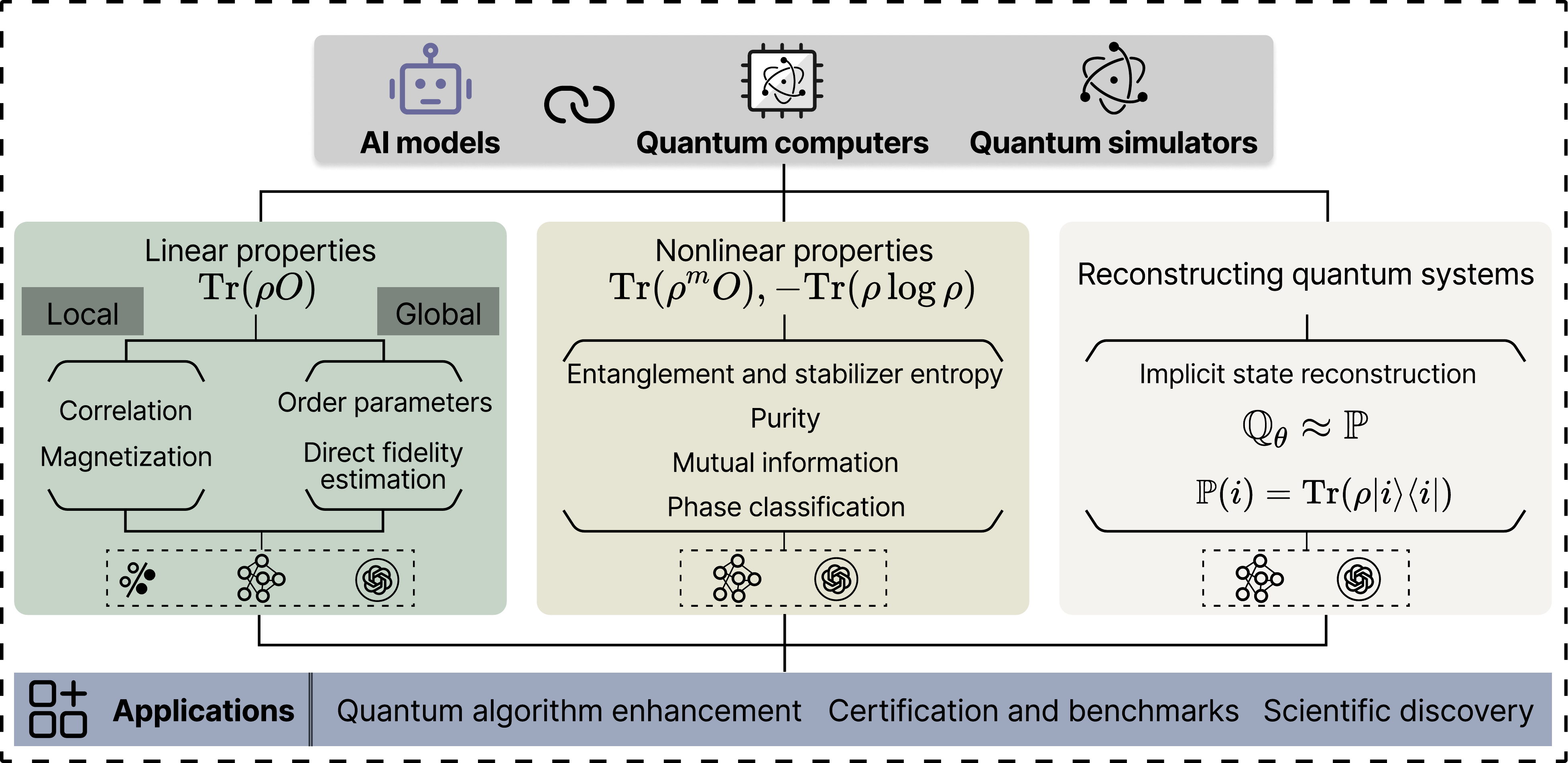}
	\caption{\small{\textbf{Overview of key tasks and applications in leveraging AI to represent and characterize quantum systems}. Representing and characterizing ground states generated by quantum analog simulators and quantum states produced by digital quantum computers can be classified into three major tasks: predicting linear properties, predicting nonlinear properties, and reconstructing quantum states and processes. Each task is further divided into subcategories reflecting specific objectives. The icons at the bottom indicate the AI paradigms, {\em i.e.}, ML models, DL models, and LM as referenced in Fig.~\ref{fig:applications}, that are typically used for each task. Current and potential applications of these approaches include quantum algorithm enhancement, certification and benchmarking of quantum devices, quantum hardware development, and scientific discovery.
}}
	\label{fig:scheme}
\end{figure*} 

\begin{figure}
     \centering
\includegraphics[width=0.97\linewidth]{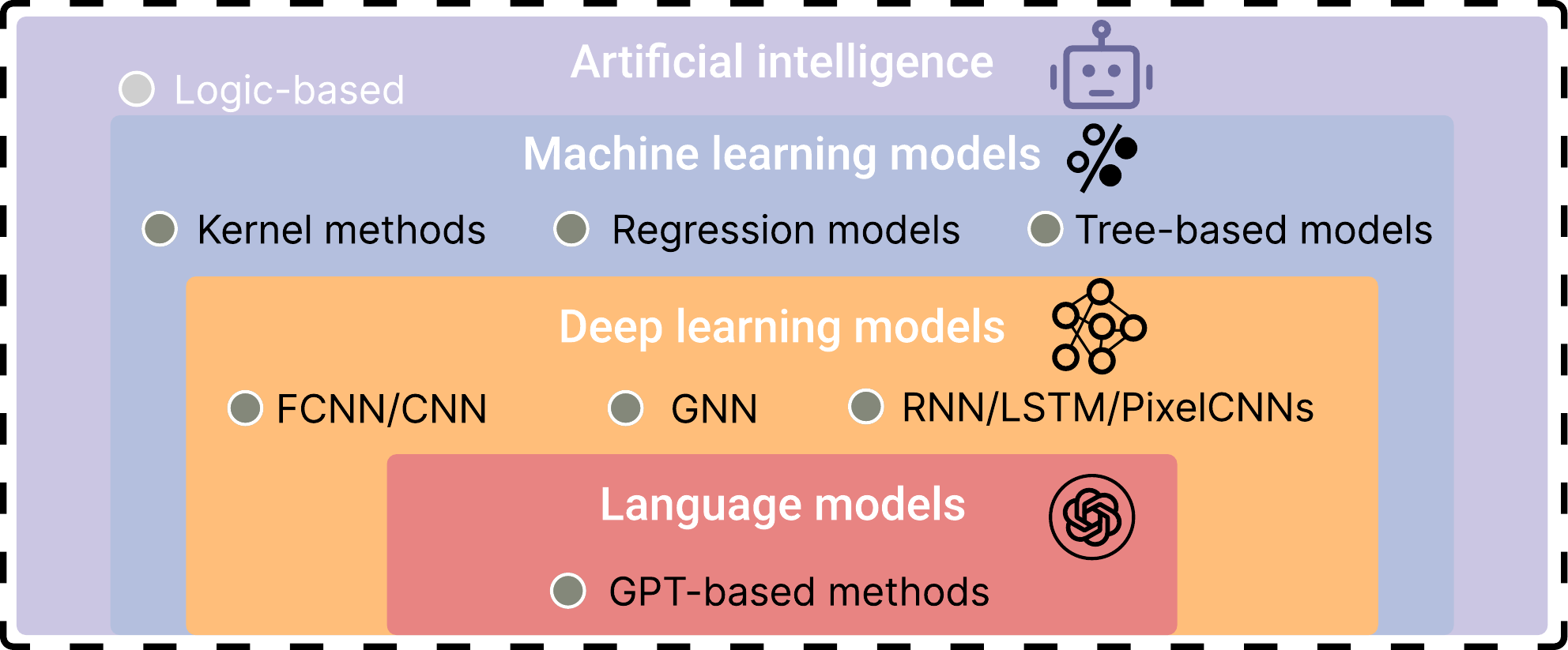}
     \caption{\small{\textbf{Overview of AI models for learning large-scale quantum systems}. The hierarchy reflects the increasing capability of AI models and their growing adaptability to handle large-scale quantum systems, progressing from the broad concept of AI to machine learning models, deep learning models, and transformer-based models. Representative strategies within each category are highlighted by green circles. The notations `NN', `NQS', and `LLM' refer to neural networks, neural quantum states, and large language models, respectively. Sequence models include recurrent neural networks, LSTM, and related architectures.}}
     \label{fig:applications}
 \end{figure}

In this review, we survey recent advances in the use of AI for representing and characterizing quantum systems since 2022, with an emphasis on theoretical foundations and algorithmic innovations. In particular,  we organize the recent progress in this context along a methodological hierarchy of AI models, including \emph{machine learning} (ML),  with \emph{deep learning} (DL) as a prominent subfield, and \emph{language models} (LMs) as a specific class of DL architectures, as illustrated in Fig.~\ref{fig:applications}.  Unlike many other fields, where newer AI approaches supplant earlier methods, progress in characterizing quantum systems has benefited from the complementary strengths of diverse AI models. To illustrate how ML, DL, and LMs are leveraged in this domain, we adopt a task-oriented perspective, deliberately focusing on three essential tasks: (1) predicting linear properties of quantum systems, (2) predicting instances of non-linear properties, and (3) reconstructing quantum states and processes. These tasks, as shown in Fig.~\ref{fig:scheme}, underpin a wide range of applications, including quantum certification and benchmarking~\cite{eisert2020quantum}, quantum hardware characterization~\cite{alexeev2024artificial}, the enhancement of variational quantum algorithms~\cite{cerezo2021variational}, and the discovery of exotic quantum phases~\cite{carleo2019machine}. On a more conceptual level, methods of tomographic reconstruction and property testing are increasingly seen as being 
part of quantum learning theory \cite{arunachalam2017guest,anshu2024survey}, reflecting a certain shift of mindset.

There have been several reviews
on the applications of AI in quantum physics, covering topics such as AI for physical sciences~\cite{carleo2019machine,wetzel2025interpretable,Giovanni2025,CUP}, AI for quantum computing and quantum technologies~\cite{alexeev2024artificial,Giovanni2025,PhysRevA.107.010101}, language models for quantum simulation~\cite{melko2024language}, quantum shadow tomography~\cite{elben2023randomized,huang2022learning,eisert2020quantum}, neural quantum states~\cite{carrasquilla2021use,CUP,lange2024architectures}, and variational quantum algorithms~\cite{cerezo2021variational,bharti2021noisy,du2025quantum}. While most of these reviews have emphasized domain-specific applications of AI models, our focus is fundamentally distinct. We analyze these applications through the lens of different AI methodologies and explore the broader spectrum of AI techniques applied to represent and characterize quantum systems produced by quantum simulators and quantum computers. This spans from ML methods with theoretical guarantees to cutting-edge foundation models such as GPTs, highlighting their advantages and limitations respectively.

In this review, we specifically focus on AI models that represent and characterize quantum systems in a data-driven manner. Accordingly,  variational neural quantum state approaches are not the primary focus~\cite{carrasquilla2021use}. Meanwhile, we emphasize scalable AI approaches designed to learn quantum systems, tackling the curse of dimensionality inherent in conventional non-ML methods. Conversely, AI approaches tailored to small-scale quantum systems, which cannot be generalized to scalable quantum systems, such as AI methods used for reconstructing the density matrix of an unknown quantum state~\cite{lange2024architectures}, are beyond the main focus of this review. Additionally, precise control of quantum systems adaptively using their extracted knowledge with AI approaches is also important, but goes beyond the scope of this review. Since we focus on theoretical and algorithmic progress, experimental work is not the primary emphasis, even though it should be clear that the methods discussed have an immediate impact there. Only those experimental results supporting either theoretical or algorithmic advancements will be covered. 
  
\section{An overview of learning paradigms}\label{sec:overview}
Leveraging AI to represent and characterize quantum systems at scale involves designing learning models capable of identifying relevant patterns and structural features across families of quantum systems. Once trained, these models can generalize to previously unseen quantum systems that exhibit similar characteristics to those encountered during training. This data-driven manner is fundamentally distinct from the conventional approaches that address each system independently without leveraging transferable insights. 
 
 \begin{figure*}
	\centering
	\includegraphics[width=0.85\textwidth]{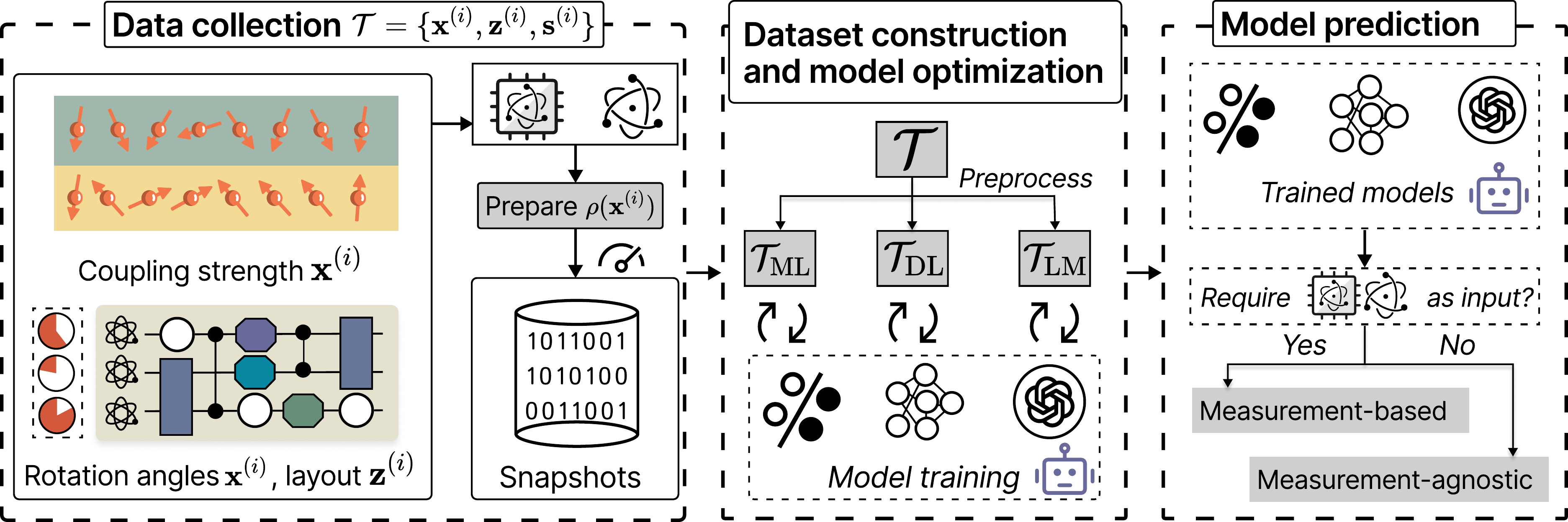}
	\caption{\small{\textbf{Framework of learning protocols for quantum systems.} Existing learning models designed to represent and characterize scalable quantum systems generally consist of three stages: data collection, model implementation and optimization, and model prediction. The left panel visualizes the data collection stage, where a quantum system is prepared with parameters $\bxi$ and auxiliary information $\bzi$. The prepared quantum state $\rho(\mathbf{x}^{(i)})$ is then measured with $T$ times to obtain the measurement outcomes $\bm{s}^{(i)}$. The middle panel illustrates the stage of dataset construction and model implementation.  Once the original dataset $\mathcal{T}$ is collected, it should be preprocessed into task-specific datasets $\TML$, $\TDL$, and $\TLM$, which are then employed to train ML, DL, and LM-based models, respectively. The right panel presents the model prediction stage. Depending on whether the prediction process requires quantum measurement data as input, the learning protocols are classified as measurement-based and measurement-agnostic. 
 }}
	\label{fig:pipeline}
\end{figure*}

\begin{figure}
\refstepcounter{mybox}
\begin{tcolorbox}[title={Box~\themybox: Linear and nonlinear property prediction}, center title, before upper={\normalsize\justifying}]
\label{mybox:linear-nonlinear-property}
 Predicting linear properties of a quantum state $\rho$ refers to estimating the expectation values of a set of observables $\mathfrak{O}$ on $\rho$, {\em i.e.}, $\{\Tr(\rho O)|O\in \mathfrak{O}\}$. Typical examples include energy, magnetization, and correlation functions.
 Predicting nonlinear properties of a state $\rho$ refers to estimating physical quantities that can be expressed as $\Tr\left(f(\rho,O)\right)$, where the function $f(.,.)$
 is nonlinear with respect to the quantum state 
 $\rho$. Typical examples include von Neumann entropy where $f(\rho, \mathds 1)=-\rho\log\rho$, and the Uhlmann fidelity where $f(\rho,\sigma)= \sqrt{ \sqrt{\rho} \sigma \sqrt{\rho} }$, with $\sigma$ denoting the mixed quantum state.
\end{tcolorbox}
\end{figure}

Existing learning protocols for scalable quantum systems have mainly investigated two classes of quantum states: (i) states in analog quantum simulation, such as Hamiltonian ground states, and (ii) states in digital quantum computing, {\em i.e.}, states produced by quantum circuits. For both classes, the learning objective is to generalize from a set of training states, which are drawn from a family with shared physical structure, to accurately predict specific physical properties (such as magnetization and fidelity) or reconstruct quantum states beyond the training data.  The core learning tasks in this context, as demonstrated in Fig.~\ref{fig:scheme}, encompass the prediction of linear and nonlinear properties (see Box~\ref{mybox:linear-nonlinear-property}), and the reconstruction of quantum systems.

For illustration, we next briefly review how the characterization and the representation of quantum systems can be reformulated as learning tasks. Taking Hamiltonian ground states as the first example, we consider a family of Hamiltonians $\{\ham(\bx)|\bx\in \mathbb{R}^d\}$, each specified by a relatively small set of real parameters $\bx$, defining
some concept class. For instance, in the transverse-field Ising model, parameters $\bx$ encode interaction strengths and external field strengths. The Hamiltonian ground state vector of $\ham(\bx)$ is denoted by $|\psi(\bx)\rangle $, where $\psi(\bx)$ denotes the pure quantum state parametrized by $\bx$. By assumption, $\ket{\psi(\bx)}$ satisfies 
\begin{equation}\label{eqn:ground_state}
	\ham(\bx) |\psi(\bx)\rangle = E_0(\bx) |\psi(\bx)\rangle,
\end{equation}
with the lowest eigenvalue $E_0(\bx)$ of $\ham(\bx)$. 
In the training phase, learning models are trained over a classical dataset comprising sampled values of $\bx$ and the associated measurement outcome data of $\ket{\psi(\bx)}$. During the prediction phase, these trained learning models are applied to predict physical properties of the state vector $\ket{\psi(\bx')}$ or reconstruct a classical representation of the state itself, for previously unseen parameters $\bx'$. 

In the regime of digital quantum computing, an example is a set of quantum states prepared by a parametrized quantum circuit $U(\bx)$, consisting of a fixed set of gates along with tunable gates parametrized by, say, $\bx\in [-\pi, \pi)^d$.  Given any $\bx$, the corresponding quantum state vector is
\begin{equation}\label{eqn:circuit_state}
|\psi(\bx)\rangle = U(\bx) |\psi_0\rangle,
\end{equation}  
where $ |\psi_0\rangle$ is a fixed $N$-qubit initial state vector. The two phases of training and prediction in this setting are similar to those for learning Hamiltonian ground states. 

Depending on the learning paradigm, existing learning protocols for characterizing and representing quantum systems can be classified into \emph{machine learning} (ML), \emph{deep learning} (DL), and \emph{language model} (LM)-based approaches. Despite differences in model architecture and application scope, these protocols share a common workflow comprising three stages:  \textit{Data collection}, \textit{model implementation and optimization}, and \textit{model prediction}, as illustrated in Fig.~\ref{fig:pipeline}.  

In what follows, we provide an overview of the connections and differences of the three learning paradigms at each stage of the protocol, with detailed discussions deferred to the subsequent content. Depending on the applications, we will either use the state vector notation  $|\psi(\bx)\rangle$ or the density matrix notation $\rho(\bx)$. 

\begin{figure*}[t]
\refstepcounter{mybox}
\begin{tcolorbox}[
title={Box~\themybox: Terminologies in AI}, center title, before upper={\normalsize\justifying}]
\label{mybox:one}

\smallskip
\centerline{\textbf{Supervised, semi-supervised, self-supervised, and unsupervised learning} }

AI learning paradigms are typically divided into \textit{supervised}, \textit{self-supervised}, \textit{semi-supervised}, and \textit{unsupervised} learning, each differing in how they use labeled and unlabeled data. Supervised learning models are trained on labeled datasets to learn the mapping between inputs and outputs. Semi-supervised learning combines a small amount of labeled data with a large amount of unlabeled data to enhance learning efficiency and generalization. Self-supervised learning constructs supervisory signals from the data itself, often by formulating pretext tasks, enabling models to learn useful representations without explicit labels. Unsupervised learning analyzes unlabeled data to identify patterns, clusters, or similarities within the data itself.

\smallskip
\centerline{\textbf{Discriminative and generative  learning} }

AI learning tasks can also be categorized as \textit{discriminative} or \textit{generative}, based on the type of relationship they model between data and labels.  Discriminative learning focuses on modeling the decision boundary between different classes by learning the conditional probability $P(\by|\bx)$, where $\bx$ is the input (for example, an image) and $\by$ is the label (such as its corresponding category). This approach is prevalent in property prediction tasks. In contrast, generative learning aims to model the joint distribution $P(\by,\bx)$ or the marginal distribution $P(\bx)$, enabling the generation of new data samples.  For example, in image generation tasks, $\bx$ may represent an image and $\by$ its category, allowing the model to generate realistic image-label pairs or synthesize new images. In the context of this review, a representative example is implicit quantum state reconstruction, generating a classical surrogate,
where the goal is to generate measurement outcomes that mimic those produced by a quantum system.
    
\smallskip
\centerline{\textbf{Feature engineering and representation learning} }
Feature engineering and representation learning are both essential for preparing data in AI models, and they often complement each other to improve learning outcomes. Feature engineering refers to the process of converting raw data into a form that can be effectively used by learning algorithms, often drawing on domain knowledge to design or select informative input features. For example, this might involve translating the classical description of a quantum system into mathematical representations suitable for neural networks. Representation learning, on the other hand, focuses on extracting useful and informative representations from the available input---either raw or feature engineered---enabling models to learn effective patterns for downstream tasks. 
\end{tcolorbox}
\end{figure*}

\subsection{Data collection}\label{sec:overview-dataset} 

In all generality, a training dataset can be expressed in the form 
\begin{equation}\label{eqn:dataset-uni}
	\mathcal{T} = \left\{\big(\bxi, \bsi, \bzi\big)\right\}_{i=1}^n,
\end{equation}
where $n$ refers to the number of training examples,  $\bxi$ and $\bzi$ refer to the classical description and the auxiliary information of each training sample, and $\bsi$ denotes the associated measurement data of the $i$-th training example $\rho(\bxi)$ with $T$ shots.  For instance, as shown in Fig.~\ref{fig:pipeline},  when $\ket{\psi(\bxi)}$ is prepared by a digital quantum computer in Eq.~(\ref{eqn:circuit_state}), $\bxi$ denotes the rotational angles in tunable gates, $\bzi$ denotes the gate layout of $U(\bxi)$, and $\bsi$ denotes the measurement outcomes of $\ket{\psi(\bxi)}$. 
Note that the dataset $\mathcal{T}$ in Eq.~(\ref{eqn:dataset-uni}) takes the most comprehensive form. Depending on the learning protocols, either auxiliary information $\bsi$ alone or the combination of auxiliary information $\bsi$ and classical description $\bxi$ can be omitted.

The quantum measurements adopted to collect $\bsi$ are versatile and problem-dependent. Recall that any quantum measurement can be described by a \emph{positive operator-valued measure} (POVM) \cite{nielsen2010quantum}. Formally,
a POVM takes the form $\mathbb M=\{M_s\}$ with $0 \preceq M_s$ and $\sum_s M_s=\mathds 1$ (or $\int \text{d}s M_s=\mathds 1 $ when the set $\{s\}$ is uncountable). Each shot of POVM measurement $\mathbb M$ yields a sample from the probability distribution $\Pr(s)= \Tr(\rho M_s)$. Thus, a finite number of shots yields finite samples of measurement outcomes, {\em i.e.}, $\bsi:=\big\{s^{(i)}_1, s^{(i)}_2, \dots, s^{(i)}_T \big\}$. When $\mathbb{M}$ is an information-complete POVM, the density matrix of $\rho$ can be reconstructed from $\bsi$ asymptotically as $T\rightarrow \infty$. An important measurement strategy widely employed in collecting $\bsi$ is to make use of a suitable randomized measurements~\cite{elben2023randomized}, with respect to a probability measure. The choice of this measure is guided both by mathematical considerations, to ensure reasonable sample complexities, and by physical constraints, such as the locality requirements of certain POVMs.

Compared to classical ML tasks, constructing high-quality datasets $\mathcal{T}$ for representing and characterizing quantum systems presents unique challenges. First, unlike conventional fields such as computer vision and natural language processing, it is nontrivial to determine which measurements provide the most relevant information for studying quantum systems. Second, the exponentially large state space associated with increasing system size makes collecting comprehensive and high-quality data (such as the data required by 
full quantum state tomography) prohibitively expensive. 

\subsection{Model implementation and optimization} 

Existing learning protocols rely on the collected training dataset $\mathcal{T}$ to perform training. However, protocols based on the ML, DL, and LM paradigms differ in how they process and extract relevant information, as well as in their optimization strategies for specific learning tasks. 

For ML models, prior studies~\cite{huang2022provably,lewis2024improved,du2025efficient} have primarily focused on predicting single or multiple \emph{linear properties}  
of quantum systems as illustrated in Fig.~\ref{fig:scheme}, with provable guarantees relating sample complexity to prediction accuracy. The collected data $\mathcal{T}$ is transformed to a \textit{supervised learning} format (see Box~\ref{mybox:one}). 
Once the labeled dataset is constructed, regression-based and kernel-based methods with task-specific feature maps are employed to complete the training~\cite{bishop2006pattern}. More precisely, the tailored feature map transforms the original input $\bxi$ into a higher-dimensional space where complex relationships between inputs and targets become linearly separable and more amenable to regression or classification.

For DL models, previous studies can be largely categorized into two subclasses based on the target learning tasks. The first subclass focuses on property prediction, aiming to infer one or more physical properties (both linear and nonlinear) of the quantum system under investigation~\cite{carrasquilla2017machine,van2017learning,PhysRevLett.120.240501,zhang2021,zhu2022,xiao2022,koutny2023deep,PhysRevLett.132.190801,PhysRevLett.132.220202,wu2023,wang2022quest,qian2023multimodal,wu2024learning}. In this scenario, the data preprocessing closely resembles that of ML models, where the collected quantum data $\mathcal{T}$ is reformatted into a supervised learning format. Given access to the prepared dataset, a wide range of DL models with diverse neural architectures and optimization strategies have been developed. A central goal in this research direction is to construct DL models capable of using limited training data to accurately predict a broader range of physical properties~\cite{wu2024learning}.  The second subclass focuses on \textit{implicit} and approximate quantum state reconstruction using \emph{neural quantum states}  (NQS)~\cite{torlai2018,carrasquilla2019reconstructing,cha2021attention,zhong2022quantum}. Unlike conventional quantum state tomography that aims at creating a complete classical description of the quantum state, this approach trains deep neural networks to approximately reproduce the measurement statistics of the target quantum state $\rho(\bx)$. Importantly, whereas property prediction is formulated as a discriminative learning task, quantum state reconstruction with NQS is fundamentally \textit{generative} (see Box~\ref{mybox:one} for explanations). In this context, the objective is to design efficient deep generative models to approximate the true measurement distribution associated with $\rho(\bx)$ using 
comparably few training examples. 

For LMs, recent efforts have explored \emph{generative pre-trained transformers} (GPTs)~\cite{brown2020language} to implement GPT-like models capable of performing a variety of tasks related to the representation and characterization of quantum systems. In contrast to ML and DL approaches, LM-based methods are typically optimized in two distinct stages: pre-training and fine-tuning. During the pre-training stage, the model learns to capture common structural patterns across a broad range of quantum states by performing generalized state reconstruction, aiming to approximate the measurement distribution corresponding to $\rho(\bx)$ conditioned on a collection of parameters 
$\bx$~\cite{wang2022,yao2024shadowgpt}. In the subsequent fine-tuning stage, the model is adapted to specific property prediction tasks~\cite{PhysRevB.107.075147,tangtowards}. This process mirrors the training strategies used in ML and DL models. The objective in this stage is to conduct supervised learning to accurately infer the desired quantum properties.

\begin{figure*}[t]
\refstepcounter{mybox}
\begin{tcolorbox}[
title={Box~\themybox: Classical shadows}, center title, before upper={\normalsize\justifying}]
\label{box:classical shadow}
 
The scheme of classical shadows for an unknown $N$-qubit state $\rho$ repeats the following procedure $T$ times~\cite{huang2020predicting}. At each time, a unitary $U$ randomly sampled from a suitably chosen predefined unitary ensemble $\mathcal{U}$ is first applied on the state $\rho$ and then each qubit is measured on the computational basis to obtain an $N$-bit string denoted by $\bm{b} \in \{0, 1\}^N$. This measurement in average yields a linear map  $\mathcal{M}(\rho) =\mathbb{E}_{U\sim \mathcal{U}}\mathbb{E}_{\bm{b}\sim \PP(\bm{b})} U^{\dagger}\ket{\bm{b}}\bra{\bm{b}} U $ with $\PP(\bm{b})=\bra{\bm{b}}U\rho U^{\dagger}\ket{\bm{b}}$. The unknown state $\rho$ can be formulated as 
\[\rho =  \sum_{\bm{b}} \int \mathsf{d}U \mathcal{M}^{-1}\Big(U^{\dagger}\ket{\bm{b}}\bra{\bm{b}} U\Big) \bra{\bm{b}}U\rho U^{\dagger}\ket{\bm{b}}.\]
It implies that $\rho$ can be estimated by sampling the snapshot $T$ times following the distribution $\PP(\bm{b})$. Define the $t$-th snapshot as $U_t^{\dagger}\ket{\bm{b}_t}\bra{\bm{b}_t}U_t$ with $t\in [T]$ and $U_t\sim \mathcal{U}$. The shadow representation of $\rho$ corresponding to these $T$ snapshots is 
\begin{equation}
	\tilderho_T = \frac{1}{T}\sum_{t=1}^T \tilderho_t,~\text{with}~\tilderho_t=\mathcal{M}^{-1}(U_t^{\dagger}\ket{\bm{b}_t}\bra{\bm{b}_t} U_t).
\end{equation}	
While the forward process is reflected by a physical quantum channel, the inverse process can be performed on the classical level. When Pauli-based random measurements are adopted, the unitary ensemble $\mathcal{U}$ amounts to the single-qubit Clifford gates, {\em i.e.}, $U_t= \otimes_{j=1}^N U_{j,t}\sim \mathcal{U}=\text{CI}(2)^{\otimes N}$ with uniform weights. In this case, the inverse snapshot takes the form  $\tilderho_t  = \mathcal{M}^{-1}(\bigotimes_{j=1}^N U_{j,t}^{\dagger} |\bm{b}_{j,t}\rangle\langle \bm{b}_{j,t}|U_{j,t})  = \bigotimes_{j=1}^N (3U_{j,t}^{\dagger} |\bm{b}_{j,t}\rangle\langle \bm{b}_{j,t}|U_{j,t} - \mathbb{I}_2 )$. This tensor product form allows an efficient estimation of the expectation values of local observables for the given state $\rho$. For such random single-qubit Clifford gates, the number of samples required scales with the Pauli weight of the observable chosen. Other common ensembles are full global random Clifford circuits \cite{huang2020predicting},
leading to the situation that the shadow is informationally complete and allows estimating expectation values of any observable. A number of important variants of the
original scheme have been suggested, in particular ones that make use of logarithmically deep Clifford circuits \cite{PhysRevLett.133.020602,ExtremelyShallow,hu2025demonstration}.

\end{tcolorbox}
\end{figure*}
 
\subsection{Model prediction} 

The diverse tasks and multiple learning paradigms in characterizing and representing quantum systems hint that no single and definitive categorization can encompass all models. Previously, we have classified them based on their implementation strategies and application domains.  Another crucial angle to distinguishing them is to check if access to quantum systems is required at the prediction phase, often referred to as ``quantum data''. From this standpoint, all learning protocols, {\em i.e.}, ML models, DL models, and LMs, can be categorized into \textit{measurement-agnostic} and \textit{measurement-based} protocols, as shown in Fig.~\ref{fig:pipeline}. In particular, the prediction of measurement-agnostic protocols relies solely on classical inputs~\cite{huang2022provably,lewis2024improved,du2025efficient,wang2022,wang2022quest,yao2024shadowgpt}, such as the parameter $\bx$ and auxiliary information $\bm z$, without requiring quantum measurements. In contrast, measurement-based protocols requires measurement outcomes $\bm{s}$ from quantum systems under consideration as input during the prediction stage~\cite{PhysRevLett.120.240501,zhang2021,zhu2022,koutny2023deep,PhysRevLett.132.190801,PhysRevLett.132.220202,wu2023, qian2023multimodal,wu2024learning,tangtowards, kim2024attention}. 
 
\section{Machine learning paradigm}\label{sec:ML}

\emph{Machine learning} (ML) models refer to a wide class of algorithms that infer patterns from data based on statistical learning principles~\cite{mohri2018foundations}. Current studies in this research line focus on devising provably efficient ML models for predicting \textit{linear properties} of quantum states (refer to Box~\ref{mybox:linear-nonlinear-property}), with particular emphasis on analyzing how the prediction error scales with the number of training examples. Despite their diversity, these learning protocols can be recast into a general scheme. For clarity, we first elucidate this general scheme of ML models, and then discuss their applications, followed by discussing their limitations and other advanced topics.

\subsection{General scheme for linear property prediction}\label{subsec:ML-scheme}

Recall that ML models for predicting linear properties involve three steps: data collection, model construction and training, and prediction. Here, we elaborate on each of these steps, complementing the high-level overview introduced earlier.
 
\smallskip
\noindent\textbf{Dataset construction}. The first step  when applying ML models to linear property prediction is transforming the raw dataset $\mathcal{T}$ in Eq.~(\ref{eqn:dataset-uni}) into a labeled dataset for supervised learning, {\em i.e.},
\begin{equation}\label{eqn:ML-dataset}
	\mathcal{T} 
	\mapsto
	\TML = \left\{\big(\bxi, \bhatyi \big)\right\}_{i=1}^n,
\end{equation}
 where $\bhatyi$ denotes the estimates of the physical properties of interest for $i$-th training example, derived from the measured outcomes $\bsi$, $\forall i \in [n]$. Given a set of observables $\mathfrak{O}$, the linear properties of state $\rho(\bx)$ refer to $\bm{y}=\{\Tr(\rho(\bx)O)\}_{O\in \mathfrak{O}}$. The estimation error between $\bm{y}$ and $\bhaty$ vanishes in the limit of an infinite number of measurements $T$, and for a number of precise settings, rigorous sample complexity bounds can be shown.

\medskip
\noindent\textbf{Model implementation and training}.  Given access to $\TML$, ML models rely on the explicitly tailored feature engineering (see Box~\ref{mybox:one}) to complete the learning. Denote the prediction of the employed ML model as $h_{\text{ML}}(\bx)$. The optimization process involves minimizing the discrepancy between the model's predictions and the labels in $\TML$. 

The majority of ML models developed for predicting linear properties of scalable quantum systems adopt linear regression frameworks~\cite{bishop2006pattern}. The mathematical expression of these models is 
 \begin{equation}\label{eqn:linear-reg-uni}
	h_{\text{ML}}(\bx;\bw)= \langle \bw, \phi(\bx)\rangle,
\end{equation} 
where $ \bw$ refers to the trainable parameters and $\phi(\bx)$ represents the engineered feature map applied to the input $\bx$. The primary focus of these ML models is manual feature engineering of $\phi(\cdot)$, which transforms the raw input $\bx$ into a higher-dimensional space so that the relationship between the input and the target property becomes approximately linear. The implementation of $\phi(\bx)$ is task-dependent and will be detailed subsequently.
 
The optimization of linear regression models in Eq.~(\ref{eqn:linear-reg-uni}) involves determining the optimal parameters $\bw^*$ that minimize the loss function, {\em i.e.},
\begin{equation}\label{eqn:obj-ML-measurement-agnostic}
	\mathcal{L}= \frac{1}{n}\sum_{i=1}^n\ell\big(h_{\text{ML}}(\bxi), \bhatyi\big) + \lambda\left(\alpha\|\bw\|_2^2 + (1-\alpha) \|\bw\|_1^2\right),
\end{equation}
where $\ell(\cdot, \cdot)$ denotes the per-sample loss, such as the mean square error, and $\lambda\geq \mathbb{R}_+$ is a hyperparameter for the regularization term. Depending on the choice of $\alpha$, there are three types of linear regression models (the value of $0$, $1$, and $(0,1)$ corresponds to \textit{Lasso}, \textit{ridge}, and \textit{elastic net} regression, respectively~\cite{zou2005regularization}). Beyond linear regression models, kernel methods form another class of ML models widely used for predicting linear properties of scalable quantum systems~\cite{bishop2006pattern}. Their underlying mechanisms and commonly employed kernels are summarized in Box~\ref{box:kernel}.   
  
\begin{figure*}[t]
\refstepcounter{mybox}
\begin{tcolorbox}[ 
title={Box~\themybox: Kernel methods}, center title, before upper={\normalsize\justifying}]
\label{box:kernel}
Kernel methods are a class of ML techniques that perform classification or regression by evaluating the similarity between data points in a high-dimensional feature space~\cite{bishop2006pattern}. In this way, kernel methods can reveal relationships in complex data that may not be immediately apparent from the original variables. Note that kernel methods are mathematically equivalent to linear regression when expressed in their dual form~\cite{scholkopf2002learning}. The generic form of a kernel machine is 
 \begin{equation}
 	h_{\text{ML}}(\bx) = \sum_{i=1}^{n} \alpha_i k\left(\bxi, \bx\right) + b, \nonumber
 \end{equation}
 where $\{\alpha_i\}$ denote the dual coefficients (or weights) associated with each training point, $b\in \mathbb{R}$ denotes the bias term, and $k\left(\bxi, \bx\right) \coloneqq
 \langle \phi(\bxi), \phi(\bx) \rangle$ is the kernel function that evaluates the similarity between $\bxi$ and $\bx$ in the feature space. As with $\phi(\cdot)$ in linear regression models, the choice of $k(\cdot, \cdot)$ is also task-dependent. Representative kernel functions developed for predicting properties of quantum systems are given below: 

\noindent $\bullet$ \textit{\underline{Dirichlet kernel}}. This kernel is designed to predict linear properties of a family of ground states. Denote  $\Lambda$ as the threshold for frequency truncation. The explicit form of Dirichlet kernel is 
\[\kappa_{\Lambda}(\bx, \bx') = \sum_{\bm{k}\in \mathbb{Z}^d, \|\bm{k}\|_2\leq \Lambda} \cos\left(\pi \bm{k}\cdot\left(\bx-\bx'\right)\right).\] 

\noindent $\bullet$ \textit{\underline{Positive good kernel}}. This kernel is also designed to predict linear properties of a family of ground states. Following the same notation with Dirichlet kernel, its explicit form is  \[\kappa_{\Lambda}(\bx,\bx')=\frac{1}{\Lambda}\prod_{j=1}^d \frac{\sin^2 (\Lambda \pi (\bx_j-\bx'_j)/2)}{\sin^2 (\pi (\bx_j-\bx'_j)/2)}.\]

\noindent $\bullet$ \textit{\underline{The kernel for topologically ordered phase classification}}. Let $\tau$ and $\gamma$ be two hyperparameters. Denote  the reduced density matrix at the $i$-th qubit for the $t$-snapshot of the classical shadow $\tilderho_T(\bx)$ as $\sigma_i^{(t)}(\bx)$. The mathematical expression of the proposed kernel is \[k(\bx, \bx') = \exp \Big(\frac{\tau}{T^2}\sum_{t,t'=1}^T\exp\big(\frac{\gamma}{n}\sum_{i=1}^N\Tr(\sigma_i^{(t)}(\bx)  \sigma_i^{(t')}(\bx'))\big)\Big).\]

\noindent $\bullet$ \textit{\underline{Truncated trigonometric monomial kernel}}.  Denote $\bomega\in \{0, \pm 1\}^d$ as the frequency vector with $d$ dimensions. Define the feature map as  $\Phi_{\bomega}(\bx)= \prod_{i=1}^{d} \alpha(\bx_i;\bomega_i)$, where $\alpha_i(\cdot;\cdot)$ contains three distinct mapping functions depending on the value $\bomega_i$. Specifically, $\alpha(\bx_i;0)=0$, $\alpha_i(\bx;1)=\cos(\bx_i)$, and $\alpha_i(\bx;-1)=-\sin(\bx_j)$, for $\forall i\in[d]$. 
Denoting the threshold of the truncation value with $\Lambda$,
 then the
kernel takes the form 
\[\kappa_{\Lambda}(\bx',\bxi)=\sum_{\bomega, \|\bomega\|_0\leq \Lambda} 2^{\|\bomega\|_0}\Phi_{\bomega}(\bx')\Phi_{\bomega}(\bxi).\]   
\end{tcolorbox}
\end{figure*}

\medskip
\noindent\textbf{Model prediction}. The trained ML models can be directly used to predict the quantum properties of interest for a new quantum state $\rho(\bx)$. As shown in Fig.~\ref{fig:pipeline}, operating in a measurement-agnostic manner, these models provide an efficient approach to characterize quantum systems without taking quantum data as input.  

A common way to assess the performance of a trained learning model is evaluating its expected risk, which measures how closely the model's predictions match the actual physical quantities of interest. Mathematically, the expected risk, {\em a.k.a.}, prediction error, is
\begin{equation}\label{eqn:off-exp-risk}
	\mathcal{R}(h)= \mathbb{E}_{\bx\sim \mathbb{D}_{\mathcal{X}}} \Big[\left|h_{\text{ML}}(\bx) - \by \right|^2 \Big], 
\end{equation}
where the input $\bx$ is sampled from the data distribution $\mathbb{D}_{\mathcal{X}}$ and $\by $ represents the true values of physical properties. In general, the data distribution $\mathbb{D}_{\mathcal{X}}$ is unknown, prohibiting the direct evaluation. An alternative is to evaluate the loss function on a test dataset, which consists of previously unseen examples drawn from the same data distribution $\mathbb{D}_{\mathcal{X}}$.

\subsection{Concrete ML models and applications}\label{subsec:ML-application}
A substantial body of work in characterizing scalable quantum systems focuses on developing efficient ML models to address various linear property prediction tasks, as well as certain nonlinear tasks, motivated by their explainability and theoretical guarantees. Here, we categorize these ML models based on their application scenarios and review their implementations and corresponding theoretical results. A summarization of representative methods in this regime is provided in Table~\ref{tab:ML-summary-work}.

\subsubsection{Linear property prediction for Hamiltonian ground states} 
The seminal work in this field 
has been established by Huang et al.~\cite{huang2022provably}, demonstrating that ML algorithms, informed with experimental data, can effectively tackle certain quantum many-body problems that are intractable for classical algorithms. Specifically, the ML model predicts the expectation values of a set of observables $\mathfrak{O}=\{O\}$ on the ground states of a family of gapped, geometrically-local Hamiltonians $\mathsf{H}(\bx)$. For example, $O$ can be a linear combination of low-weight Pauli operators. The learner adopts the Pauli-based classical shadow \cite{huang2020predicting} to acquire raw data $\mathcal{T}$~(\ref{eqn:dataset-uni}), where $\bsi$ corresponds to the randomized measurement outcomes of $\rho(\bxi)$ while 
auxiliary 
information $\bzi$ is not required. From these measurements, the learner reconstructs shadow representation $\tilderho_T(\bxi)$ following the procedure in Box~\ref{box:classical shadow}. The training dataset $\TML$~(\ref{eqn:ML-dataset}) is built by computing label $\bhatyi$ via shadow estimation. Given access to $\TML$, 
for each observable $O_j\in \mathfrak{O}$, the kernel machine takes the explicit form \[h_{\text{ML}}(\bx)=  \frac{1}{n}\sum_{i=1}^n\kappa_{\Lambda}\left(\bx, \bxi\right)\bhatyi,\] where the $j$-th entry of $\bhatyi$ refers to the estimation of $\braket{O_j}$ and $\kappa_{\Lambda}(\cdot,\cdot)$ denotes the truncated Dirichlet kernel (see Box~\ref{box:kernel}).  

The proposed ML model is provably efficient in many practical scenarios.  Specifically, when $\bx$ is sampled from a uniform distribution, the average gradient norm is bounded, i.e.\ $\mathbb{E}_{\bx}\|\nabla_{\bx} \Tr(O\rho(\bx))\|^2_2 \leq C$, and the observable is well-bounded, the proposed model achieves an $\epsilon$-prediction error in Eq.~(\ref{eqn:off-exp-risk}) with high probability. In addition, both the classical training time for model implementation and the prediction time are upper bounded by $\mathcal{O}\left(d^{\mathcal{O}(C/\epsilon)}\right)$. The efficacy of this ML model is  
suggested
through predicting local expectation values for the ground states of one-dimensional 51-atom Rydberg atom systems and predicting two-point correlation functions for the ground states of two-dimensional $25$-qubit antiferromagnetic Heisenberg models.

Follow-up works in this research direction focus on how to further reduce the sample and runtime complexities by exploiting different conditions of Hamiltonians. When the geometry of the explored family of $N$-qubit Hamiltonians $\{\ham(\bx)\}$ is known, the feature map $\phi(\cdot)$ can incorporate the geometric inductive bias and the resulting Lasso enables an efficient and accurate prediction, requiring only $\mathcal{O}(\log(N))$ samples and $\mathcal{O}(N\log(N)\text{poly}(\epsilon^{-1}))$ runtime~\cite{lewis2024improved}.  Moreover, when the observable set $\mathfrak{O}$ only contains a single element with the known decomposition and the family of Hamiltonians is geometrically local and gapped, ridge regression achieves a sample complexity of $\mathcal{O}\left(2^{\text{polylog}(1/\epsilon)}\right)$ (independent with $d$) and a runtime linear in~$N$. The employed feature map, as with Ref.~\cite{lewis2024improved}, also encodes the geometry of Hamiltonian with a slight modification~\cite{wanner2024predicting}.

When the number of classical parameters $d$ is constant, or independent of the qubit count $N$, the positive good kernel defined in Box~\ref{box:kernel} can be replaced with the Dirichlet kernel to reach an improved sample complexity of $\mathcal{O}(\text{poly}(1/\epsilon, N))$~\cite{che2024exponentially}. In addition, efficient ML models exist for predicting properties of (equivariant) long-range Hamiltonians. As with Ref.~\cite{lewis2024improved}, the Lasso associated with geometry-informed feature map enables accurate prediction of linear properties of ground states with long-range interactions, achieving sample complexity that scales logarithmically with system size $N$~\cite{vsmid2025efficient}. Besides, ML models that preserve equivariance under the automorphism group of the interaction hypergraph can further reduce sample complexity~\cite{vsmid2024accurate}. The effectiveness of these ML models has been validated on a 127-qubit IBM quantum computer, demonstrating successful prediction of two-point correlation functions in both the random hopping system and Su-Schrieffer-Heeger system~\cite{cho2024machine}. 
 
Beyond ground states of gapped Hamiltonians, initial efforts have been made to predict linear properties of thermal states within stationary states of Liouvillians capturing Markovian open quantum systems~\cite{onorati2023provably,rouze2024efficient}. Instead of employing conventional ML models, the key strategy for predicting these properties relies on computing empirical averages of classical shadow protocols. In this way, $\mathcal{O}(\log(N/\delta)2^{\text{polylog}(1/\epsilon)})$ samples suffice to learn local expectation values for quantum systems within a phase with prediction errors less than $\epsilon$ and failure probability at most $\delta$.  
   
\subsubsection{Phase classification for Hamiltonian ground states}
Classifying quantum phases of matter is another key application of ML in characterizing scalable quantum systems. A seminal contribution in this research line has also been made by Huang et al.~\cite{huang2022provably}, who have proposed two ML models tailored for different quantum phase classification tasks: distinguishing symmetry-breaking phases and identifying topologically ordered phases. 

In the task of symmetry-breaking phase classification, suppose there are two phases, denoted by $A$ and $B$. There exists an observable $O$ consisting of multiple local observables, called the local order parameter, such that the relevant ground states satisfy $\Tr(\rho(\bx)O)\geq 1$ if $\rho(\bx)\in \text{phase $A$}$ and $\Tr(\rho(\bx)O)\leq -1$ if $\rho(\bx)\in \text{phase $B$}$. In this regard, the truncated Dirichlet kernel in Box~\ref{box:kernel} can be employed to implement an ML-based classifier to attain a satisfactory classification accuracy. 

In the task of topologically ordered phase classification, no linear function with respect to $\rho(\bx)$ can be used to complete the accurate prediction. Nevertheless, Ref.~\cite{huang2022provably} has demonstrated that a nonlinear classifier can solve this task with a provable guarantee. Specifically, they designed a feature map that takes the classical shadow to a feature vector that incorporates arbitrarily large reduced density matrices, with the corresponding kernel function specified in Box~\ref{box:kernel}. The proposed ML model provides a rigorous guarantee: if a nonlinear function of few-body reduced density matrices can classify phases, then the proposed ML model can accurately learn to perform this classification. Both the required amount of training data $n$ and computational resources scale polynomially with system size $N$. The effectiveness of this ML model has been demonstrated by distinguishing a topological phase from a trivial phase in a $200$-qubit system.

\subsubsection{Linear property prediction for gate-based states}

There are two distinct learning settings for applying ML models to predict linear properties of quantum states output by digital quantum computers, determined by the flexibility of observables. Each setting introduces unique challenges and necessitates different learning strategies, which will be discussed separately.

\begin{figure}[h!]
	\centering
	\includegraphics[width=0.48\textwidth]{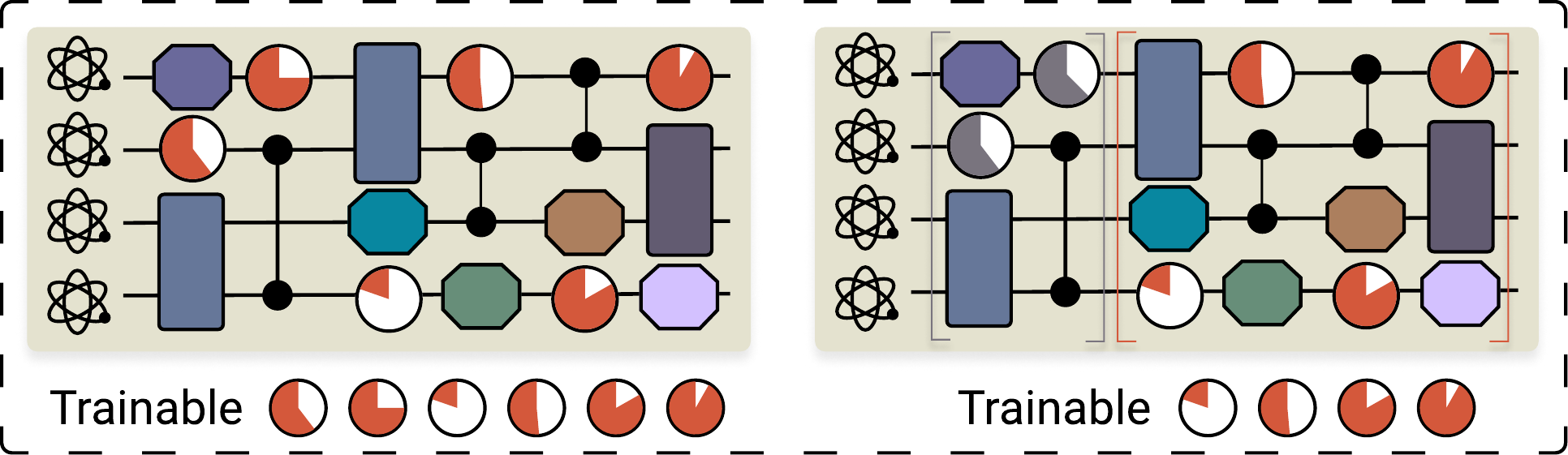}
	\caption{\small{\textbf{Scenarios for predicting properties of states from digital quantum computers}}. The left panel shows a typical VQE circuit, where all tunable gates (indicated by orange circles) serve as trainable parameters. The right panel depicts a QNN architecture, where a subset of tunable gates is allocated for encoding classical data, while the remaining gates function as trainable parameters for optimization. The hexagonal and rectangular gates represent Clifford gates.  
	}
	\label{fig:vqe-and-qnn}
\end{figure}

\smallskip
\noindent\textbf{Fixed input state and observable}. The first setting considers the case where both the input state $\rho_0$ and observable $O$ are fixed.  The primary motivation for this setting arises from developing classical surrogates of variational quantum algorithms~\cite{cerezo2021variational}, including the \emph{variational quantum eigensolver} (VQE) and variants thereof as well as \emph{quantum neural networks} (QNNs). In other words, the objective is to design an ML model to accurately predict the expectation value $\Tr(\rho(\bx)O)$ with $\rho(\bx)=U(\bx)\rho_0 U(\bx)^{\dagger}$ and $U(\bx)$ being an $N$-qubit parametrized circuit defined in Eq.~(\ref{eqn:circuit_state}). 

As shown in Fig.~\ref{fig:vqe-and-qnn}, the classical input in QNNs consists of two components, {\em i.e.}, $\bx \coloneqq
[\tilde{\bx}, \btheta]$, where  $\tilde{\bx}$ refers to the classical data such as images and texts, and  $\btheta$ comprises the trainable parameters. The generic form of QNNs is 
\[f(\tilde{\bx}; \btheta) = \Tr(U(\tilde{\bx};\btheta)\rho_0 U(\tilde{\bx};\btheta)^{\dagger} O).\]  Prior studies~\cite{Schuld2021Effect,vidal2020input} have proven that the trained QNN can be expanded as a truncated Fourier series, where the threshold of the truncation value depends on the adopted  $U(\tilde{\bx};\btheta)$. See Box~\ref{mybox:QNN} for more details.

Schreiber et al.~\cite{schreiber2022classical} have employed this Fourier-based formulation to design a linear regression model for predicting the QNN output $f(\tilde{\bx}; \hat{\btheta})$. The proposed model takes the form $h_{\text{ML}}(\tilde{\bx})=\sum_{\bomega \in \Omega}c_{\bomega}\exp({-\imath \langle\bomega, \tilde{\bx}\rangle})$, where $\{c_{\bomega}\}$ are trainable Fourier coefficients. The learning objective is to optimize these coefficients by minimizing the loss function in Eq.~(\ref{eqn:obj-ML-measurement-agnostic}), using a labeled dataset $\TML = \{\tilde{\bm{x}}^{(i)}, \bhatyi\}$. The label $\bhatyi$ refers to the estimation of $f(\tilde{\bx}; \hat{\btheta})$ derived from the measurement outcomes $\bsi$.  In the worst-case scenario, the proposed ML model achieves prediction error bounded by $\|h_{\text{ML}}(\tilde{\bx}) - f(\tilde{\bx};\hat{\btheta})\|\leq \epsilon$ with high probability for any $\tilde{\bx}$ from the data domain $ \mathcal{X}$, provided that the number of training examples satisfies $n\sim \mathcal{O}(\bomega_{\max}^d/\epsilon^2)$ with $\bomega_{\max}=\max\{|\bomega_i|:\bomega\in \Omega\}$ and $d$ being the dimension of $\tilde{\bx}$. The effectiveness of the proposed model has been verified on a standard classical ML dataset and synthetic datasets up to eight qubits. 

\begin{figure}
\refstepcounter{mybox}
\begin{tcolorbox}[ 
title={Box~\themybox: Fourier analysis of QNNs}, center title, before upper={\normalsize\justifying}]
\label{mybox:QNN}
Without loss of generality, the parametrized quantum circuit employed in QNNs can always be written as
\[U(\tilde{\bx};\btheta)=\prod_{l=1}^L W^{(l)}(\btheta)S^{(l)}(\tilde{\bx}),\] 
where $L$ refers to the layer number, and the trainable unitaries $\{W^{(l)}(\btheta)\}_l$ interleave with data-encoding unitaries $\{S^{(l)}(\tilde{\bx})\}_l$~\cite{du2023problem}. Following the above notation, prior studies have proven that QNNs can be expanded as a truncated Fourier series~\cite{Schuld2021Effect,vidal2020input}. The mathematical expression yields  \[f(\tilde{\bx}; {\btheta})=\sum_{\bomega\in \Omega}c_{\bomega}^*({\btheta})\exp({-\imath \langle\bomega, \tilde{\bx}\rangle}),\] where $\{c_{\bomega}^*({\btheta})\}$ refer to the optimal Fourier coefficients, and the frequency set $\Omega$ depends only on the structure of $\{S^{(l)}(\tilde{\bx})\}_l$ and layer number $L$. The Fourier expansion of QNN outputs provides a theoretical foundation for employing random Fourier features to predict their behavior. The mathematical form of the classical learning model is $h_{\text{ML}}(\bx)=\langle \bw, \phi(\tilde{\bx}) \rangle$, where the random Fourier feature~\cite{rahimi2007random} is defined as 
 \[\phi(\tilde{\bx})=\frac{1}{D}[\psi(\tilde{\bx},\nu_1), \dots, \psi(\tilde{\bx},\nu_D)]\] with $\psi(\tilde{\bx},\nu)=\sqrt{2}\cos(\langle \bm{a}_i, \tilde{\bx}\rangle+b_i)$. Here, $D$ is a hyperparameter and the feature $\nu_i= (\bm{a}_i,b_i)$ is sampled from a prior distribution. 
\end{tcolorbox}
\end{figure}

The classical surrogate may suffer from an efficiency bottleneck, as its runtime complexity grows exponentially with the size of the frequency set---or equivalently, with the dimension of the classical input vector $\tilde{\bx}$. To mitigate this issue, one promising solution is to employ random Fourier features for model construction~\cite{landman2022classically,sweke2025potential}. Specifically, the original feature map $\phi(\cdot)$ in $h_{\text{ML}}(\tilde{\bx})$ is replaced by random Fourier features with $D$ dimensions (see Box~\ref{mybox:QNN}). 

This alternative enables significant computational benefits: both the space and time complexities of training the model scale as $\mathcal{O}(nD^2)$ and $\mathcal{O}(nD^2 + D^3)$, which are independent of the dimension of the training data. Moreover, when $f(\tilde{\bx}, \hat{\btheta})$ is well-bounded and $\{\bsi\}$ is collected by a sufficiently large $T$, $n\sim \mathcal{O}(d/\epsilon^2)$ guarantees good prediction performance, {\em i.e.}, $ |h_{\text{ML}}(\tilde{\bx})-f(\tilde{\bx}, \hat{\btheta})|\leq \epsilon$ for $\forall \bx \in \mathcal{X}$.

The relation between the number of training examples $n$ and the snapshot $T$ for each example is further explored in Ref.~\cite{gan2024concept}. In particular, they proposed a new ML model by incorporating an $L_1$-Lipschitz nondecreasing function $u(\cdot)$, {\em i.e.}, $h_{\text{ML}}(\tilde{\bx})=u(\sum_{i=1}^n\bw_ik(\tilde{\bx}, \bxi))$ with the kernel $k$ corresponding to the Fourier feature map. Through optimizing $h_{\text{ML}}(\tilde{\bx})$ in an iterative method based on $\TML$, the prediction error in the measure in Eq.~(\ref{eqn:off-exp-risk}) is upper bounded by ${c_1}/{\sqrt{n}}+ {c_2}/{\sqrt{nT}}$. Here the parameters $c_1$ and $c_2$ polynomially scale with $L_1$, the infinite norm of the observable $O$, and the norm of the Fourier coefficients. The results highlight the dominant role of $n$ compared to $T$, as a limited $n$ leads to a high prediction error.

An orthogonal research line involves employing Fourier series expansions, the structure of parametrized quantum circuits, and Heisenberg evolution to design efficient classical simulators of VQAs rather than learning-based models~\cite{cerezo2023does, goh2023lie,Nemkov2023Fourier,rudolph2023classical,fontana2023classical,beguvsic2024fast, cirstoiu2024fourier}. As efficient simulation of quantum circuits falls beyond the scope of this review, it will not be discussed in detail.

\smallskip
\noindent\textbf{Fixed input state and varied observables}. This setting considers a broader scenario such that the input state $\rho_0$ is fixed, while the observables in $\mathfrak{O}$ can be varied. An immediate observation is that the first setting is a special case of the latter. In this scenario, an efficient ML model serves as a shadow representation predictor $\hat{\sigma}(\bx)$, meaning that given any new input $\bx'$, it can accurately predict its shadow representation with $\hat{\sigma}(\bx')\approx \tilderho_T(\bx')$. Consequently, the ML model enables accurate prediction of expectation values for many observables, {\em i.e.}, $h_{\text{ML}}(\bx',O) \coloneqq
\Tr(\hat{\sigma}(\bx')O)\approx \Tr(\rho(\bx')O)$.

The first approach toward this goal was introduced in Ref.~\cite{du2025efficient}, aiming to predict many linear properties of quantum states, where the quantum circuit $U(\bx)$ in Eq.~(\ref{eqn:circuit_state}) consists of $d$ rotational-Pauli gates and $G-d$ Clifford gates and the input state $\rho_0$ is arbitrary. The learning procedure follows the standard three-stage supervised learning pipeline. Conceptually, the learner draws an input control parameter $\bxi$ from a prior distribution and feeds it to the explored circuit. For each example $\bxi$, the learner adopts Pauli-based classical shadow~\cite{huang2020predicting} to collect measurement outcomes $\bsi$. By repeating this process $n$ times, the raw data $\mathcal{T}$ is collected. 

Given access to $\mathcal{T}$, the explicit form of the shadow representation predictor for any new input $\bx$ yields \[\hat{\sigma}_{\Lambda}(\bx)= \sum_{i=1}^n \kappa_{\Lambda}(\bx,\bxi)\tilderho_T(\bxi),\] where $\kappa_{\Lambda}(\bx,\bxi)$ refers to the truncated trigonometric monomial kernel defined in Box~\ref{box:kernel}. For any observable $O\in \mathfrak{O}$, the ML model is $h_{\text{ML}}(\bx,O)=\Tr(\hat{\sigma}_{\Lambda}(\bx)O)$, which can be efficient implemented on the classical side when $n$ and $\Lambda$ are not too large.  

 When the input data is sampled from a uniform distribution and $\Lambda = d$, the shadow predictor is an unbiased estimator of $\rho(\bx)$. Moreover, when the averaged gradient norm $\mathbb{E}_{\bx} \|\nabla_{\bx}\Tr(\rho(\bx)O)\|_2^2$ is upper bounded by a small $C$, the proposed ML model is both sample- and computationally efficient. To achieve an $\epsilon$-prediction error in Eq.~(\ref{eqn:off-exp-risk}),  the required sample complexity is $\mathcal{O}(|\mathfrak{E}(\Lambda)|\epsilon^{-1})$ with $\Lambda=4C/\epsilon$ and $\mathfrak{E}(\Lambda)=\{\bomega|\bomega\in \{0, \pm 1\}^d,~s.t.~\|\bomega\|_0\leq \Lambda\}$. In addition, the polynomial sample complexity ensures an overall polynomially computational complexity. Numerical simulations demonstrate the effectiveness of the proposed method in predicting two-point correlation functions of $60$-qubit rotational GHZ states, predicting the magnetization of a $60$-qubit global Hamiltonian, and pre-training a $50$-qubit VQE for transverse-field Ising models.
  
\subsection{Fundamental limitations}\label{subsec:ML-limitation}

A common characteristic of existing ML models is that they all adhere to the measurement-agnostic learning protocols in Fig.~\ref{fig:pipeline}. These models adopt a strategy in which measurements are performed first, followed by classical processing of the collected data. This raises a fundamental question in understanding the computational separation between these classical ML models and quantum learning models~\cite{servedio2004equivalences,gyurik2022establishing}, where the learning process is executed fully or partially on a quantum device. Resolving this question not only enriches quantum learning theory but also provides concrete guidance in identifying quantum utility and quantum advantages. For instance, hypothetically speaking, if every quantum problem could be efficiently solved by classical ML models, the practical advantage of quantum computing would be confined to the data acquisition stage. Recent research, however, indicates that while current ML models reviewed here are effective in many practical scenarios, they also exhibit fundamental limitations, resulting in substantially worse performance compared to their quantum counterparts. 

The first study exploring computational hardness of ML models in characterizing scalable quantum systems has been conducted by Gyurik et al.~\cite{gyurik2023exponential}. They construct a family of Hamiltonians whose ground-state properties cannot be predicted by any classical ML method, assuming standard cryptographic assumptions. These results demonstrate that the conditions required for efficient prediction, {\em e.g.},  smoothness and geometric locality \cite{huang2022provably}, cannot be significantly relaxed. Following the same routine, subsequent work further strengthens these results by demonstrating a classical-quantum separation in predicting expectation values of an unknown observable from measurements on ground states under the mild assumption $\mathsf{BQP}\not\subseteq \mathsf{P/poly}$~\cite{molteni2024exponential}. In addition, the relevant results can be effectively extended to establish the computational hardness of ML models in predicting the linear properties of bounded-gate quantum states~\cite{du2025efficient}. For nonlinear tasks, it has been shown that learning `gapless' quantum phases of matter is computationally hard under
standard cryptographic assumptions~\cite{bouland2023public,bouland2024hardness}.
 
In parallel with general theoretical analyses, an independent research direction explores the fundamental limitations of specific classes of ML models. A notable example is understanding the potential but also the limitations of random Fourier features in dequantizing QNNs. Specifically, a recent study has provided counterexamples demonstrating that classical surrogates based on random Fourier features fail to achieve reliable predictive performance~\cite{sweke2025potential}, building upon the findings of
Ref.\ \cite{schreiber2022classical}. Furthermore, a follow-up study established the necessary conditions for regression models to serve as classical surrogates of QNNs, showing that classical-quantum separation emerges when the parameters optimized by quantum models approach their optimal values~\cite{thabet2024quantum}. Besides, Refs.~\cite{Dequantization,Dequantization2} revealed that in regimes in which variational quantum learning models can be trained, one can at the same time find an efficient classical algorithm for the setting at hand and hence ``dequantize'' the setting.

\begin{table*}[]
\centering
\resizebox{0.9\textwidth}{!}{%
\begin{tabular}{@{}cccccc@{}}
\toprule
 &
  Model &
  Tasks &
  Conditions &
  Runtime complexity &
  Simulation/experiment scale \\ \midrule
\multicolumn{1}{c|}{\multirow{4}{*}[-9.5ex]{$\mathsf{G.S.}$}} &
  \multicolumn{1}{c|}{Truncated Dirichlet kernel~\cite{huang2022provably} } &
  \multicolumn{1}{c|}{$\mathsf{M.L.P}$} &
  \multicolumn{1}{c|}{$\mathbb{E}_{\bx}\|\nabla_{\bx} \Tr(O\rho(\bx))\|^2_2 \leq C$} &
  \multicolumn{1}{c|}{$\mathcal{O}(d^{\mathcal{O}(C/\epsilon)})$} &
  \begin{tabular}[c]{@{}c@{}}Simulation\\ 2D Heisenberg models (25Q)\\ Rydberg atom systems (51Q)\end{tabular} \\ \cmidrule(l){2-6} 
\multicolumn{1}{c|}{} &
  \multicolumn{1}{c|}{Lasso~\cite{lewis2024improved}} &
  \multicolumn{1}{c|}{$\mathsf{M.L.P}$} &
  \multicolumn{1}{c|}{Gapped geometrically local $\ham(\bx)$} &
  \multicolumn{1}{c|}{$\mathcal{O}(N\log(N)\text{poly}(\epsilon^{-1}))$} &
  \begin{tabular}[c]{@{}c@{}}Simulation\\ 2D Heisenberg models (45Q)\end{tabular} \\ \cmidrule(l){2-6} 
\multicolumn{1}{c|}{} &
  \multicolumn{1}{c|}{\multirow{2}{*}{Benchmark~\cite{cho2024machine} }} &
  \multicolumn{1}{c|}{$\mathsf{M.L.P}$} &
  \multicolumn{1}{c|}{-} &
  \multicolumn{1}{c|}{-} &
  \begin{tabular}[c]{@{}c@{}}Experiments\\ Random hopping system (12Q)\\ Su-Schrieffer-Heeger system (12Q)\end{tabular} \\ \cmidrule(l){6-6} 
\multicolumn{1}{c|}{} &
  \multicolumn{1}{c|}{} &
  \multicolumn{1}{c|}{$\mathsf{P.C}$} &
  \multicolumn{1}{c|}{-} &
  \multicolumn{1}{c|}{-} &
  \begin{tabular}[c]{@{}c@{}}Experiments\\ Topologically ordered phase (25Q)\\ Symmetry protected topological phase (41Q)\end{tabular} \\ \midrule
\multicolumn{1}{c|}{\multirow{2}{*}[-3.5ex]{$\mathsf{Q.C.}$}} &
  \multicolumn{1}{l|}{Fourier-based classical surrogates~\cite{schreiber2022classical}} &
  \multicolumn{1}{c|}{$\mathsf{S.L.P}$} &
  \multicolumn{1}{c|}{$\bomega_{\max}=\max\{|\bomega_i|:\bomega\in \Omega\}$ is small} &
  \multicolumn{1}{c|}{$\mathcal{O}(\bomega_{\max}^d |\Omega|^2 \epsilon^{-2} + |\Omega|^3\epsilon^{-2})$} &
  \begin{tabular}[c]{@{}c@{}}Simulation\\ California housing dataset (8Q)\end{tabular} \\ \cmidrule(l){2-6} 
\multicolumn{1}{c|}{} &
  \multicolumn{1}{c|}{\begin{tabular}[c]{@{}c@{}}Truncated trigonometric\\ monomial kernel~\cite{du2025efficient} \end{tabular}} &
  \multicolumn{1}{c|}{$\mathsf{M.L.P}$} &
  \multicolumn{1}{c|}{$\mathbb{E}_{\bx}\|\nabla_{\bx} \Tr(O\rho(\bx))\|^2_2 \leq C$} &
  \multicolumn{1}{c|}{$\mathcal{O}(T9^K|\mathfrak{E}(4C/\epsilon)|^2\epsilon^{-1})$} &
  \begin{tabular}[c]{@{}c@{}}Simulation\\ Pre-train VQE on TFIM (50Q)\\ Two-point correlation of GHZ (60Q)\end{tabular} \\ \bottomrule
\end{tabular}%
}
   \caption{\small{Summary of representative results on characterizing scalable quantum systems with ML models. The types of the explored quantum systems are denoted by $\mathsf{G.S.}$ for ground states and $\mathsf{Q.C.}$ for states prepared by digital quantum computers. Multiple linear property prediction and single linear property prediction are denoted $\mathsf{M.L.P}$ and $\mathsf{S.L.P}$, respectively. The task of phase classification is denoted by $\mathsf{P.C}$. The notation $a$Q denotes the number of qubits is $a$.}}
\label{tab:ML-summary-work}
\end{table*}

\subsection{Advanced topics}\label{subsec:ML-advance}
Existing ML models for representing and characterizing scalable quantum systems predominantly follow the supervised learning paradigm. However, in phase classification tasks, a distinct line of research employs \textit{unsupervised learning} algorithms (see Box~\ref{mybox:one}) to accomplish the learning objective~\cite{huang2022provably,sadoune2023unsupervised,che2025quantum}. For example, principal component analysis has been applied to identify different 
quantum phases of matter, inspired by
research questions in condensed matter physics, in a $300$-qubit bond-alternating XXZ model~\cite{huang2022provably}, while tensorial-kernel support vector machines have been used to reconstruct the phase diagram of a cluster-Ising model~\cite{sadoune2023unsupervised}. For a comprehensive review of unsupervised learning methods in phase classification, refer to Refs.~\cite{carleo2019machine,CUP}.

On par with the application of ML models for predicting linear properties of quantum states generated by digital quantum computers, another research direction is developing provably efficient learning approaches for specific quantum states, unitary operations, and quantum processes. For quantum state learning, certain restricted classes of states, such as stabilizer states~\cite{rocchetto2017stabiliser}, $t$-doped stabilizer states~\cite{grewal2023efficient,leone2024learning}, and states prepared by shallow circuits~\cite{landau2024learning} can be efficiently learned within a polynomial runtime. For unitary learning, a polynomial-time classical algorithm can reconstruct the description of an arbitrary unknown $N$-qubit shallow quantum circuit~\cite{huang2024learning}. Last, for quantum processes, efficient ML models can learn to predict any local property of the output of an unknown process, with a small average error over input states drawn from some specific prior distributions~\cite{huang2023learning,chen2024predicting}.   
 
Although the aforementioned works are partially related to the application of ML in scalable quantum systems, we do not elaborate on them here for two reasons. First, these problems can be seen as simplified variants of quantum state/process reconstruction, either without classical controls or with controls limited to the input state. For instance, while reconstructing an unknown quantum state typically requires exponential runtime, the problem can be reformulated into more tractable learning scenarios with efficient algorithms, such as shadow tomography~\cite{aaronson2018shadow}. Second, Ref.~\cite{anshu2024survey} has already provided a comprehensive review of these approaches.

\section{Deep learning paradigm}

The emergence of \emph{deep learning} (DL) in the early 2010s~\cite{pouyanfar2018survey} has opened new avenues for representing and characterizing scalable quantum systems. Through harnessing the representational power of deep neural networks, DL models can implicitly capture complex patterns and structures from data, exhibiting strong empirical performance across a broad spectrum of tasks. To better elaborate on the progress of DL models in the field, here we first recap the general framework when applying DL to scalable quantum systems, followed by presenting the major applications of current DL models and advanced topics. 

\subsection{General schemes}\label{subsec:DLscheme}

DL models employ \emph{deep neural networks} (DNNs) to automatically and implicitly extract meaningful representations from training data~\cite{goodfellow2016deep}. Existing DL models have been employed across diverse tasks in property prediction and reconstruction tasks, as summarized in Fig.~\ref{fig:scheme}. This stands in sharp contrast to prior ML approaches, which are typically tailored for measurement-agnostic prediction of linear properties. Depending on the specific task and learning objective, the implementation of DL models can differ markedly---ranging from measurement-agnostic to measurement-based protocols, and varying in the incorporation of auxiliary information. In the following, we delineate these distinctions, outlining how DL models are adapted to the diverse tasks of representing and characterizing scalable quantum systems.

\smallskip
\noindent\textbf{Dataset construction}. From the perspective of learning paradigms, property prediction falls under discriminative learning, while state reconstruction is typically framed as generative learning, as summarized in Box~\ref{mybox:one}. This distinction results in different approaches to training dataset construction: property prediction tasks employ a variety of data preprocessing strategies tailored to specific models and objectives, whereas reconstruction tasks generally follow a more standardized and unified methodology.

\smallskip
\noindent\textit{Property prediction.} 
The most general way in this context involves employing a single DL model to predict multiple linear and nonlinear properties of a given family of quantum states. To enable this, the raw data $\mathcal{T}$ in Eq.~(\ref{eqn:dataset-uni}) is reformatted into a labeled training dataset $\TDL$, consistent with the multi-task discriminative learning framework. The construction of  $\TDL$ can be categorized based on whether the DL model operates under a measurement-agnostic or measurement-based protocol, as illustrated in  Fig.~\ref{fig:pipeline}.

Most DL-based approaches fall into the category of measurement-based protocols, where the training dataset $\TDL$ incorporates measurement outcomes $\bsi$ from $\mathcal{T}$ as part of the input. Two commonly studied scenarios arise in this setting. In the first scenario~\cite{zhang2021,zhu2022,xiao2022,wu2024learning,koutny2023deep,PhysRevLett.132.220202}, only the measurement data is available, while the underlying physical parameters $\bxi$ are either unknown or inaccessible. In this case, the dataset is constructed as $\TDL=\{(\tilde{\bm s}^{(i)}, \bhatyi)\}$, where $\bhatyi$ denotes the estimates of physical properties of interest calculated from the measurement outcomes $\bsi$. Notably, the raw measurement outcomes $\bsi$ must be appropriately processed into a representation compatible with DL architectures, denoted by $\tilde{\bm{s}}^{(i)}$. In the second scenario~\cite{mohseni2022deep,qian2023multimodal}, the physical parameters $\bx$ are controlled by the learner, and the training dataset takes the form $\TDL=\{(\bxi, \tilde{\bm s}^{(i)}, \bhatyi)\}$. 

For measurement-agnostic protocols~\cite{wang2022quest,mohseni2024deep}, the training dataset excludes explicit measurement information. Given access to the raw data $\mathcal{T}$, the preprocessed training dataset takes the form $\TDL=\{\bxi, \bzi, \bhatyi\}$, where the construction of the label $\bhatyi$ follows the same procedure as in measurement-based DL models. A key distinction among DL models in this setting lies in whether (and how) 
auxiliary information $\bzi$ is incorporated, which is either omitted or used to encode system-specific details, such as gate layouts or noise features of the quantum system~\cite{wang2022quest}.

\smallskip
\noindent\textit{Single state reconstruction.} Unlike property prediction tasks, most DL models for quantum state reconstruction~\cite{torlai2018,carrasquilla2019reconstructing,cha2021attention,zhong2022quantum,ahmed2021quantum,smith2021,schmale2022efficient} follow a simple and standardized approach to dataset construction. Since this task falls under the generative modeling paradigm, the training dataset is typically unlabeled and takes the form
\begin{equation}\label{eqn:dataset-NQS}
    \mathcal{T}
    \mapsto
    \mathcal{T}_{\text{DL}} = \left\{\bm{s}_t\right\}_{t=1}^T.
\end{equation}
For a POVM measurement $\{M_{\bm{s}}\}$, the corresponding outcomes $\{\bm{s}_t\}$ are sampled from the probability distribution $\PP(\bm{s})= \Tr(\rho M_{\bm{s}})$. 
 
\smallskip
\noindent\textbf{Model implementation and training}. DL models for property prediction are typically formulated within a discriminative learning framework. Depending on the learning protocol, the employed DNN is denoted by $h_{\text{DL}}(\bm{x}, \tilde{\bm{s}}; \btheta)$ for measurement-based protocols, and $h_{\text{DL}}(\bm{x}, \bm{z};\btheta)$ for measurement-agnostic models, where $\btheta$ represents the trainable parameters. The objective is to optimize these parameters by minimizing the empirical loss function 
\begin{equation}\label{eqn:loss_DL}
	\mathcal{L}(\btheta) = \frac{1}{n} \sum_{i=1}^n \ell\big(h_{\text{DL}}(\bxi, \bzi, \tilde{\bm{s}}^{(i)};\btheta), \hat{\bm{y}}^{(i)}\big),
\end{equation}
where each component $\bxi, \bzi$ and  $\tilde{\bm s}^{(i)}$ is included or not depending on the available $\TDL$, and $\ell(\cdot, \cdot)$ denotes a task-specific loss, {\em e.g.}, the mean squared error for regression or cross-entropy for classification. The optimization is typically performed using gradient-based optimizers.
 
While prior work on property prediction largely follows the discriminative learning paradigm, the implementation of DL models diverges along two key directions. First, numerous studies~\cite{zhu2022,qian2023multimodal,tang2024ssl4q} focus on developing specialized neural architectures and optimization strategies to improve data efficiency, enabling models to accurately predict a broader range of physical properties ({\em i.e.}, a high-dimensional $\bhaty$) from a limited number of training examples $n$. These architectures are typically composed of modular components, such as fully connected layers, convolutional layers, or \emph{graph neural networks} (GNNs), tailored to the specific structure and modality of the dataset $\TDL$. Second, a growing body of work~\cite{mello2024retrieving, chen2023certifying} expands DL applications beyond standard benchmarks by tackling property prediction tasks unexplored in previous literature.  
 
The DL models for reconstructing a single quantum state are commonly referred to as \emph{neural network quantum states} (NQS)~\cite{lange2024architectures}. Existing approaches in this regime can be categorized into two classes: \textit{explicit} reconstruction and \textit{implicit} reconstruction. The primary distinction lies in the output representation of DNNs. In explicit reconstruction, the DNN directly outputs the full classical description of the
density matrix of the target quantum state~\cite{cha2021attention,ahmed2021quantum,du2023shadownet}. However, this approach suffers from exponential memory scaling with qubit number $N$, making it impractical for scalable systems. We, therefore, do not emphasize explicit reconstruction in this review. 

In contrast, and importantly, implicit reconstruction methods emulate the behavior of a quantum state without explicitly reconstructing its full density matrix, the formal description of which is provided in Box~\ref{mybox:state-reconstruction}. These approaches can be further categorized into two paradigms. The first develops DL models that take a measurement basis as input and output the corresponding measurement outcome probabilities~\cite{schmale2022efficient,smith2021}. The second paradigm, which has been more widely studied, treats DNN as a generative model, particularly in the form of autoregressive architectures, like \emph{recurrent neural networks} (RNNs)~\cite{goodfellow2016deep} and transformers~\cite{vaswani2017attention} (see Box~\ref{mybox:Terms-DL}). 

\begin{figure}
\refstepcounter{mybox}
\begin{tcolorbox}[
title={Box~\themybox: Implicit state reconstruction}, center title, before upper={\normalsize\justifying}]
\label{mybox:state-reconstruction}
Implicit state reconstruction refers to the task of learning a generative model that acts as a parametrized distribution $\mathbb{Q}(\bs; \btheta)$, with the goal of optimizing $\btheta$ so that $\mathbb{Q}(\bs; \btheta)$ closely approximates the target distribution $\mathbb{P}(\bs) = \Tr(\rho(\bx) M_s)$ over measurement outcomes $\bs$. Here, $\mathcal{M} = \{M_s\}$ denotes a predefined set of POVM elements, such as the one corresponding to computational basis measurements. This approach enables the model to reproduce the measurement statistics of the quantum state $\rho(\bx)$ without explicitly reconstructing its density matrix.
\end{tcolorbox}
\end{figure}

When an autoregressive model $h_{\text{DL}}(\btheta)$ is employed for quantum state reconstruction, it factorizes the joint probability distribution of measurement outcomes into a product of conditional probabilities using the chain rule. Specifically, the distribution is expressed as $\QQ(\bm{s}) = \prod_{i=1}^N Q(\bm{s}_i|\bm{s}_{<i};\btheta)$, where $\bm{s}_{<i}$ denotes the sequence of bits preceding index $i$. After this reformulation, the model parameters $\btheta$ in $h_{\text{DL}}(\btheta)$ are optimized by minimizing the negative log-likelihood loss
\begin{equation}\label{eqn:state-reconstruction}
	\mathcal{L}(\btheta) = -\frac{1}{|\TDL|} \sum_{\bm{s}\in \TDL} \log \Bigg(\prod_{i=1}^N Q(\bm{s}_i|\bm{s}_{<i};\btheta)\Bigg).
\end{equation}
This loss function incentivizes the model to assign higher probability to configurations that align closely with the observed measurement outcomes.
 
When DL models are used in quantum state reconstruction, their performance is evaluated by the similarity between the learned distribution $\QQ$ and the true measurement outcome distribution $\PP$. Unlike property prediction tasks that employ a standardized accuracy metric (Eq.~(\ref{eqn:off-exp-risk})), there is no standard metric in state reconstruction. Common performance measures include \emph{Kullback-Leibler} (KL) divergence, total variation distance, and Wasserstein distance. The learning model employed is considered to be efficient if the number of training examples, the total queries of quantum systems, and the computational complexity scale at most polynomially with the qubit count $N$ to achieve an $\epsilon$-accurate estimation between $\QQ$ and $\PP$.

\smallskip
\noindent\textbf{Model prediction}. Once trained, DL models can be applied to downstream prediction tasks based on their learning objectives. For property prediction, they are used to infer physical properties of previously unseen quantum states. For quantum state reconstruction, the trained DL model serves as a sampler, generating bit-string samples that faithfully reproduce the statistics of the target quantum state under the same measurement setting used during training. This enables efficient sampling from distributions over measurement outcomes 
without requiring direct access to the physical system.

\begin{figure}
    \centering
    \includegraphics[width=0.97\linewidth]{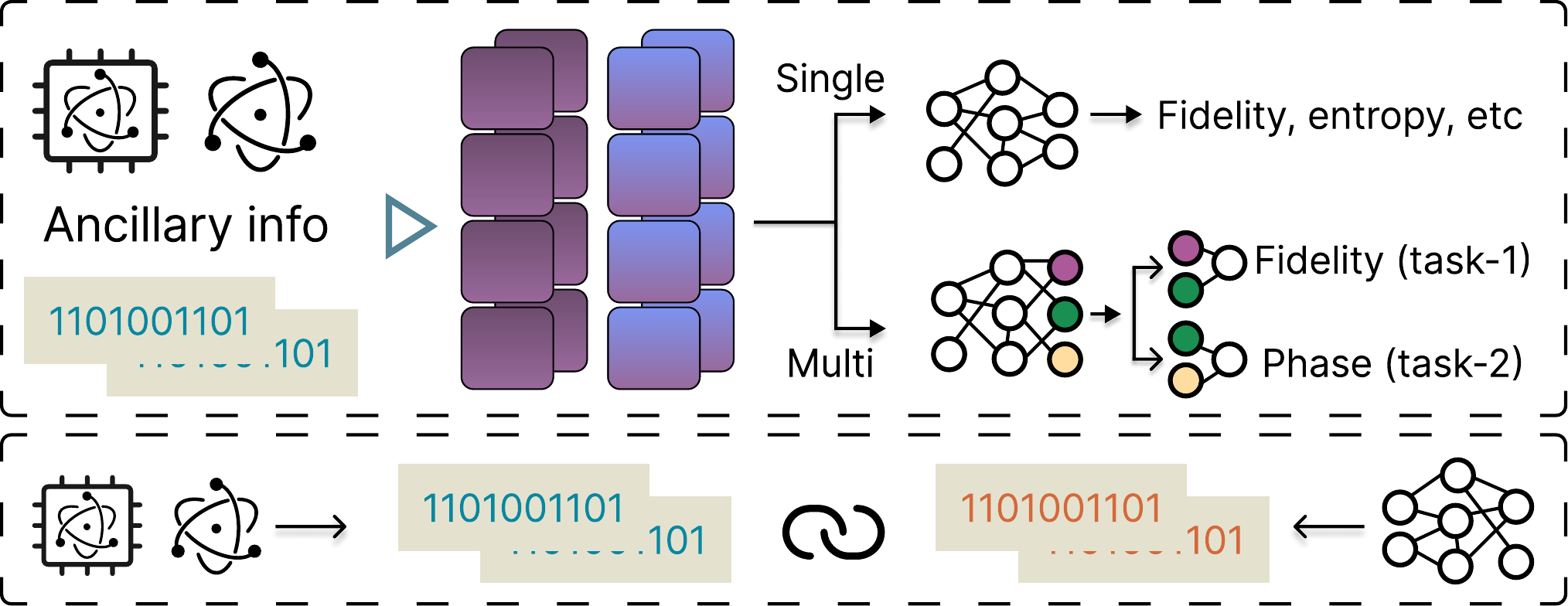}
    \caption{\small{\textbf{Schematic illustration of DL models for characterizing and representing quantum systems}. The upper panel illustrates the workflow of applying deep learning models to predict properties of quantum systems in both single-task and multi-task settings. In both cases, 
    auxiliary information and measurement outcomes are first preprocessed to ensure compatibility with neural network architectures. For single-task prediction, DL models are used to estimate specific properties, such as fidelity or entropy. In multi-task scenarios, a common strategy involves learning latent representations that enable multiple downstream tasks, including fidelity prediction and phase classification. The lower panel depicts the use of DL models for implicit state reconstruction, where neural networks are trained to generate samples that closely replicate the measurement outcomes produced by quantum systems.}}
    \label{fig:DL}
\end{figure}

\begin{figure*}
\refstepcounter{mybox}
\begin{tcolorbox}[ 
title={Box~\themybox: Terminologies in DL models}, center title, before upper={\normalsize\justifying}]
\label{mybox:Terms-DL}

\centerline{\textbf{Latent representation} }

The latent representation refers to a set of internal variables or features---often learned automatically by deep learning models---that summarize the most relevant information for downstream tasks. By operating on these latent representations, models can perform property prediction or classification more efficiently and with improved generalization.

\centerline{\textbf{Transfer learning  and few-shot learning} }

These paradigms aim to predict properties of new quantum systems with minimal quantum computational cost by leveraging knowledge from previously learned tasks or models. In particular, \textit{transfer learning} adapts a model trained on one task to improve performance on a related task, while \textit{few-shot learning} enables models to generalize from only a handful of labeled examples. Both approaches are crucial for reducing the number of required accesses to quantum systems in scenarios where data acquisition is expensive.

\smallskip

\centerline{\textbf{Multimodal learning} }
Multimodal learning refers to the ability of models to process and combine information from multiple distinct sources or data types~\cite{baltruvsaitis2018multimodal}. Multimodal learning models often employ neural architectures with dedicated modules for each data modality---such as CNNs for image-like data and RNNs for sequential data---which are subsequently fused through joint layers to enable effective information integration and cross-modal reasoning. When applied to characterize quantum systems, this approach integrates diverse types of data, such as quantum measurement outcomes and physical control parameters, to construct more comprehensive representations of quantum systems.

 \smallskip

\centerline{\textbf{Deep generative models} }
Deep generative models are deep neural networks that learn to generate new data samples similar to those seen during training. When applied to implicit quantum state reconstruction tasks, these models aim to learn the underlying probability distribution of quantum measurement outcomes. Two major realizations are:
\begin{itemize}
	\item Autoregressive models. These factorize the joint probability distribution of measurement outcomes into a product of conditional probabilities using the chain rule. 
	Examples include \emph{recurrent neural networks} (RNNs), PixelCNNs, and transformers for neural quantum state sampling.
	\item Energy-based models. These assign an unnormalized energy to each possible configuration and define probabilities through an energy function. The model is trained to lower the energy for observed samples, with normalization performed over all configurations. Unlike autoregressive models, energy-based models do not inherently require sequential sampling.  Notable examples include \emph{restricted Boltzmann machines} (RBMs) and \emph{deep Boltzmann machines} (DBMs). 
\end{itemize}
\end{tcolorbox}
\end{figure*}

\subsection{Concrete DL models and applications}\label{subsec:DLapplications}

In the following, we present recent advancements in DL models for quantum property prediction, quantum system reconstruction and quantum computing. For clarity, we further distinguish between single-property and multi-property prediction tasks within the property prediction category. Some representative works in this regime are summarized in Table~\ref{tab:DL-LM-summary-work}.

\subsubsection{Predicting specific quantum properties}
\label{subsubsec:specific}
Unlike ML models, which are primarily designed for linear property prediction, the expressive capacity of neural networks enables the development of DL models tailored to predict a given property, whether linear or nonlinear, as illustrated in Fig.~\ref{fig:scheme}.	

\smallskip
\noindent\textbf{Quantum state similarity}. Quantum state similarity, like quantum fidelity, refers to a quantity to describe the closeness between quantum states~\cite{nielsen2010quantum}. Detecting quantum state similarity is important for verifying the reliability of scalable quantum processors. Two important tasks of quantum similarity detection include direct quantum fidelity estimation~\cite{PhysRevLett.106.230501} and cross-platform quantum verification~\cite{elben2020cross}. In the former task, a learner performs measurements on copies of an experimental state to estimate its quantum fidelity with respect to a theoretical target state. In the latter one, the learner aims to estimate the quantum similarity between two uncharacterized experimental quantum states via performing local measurements on each state. 

Direct fidelity estimation seeks to quantify the similarity between an experimentally prepared state and a target pure state while minimizing measurement overhead. An initial approach to this task employs a simple \emph{fully connected neural network} (FCNN) to perform multi-class classification~\cite{zhang2021}. Intuitively, the FCNN takes as input a subset of statistical estimates of Pauli operator expectation values measured from quantum states and outputs a fidelity prediction. The training labels are generated by discretizing the fidelity into predefined intervals, which is computed by using an idealized infinite measurement scenario.  Rather than formulating fidelity estimation as a classification problem, a follow-up study proposed to employ regression models to estimate the fidelity of experimental states from measurement data~\cite{PhysRevLett.132.190801} as well as from physical parameters of the explored quantum system~\cite{du2023shadownet}. 

Cross-platform quantum verification directly compares quantum states experimentally produced on separate devices. There are several different approaches for this task. The first formats the resulting measurement outcome distributions on different bases into structured input tensors, which are then processed by a \emph{convolutional neural network}  (CNN) for feature extraction and similarity comparison~\cite{wu2023}. The second approach utilizes quantum circuit layouts as inputs, employing GNNs to analyze circuit structures and predict quantum fidelity~\cite{wang2022quest}. In light of multi-modal learning (refer to Box~\ref{mybox:Terms-DL} for explanation of multi-modal learning), a recent study advocates for synthesizing the circuit layouts with their measurement data to estimate the overlap between two states. Simulation results up to $50$ qubits exhibit the effectiveness of this approach~\cite{qian2023multimodal}. Another attempt has been made to estimate the quantum fidelity between two arbitrary quantum states. The proposed approach employs a CNN, which takes a one-hot encoded representation of quantum circuits as input to predict fidelity~\cite{vadali2024quantum}.
 
\smallskip
\noindent\textbf{Entanglement and other nonclassical features}. Quantum entanglement~\cite{horodecki2009quantum} is a fundamental feature that sets quantum mechanics apart from classical physics. Yet,  detecting and quantifying entanglement in an arbitrary quantum system is not only quantum resource-intensive but also computationally intractable~\cite{gurvits2003classical}. Recent advances in DL models provide a promising alternative approach, enabling efficient predictions of entanglement for particular classes of quantum states without the need for direct estimation from first principles.

 An early effort toward solving this task utilizes a simple FCNN to classify entangled and separable bipartite states~\cite{PhysRevLett.120.240501}. Conceptually, the FCNN takes statistical estimates of Pauli operator expectation values as input and outputs a prediction. Later research developed more advanced DL models to detect multipartite entanglement in multi-qubit states~\cite{harney2020entanglement,chen2021detecting,chen2021entanglement,roik2021accuracy,chen2023certifying,asif2023entanglement,luo2023detecting,kookani2024xpookynet,pawlowski2024identification}. A recent work has employed LSTM to predict the entanglement entropy of subsystems in a dynamically evolving quantum system, scaling up to $100$ qubits, using only single- and two-qubit measurements~\cite{huang2025direct}. In addition, specialized DL models have been designed for entanglement detection in continuous-variable quantum systems~\cite{PhysRevLett.132.220202}. Alongside entanglement detection, recent studies began to explore DL models capable of quantifying entanglement~\cite{roik2022entanglement,koutny2023deep,denis2023estimation,lin2023quantifying,rieger2024sample}. Extending beyond entanglement, studies have begun to design DL models for quantifying broader types of quantum features, such as nonclassicality~\cite{PhysRevLett.125.160504}, quantum discord~\cite{taghadomi2024effective,krawczyk2024data}, and non-stabilizerness~\cite{mello2024retrieving,sinibaldi2025non}.
 
\smallskip \noindent\textbf{Phase classification}. Phase transitions are fundamental phenomena prevalent in many-body physics, where small perturbations of specific physical parameters can cause significant changes in a system's behavior. Phase classification does not forcibly require access to Hamiltonian parameters $\bm x$. However, when $\bm x$ is accessible, classifying different phases enables the identification of phase transition critical points. When the collected raw data $\mathcal{T}$ corresponds to a set of example ground state vectors 
$\{\ket{\psi(\bxi)}\}_{i=1}^n$, the optimized learning model can identify the critical point $\bx^*$, where $\ket{\psi(\bx^*)}$ undergoes a quantum phase transition. 

DL models have been widely employed to identify critical points or phase boundaries in parameter spaces for both classical and quantum phase transitions. In the supervised learning paradigm, a sufficient number of measurements $T$ or auxiliary information $\bzi$ is necessary for each training example, indicating which phase of matter this example belongs to~\cite{carrasquilla2017machine}. In contrast, both learning by confusion~\cite{van2017learning,liu2018} and learning by prediction~\cite{PhysRevE.99.062107,greplova2020unsupervised,PhysRevResearch.3.033052} models do not rely on any prior auxiliary information.  
 
For the purpose of discovering unknown phases, a DL model based on anomaly detection~\cite{PhysRevLett.125.170603} identifies potential new phases of matter with minimal or no prior data from these unknown phases. Rather than testing on simulated data, DL models have also been applied for classifying various quantum phases and generating entire phase diagrams from experimental data~\cite{miles2021correlator,PhysRevLett.127.150504,PhysRevResearch.5.013026,kim2024attention}. Reliability of DL models for classifying phases of matter, especially the existence of adversarial examples, has also been investigated~\cite{zhang2022,jiang2023adversarial}.

Phase transitions also occur in random quantum circuits with mid-circuit measurements~\cite{PhysRevX.9.031009}. In such quantum systems, slight fluctuations in the randomized measurement rates near the critical point induce drastic changes in the entanglement entropy of the output quantum state. Learnability of such quantum systems has become an important probe for detecting measurement-induced phase transitions~\cite{PhysRevLett.129.200602,PRXQuantum.5.020304,PhysRevB.109.094209,PhysRevX.14.041012}.  By training a CNN to predict Pauli expectation values of a reference qubit from mid-circuit measurement snapshots, the prediction accuracy effectively identifies the phase transition~\cite{dehghani2023neural}. A recent work applies an attention-based model to identify measurement-induced phase transition by distinguishing two different scrambled states from their measurement trajectories~\cite{kim2025learning}. 

In addition to the works discussed above, DL models have been developed to classify phases of matter by leveraging classical prior knowledge of quantum systems, such as entanglement spectra~\cite{PhysRevB.95.245134}, rather than based on measurement data. Besides the supervised learning paradigm, classifying different phases of matter can be achieved in an unsupervised manner by performing dimension reduction of raw data with a clustering algorithm~\cite{PhysRevB.94.195105}, and analysis of the bottleneck of an autoencoder~\cite{PhysRevE.96.022140}.  

\subsubsection{Predicting multiple quantum properties}
\label{subsubsec:multi}
Quantum shadow tomography enables the effective estimation of the expectation values of many observables without performing full state tomography~\cite{aaronson2018shadow}. Inspired by this idea, DL models have been developed to predict multiple physical properties simultaneously. These models learn transferable \textit{latent representations}~\cite{bengio2013representation} from the dataset $\TDL$ (refer to Box~\ref{mybox:Terms-DL} for a detailed explanation of latent representations and transfer learning), capturing structural patterns and physical correlations within quantum states. Once trained, these latent representations can be used to infer diverse physical properties beyond those seen during training. In this sense, the latent representations serve an analogous role to classical shadows, acting as compressed but information-rich summaries of quantum states for downstream prediction tasks.

Depending on how these latent representations are constructed, existing DL models can be classified into \textit{supervised}, \textit{semi-supervised}, and \textit{self-supervised representation learning} (refer to Box~\ref{mybox:Terms-DL} for their different concepts). We next elucidate the first two categories, while the topic of self-supervised learning will be addressed later in the context of language model paradigms. 
 
Existing DL models for supervised and semi-supervised learning typically follow the measurement-based protocol, where measurement outcomes $\{\bsi\}$ serve as input data in $\TDL$. The primary differences among these DL models lie in how they construct latent representations and the associated cost of acquiring labeled examples.

In the category of supervised representation learning, the seminal work is leveraging the concept of \emph{generative query neural networks} (GQNN) to learn data-driven representations of quantum states~\cite{zhu2022}.  The learned latent representations enable predictions of measurement statistics over those measurement bases that have not been performed yet and clustering of distinct classes of quantum states. Follow-up works have explored how to learn generalizable representations that can be transferred to other tasks. Specifically, a classifier trained to distinguish quantum phases can reuse the learned latent representations to predict other physical properties, including entanglement entropy and quantum state overlap~\cite{xiao2022}. Moreover, DL models trained on predicting Pauli expectation values can be transferred to predicting entanglement entropy in a dynamical quantum system~\cite{mohseni2025transfer}.

A complementary approach within the supervised representation learning paradigm involves multi-task learning, in which DL models are trained to predict multiple quantum properties simultaneously. In particular, each training example is annotated with multiple labels, {\em i.e.}, the dimension of $\bhatyi$ in $\TDL$ is larger than 1, and the loss function in Eq.~(\ref{eqn:loss_DL}) calculates prediction errors across all target properties. A representative example has been
given by Wu et al.~\cite{wu2024learning}. They have demonstrated that for ground states of bond-alternating XXZ models, a DL model trained for predicting spin correlations and entropic mutual information from short-range measurements can also distinguish \emph{symmetry protected topological} (SPT) and trivial phases. This is achieved by applying dimensionality reduction on the learned representations, revealing phase distinctions without explicit phase supervision.

Semi-supervised representation learning leverages a small fraction of labeled data alongside a larger pool of unlabeled data~\cite{van2020survey}. A recent study has shown that reliable quantum property prediction, such as phase classification, remains achievable within this paradigm~\cite{tang2024ssl4q}. The key idea is to employ a hybrid loss function that combines a supervised loss on labeled data with an unsupervised contrastive loss. The contrastive loss enforces similarity in latent representations for quantum states with comparable measurement statistics~\cite{wu2023}.

\subsubsection{Quantum system reconstruction}
\label{subsubsec:reconstruct}

Although full reconstruction of an arbitrary quantum state in an explicit manner is implausible for large-scale quantum systems, numerous DL models have been developed to efficiently reconstruct well-structured quantum systems implicitly.
This includes not only the implicit reconstruction of quantum states with generative models, but also the prediction of output states of quantum dynamics or their physical properties, and learning to predict the Hamiltonians of quantum systems.

\noindent\textbf{Implicit quantum state reconstruction}.  
For implicit state reconstruction, current studies on NQS, which formulate DNNs as generative models, adopt two approaches: data-driven and variational. The data-driven NQS approaches typically employ either autoregressive models or energy-based models.

As introduced in the model implementation part, one widely studied class of generative models for NQS is based on autoregressive architectures~\cite{carrasquilla2019reconstructing, PhysRevLett.122.080602, PhysRevLett.124.020503}. For example, RNNs and their variants have been successfully employed to implicitly reconstruct unknown quantum states from measurement data~\cite{carrasquilla2019reconstructing,PhysRevA.104.012401,niu2020learnability}. These DL models are valued for their strong expressive capacity in capturing complex quantum correlations. A notable recent development is the use of transformer architectures, which excel at modeling long-range dependencies in sequential data. For instance, Refs.~\cite{cha2021attention,zhong2022quantum} proposed transformer-based learning models to reconstruct GHZ states and the ground states of transverse-field Ising models.  

Another major approach to building NQS, developed earlier than autoregressive models, is based on energy-based generative models. These models assign an unnormalized energy to each possible configuration, with sampling guided by the principle that configurations with lower energy correspond to higher probability. Sampling is typically performed using techniques such as Markov Chain Monte Carlo. A seminal example in this category is the \emph{restricted Boltzmann machine} (RBM)~\cite{hinton2012practical}, which was among the first architectures applied to quantum state reconstruction~\cite{torlai2018,tiunov2020experimental}. 

For variational methods, the employed DNNs are treated as variational ans{\"a}tze to approximate ground states of Hamiltonians~\cite{carleo2017solving}. Unlike the data-driven approaches, this method is specifically tailored for ground-state estimation and does not require access to quantum measurement data. The training objective is to minimize the expectation value of a target Hamiltonian $\mathsf{H}(\bx)$, thereby searching for the lowest-energy state within the expressive function space defined by the DNN. After training, the output of the DNN mimics the measurement statistics of the ground state vector $\ket{\psi(\bx)}$. 

This variational method can be combined with the above autoregressive models trained with measurement data to improve the accuracy of quantum simulations~\cite{PhysRevB.105.205108,moss2023enhancing,lange2025transformer}. In this approach, an autoregressive model is first trained to reconstruct an approximated ground state, prepared by a noisy quantum simulator, from its experimental measurement data. After that, variational optimization is performed to further approach the ideal ground state. This hybrid approach has been demonstrated to be more efficient than variational methods and robust to experimental errors. 

Since this review focuses on  AI applications for learning quantum systems from measurements, it does not provide comprehensive coverage of all works on NQS. For a complete overview of NQS, readers are referred to Refs.~\cite{carrasquilla2021use,CUP,lange2024architectures}.

An intermediate approach between explicit and implicit recovery, which merits brief mention here, is the explicit construction of a quantum state under specific structural assumptions. Quantum states that feature low degrees of entanglement over all cuts of the system in a precise sense \cite{eisert2010colloquium} can be well approximated by \emph{tensor network states} \cite{RevModPhys.93.045003}. Once a tensor network state has been learned, one can sample from a distribution that closely approximates the target distribution
$\mathbb{P}(\bs)$. The situation at hand is particularly clear in one spatial dimension
and pure states, referred to as \emph{matrix product states}, but generalizations to locally purified mixed states have also been considered. Early approaches have focused
on the practical recovery from local or suitable randomized global measurements \cite{Cramer-NatComm-2010,Efficient}, while newer work provides sample complexity bounds for the
rigorous learning of tensor networks from 
appropriate -- usually randomized -- data
\cite{RigorousTensorNetworkTomography1,RigorousTensorNetworkTomography2,anshu2024survey}.
Such methods are closely related to classical shadows as well.

\smallskip
\noindent\textbf{Predicting quantum dynamics}. In quantum dynamics, the relevant parametrized quantum state vectors become $\ket{\psi(\bx; t)} \coloneqq e^{-\imath H(\bm x) t}\ket{\psi_0}$, where $\bx$ denotes the classical controls of the explored Hamiltonian and $t$ refers to the evolution time. Given a measurement dataset of $\ket{\psi(\bx; t)}$ at multiple times corresponding to different control parameters, the objective is to use DL models to predict the state vector $\ket{\psi(\bx'; t')}$ or its physical properties at a future time $t'$. Due to the time-series nature of the problem, sequential DL models with the ability to capture temporal dependencies are naturally suited to address such tasks. 

An early effort towards addressing this task utilizes RNN and LSTM models to predict the expectation values of observables for time-evolved state vectors $\ket{\psi(\bx; t)}$ in spin models~\cite{mohseni2022deep}. By taking both the spin-system parameters and a sequence of past measurement data as input, the models output the predicted expectation values at future time steps. Following the same routine, sequential DL models have been applied to predict the expectation values of observables for time-evolved states generated by quantum circuits~\cite{mohseni2024deep}.  A recent bidirectional DL model can not only predict the expectation values of observables in a dynamical evolution based on its Hamiltonian, but also predict the time-dependent Hamiltonian parameters from the associated dynamical observation data~\cite{PhysRevLett.134.120202}. Different from the above research line, inspired by ML models~\cite{huang2023learning}, a DL model is developed to emulate a quantum process by predicting the properties of the unknown quantum process outputs corresponding to any input state randomly drawn from a predetermined ensemble~\cite{zhu2023predictive}.

Besides predicting the dynamics of closed quantum systems, an active research direction focuses on employing DL models to model the dynamics of open quantum systems, with broad applications in quantum chemistry and drug discovery. As these topics fall beyond the scope of this review, refer to Refs.~\cite{hartmann2019neural,carrasquilla2020machine,harrington2022engineered,campaioli2024quantum} for further details.

\smallskip \noindent\textbf{Hamiltonian learning}. The Hamiltonian of a system lies at the heart of quantum physics, governing both the structure of quantum states and their dynamical evolution. Given its central role, a key research direction is Hamiltonian learning~\cite{wiebe2014hamiltonian,wang2017experimental}, which seeks to infer the underlying structure and estimate the coupling strengths of the Hamiltonian from measurement data. The learned model parameters not only provide insight into system dynamics but also support the characterization and certification of scalable quantum systems~\cite{eisert2020quantum,PRXQuantum.2.010102} and allow to enhance quantum simulations
with stronger predictive power.

Recent studies have introduced dedicated DL models to complete different Hamiltonian learning tasks with a minimal measurement overhead under certain structural assumptions on the Hamiltonians to be learned, most often including geometric locality. 
To name a few, FCNN trained on a small set of local measurement outcomes have been employed to learn and verify the structure of instances of stabilizer Hamiltonians, with potential applications in quantum error correction~\cite{PhysRevResearch.1.033092}. In the out-of-equilibrium regime, FCNNs have also been used to reconstruct Hamiltonians from dynamical measurement data~\cite{PhysRevA.105.023302,PhysRevResearch.6.023160}. 
Beyond feedforward architectures, RNNs have demonstrated promise in learning time-dependent Hamiltonian parameters, such as those of driven Ising models, by processing time-series data from single-qubit measurements~\cite{PhysRevResearch.3.023246}. Besides, a large-scale experimental study of Hamiltonian learning involving up to 27 qubits has been performed to learn the Hamiltonian of a 
superconducting quantum processor \cite{HamiltonianLearningGoogle}, using ideas of super-resolution such as tensorESPRIT and constrained manifold optimization. Such large-scale experimental studies also showcase how important it is
to develop robust methods of Hamiltonian learning that can accommodate state preparation and measurement errors.
 
\subsubsection{Applications in quantum computing}
Beyond the applications discussed earlier, initial efforts have begun to explore the potential of DL models across diverse quantum computing tasks. 

\smallskip
\noindent\textbf{Quantum system benchmarking}.  Benchmarking quantum processors at scale is crucial for realizing reliable quantum computing and quantum simulation~\cite{proctor2025benchmarking,eisert2020quantum}. 
Benchmarking methods build trust in the correct preparation of a quantum state or the accurate implementation of a quantum circuit or parts thereof. To enhance efficiency and scalability, a promising approach is employing DL models to predict the performance of specific quantum processors~\cite{hothem2024my,hartnett2024learning,hothem2024learning}. A concrete example involves employing an ensemble of DL models that take time, bond dimension of matrix product states, and system size as inputs to estimate the fidelity between experimental simulations and their classical counterparts~\cite{shaw2024benchmarking}. Once a system has been benchmarked, one can also often obtain actionable advice on how to improve the experimental setting at hand. Again, notions of benchmarking are increasingly being seen as belonging to quantum learning theory.

\smallskip
\noindent\textbf{Quantum error mitigation}. Notions of \emph{quantum error mitigation} (QEM) play a critical role in suppressing estimation errors arising from noise in quantum systems~\cite{RevModPhys.95.045005}. It is not a single method, but rather a portfolio of methods that operate largely on the classical level aimed at undoing
parts of the quantum noise. Based on their algorithmic strategies, existing QEM techniques can be categorized into non-learning and learning-based approaches. The non-learning category includes methods such as zero-noise extrapolation~\cite{temme2017error} and virtual distillation~\cite{huggins2021virtual}. In contrast, learning-based methods can be further divided into those employing traditional ML~\cite{czarnik2021error, czarnik2022improving, strikis2021learning} and those based on DL, with this discussion focusing primarily on the latter. Such methods play a crucial role in mitigating quantum noise in near-term experimental implementations, but face substantial obstructions in their
scalability, requiring a more than exponential sampling complexity in the size of the circuit \cite{ErrorMitigationObstructions,ErrorMitigationObstructionsOld}.
 
Current DL models applied to QEM belong to the measurement-based protocol, where the noisy measurement outcomes $\{\bsi\}$ are used as models' inputs, and the corresponding estimated expectation values serve as labels. Following this paradigm, FCNN, augmented with task-specific auxiliary information $\bzi$, has been developed to mitigate errors in various settings, including qubit noise spectroscopy~\cite{PRXQuantum.2.010316}, small-scale quantum circuits~\cite{kim2020quantum}, quantum approximate optimization algorithms~\cite{PhysRevResearch.6.013223}, and Hamiltonian simulation~\cite{zhukov2022quantum}. Beyond these applications, a DL model empowered by data augmentation has been introduced to enable training directly on hardware-generated data, improving adaptability to device-specific noise~\cite{liao2025noise}.  Moreover, non-message-passing graph transformers have been proposed to improve performance across circuit architectures and noise types~\cite{bao2025beyond}. In parallel with the direct application of DL models to predict error-mitigated observables, an alternative solution in ground state estimation involves using NQS to reconstruct an approximate ground state from noisy measurement data, which is then further optimized classically to minimize the energy with respect to the target Hamiltonian~\cite{bennewitz2022neural}.
 
\smallskip
\noindent\textbf{Quantum error correction}.
Although quantum error mitigation is crucial for modern quantum processors as an intermediate step, \emph{quantum error correction} (QEC) remains the definitive pathway toward fault-tolerant quantum computing. DL models have been successfully applied to QEC, particularly in the decoding procedure, where they in instances outperform conventional approaches. 
These applications can be categorized into identifying the occurrence of specific quantum errors from which the actual correcting operation can be deduced~\cite{Reinforcement,zhou2025learning,wang2023transformer,bausch2024learning} and generating quantum decoding configurations~\cite{cao2023qecgpt}, in a prescription in which after the training, the model can efficiently compute the likelihood of logical operators for any given syndrome. While both approaches use measurement syndromes as input to DL models, the former is typically framed as a classification task, similar to property prediction, whereas the latter falls under generative learning, analogous to implicit state reconstruction. A specific challenge to such learning-based decoders is constituted by faulty measurements in the context of fault-tolerant quantum memories, as this leads to a wrong identification of the syndrome. Notions of deep learning are also routinely being used to identify new quantum error correcting codes, but this application goes beyond the scope of this review.

\smallskip
\noindent\textbf{Enhanced variational quantum algorithms}. \emph{Variational quantum algorithms} (VQAs) remain an active area of research, with extensive theoretical analyses and a wide range of practical applications. Given the existence of comprehensive reviews on this topic~\cite{cerezo2021variational,bharti2021noisy,tian2022recent,tilly2022variational,du2025quantum}, we do not attempt an exhaustive survey here. Instead, we focus on recent advances in leveraging DL models to enhance the performance and scalability of VQAs. 

Different from directly predicting properties of quantum states output by digital quantum computers, DL-enhanced VQAs aim to improve optimization efficiency and circuit deployment. In terms of optimization, DL models have been developed to identify high-quality initial parameters~\cite{sauvage2021flip,jain2022graph,friedrich2022avoiding,lee2025q} and act as surrogate optimizers that predict gradient trajectories~\cite{verdon2019learning,khairy2020learning,cervera2021meta,huang2022learningto,luo2023quack}. To enhance performance, DNNs have been exploited to design data encoders and the gate layout of ans{\"a}tze~\cite{zhang2021neural,he2023gnn,qian2024mg}. Besides, reinforcement learning and diffusion models have been explored to discover compact gate sequences for circuit compilation, further facilitating the practical deployment~\cite{fosel2021quantum,furrutter2024quantum,ruiz2025quantum}.

\subsection{Advanced topics}

The black-box nature of deep learning poses great challenges in understanding the behavior of DL models when applied to scalable quantum systems. As a result, in contrast with ML models, the underlying principles of DL models remain largely unexplored. However, recent years have seen substantial progress in developing a theory of explainable artificial intelligence and allows for an enhanced understanding, visualizing and 
interpretation of DL models \cite{ExplainableAI}. A notable example is in the prediction of linear properties of ground states, where a DNN has been shown to offer provable guarantees on prediction accuracy~\cite{wanner2024predicting}. Moreover, in phase classification, when the employed DNN is sufficiently expressive, its output can be effectively replaced by a surrogate function with a closed-form solution, eliminating the need for explicit training~\cite{arnold2022replacing}. In the context of state reconstruction, several studies have investigated the expressivity of NQS, {\em i.e.}, the classes of quantum states that can be effectively represented by the underlying neural architecture. The achieved results indicate that the conditional correlation and entanglement entropy are dominant factors~\cite{gao2017efficient1,PhysRevB.106.205136,zhao2023provable,yang2024can}. Despite these advances, the inner workings of most DL models remain elusive, necessitating further research to uncover their underlying mechanisms.   

An alternative to address this issue is to develop interpretable DL models. A transparent and explainable DL model can not only enhance the reliability of predictions regarding quantum properties but also build greater confidence among physicists in leveraging it as a powerful tool for advancing scientific knowledge. Some progress has been made in this research direction~\cite{wang2019emergent,iten2020,PhysRevA.105.042403,flam2022learning,PhysRevB.106.L041110,frohnert2024explainable,zhang2024observing,cybinski2025characterizing,de2025interpretable}. The core principle of these works is revealing the relation between input data and the latent representations. Assisted by the dimension reduction algorithms, like the t-SNE algorithm~\cite{van2008visualizing}, one can visualize how data representations are distributed in the high-dimensional representation space. More recent progress on interpretable DL models for learning quantum systems can be found in this review~\cite{wetzel2025interpretable}.

Beyond enhancing interpretability, an equally critical research direction is establishing the transferability of DL models---extending their applicability from simple to complex systems and from small-scale to large-scale quantum regimes. Advancements in this area hold the potential to substantially reduce the quantum resources required for data collection, training, and prediction, enabling more scalable and efficient learning frameworks. Some initial efforts have been made in this direction. For instance, a DL model trained on data from one-dimensional Rydberg atoms of varying sizes has exhibited good performance in predicting the phase diagram of a larger system not seen during training, even though the true phase diagram is highly size-dependent~\cite{wang2022}.  
 
\begin{figure*}[t!]
\refstepcounter{mybox}
\begin{tcolorbox}[ 
title={Box~\themybox: Terminologies in LMs}, center title, before upper={\normalsize\justifying}]
\label{mybox:Terms-LM}
\centerline{\textbf{Foundation models} }
Foundation models are large-scale, general-purpose AI models trained on broad and diverse datasets using self-supervised learning objectives. These models, such as GPT, acquire versatile representations and capabilities that can be adapted to a wide range of downstream tasks with minimal additional training. Foundation models typically employ advanced neural architectures, such as deep transformers, that enable them to capture complex patterns and relationships across different modalities. Their flexible and transferable knowledge has made them a cornerstone of recent advances in artificial intelligence.

\smallskip
 
\centerline{\textbf{Transformer architecture} }

The transformer is a deep learning architecture based on self-attention mechanisms, designed to model dependencies within sequences regardless of their length or position. At its core, the transformer replaces recurrent structures with multi-head self-attention, enabling efficient parallelization and the capture of long-range correlations. For an input sequence, the self-attention module computes representations as weighted sums
\[
\mathrm{Attention}(\mathbf{Q}, \mathbf{K}, \mathbf{V}) = \mathrm{softmax}\left(\frac{\mathbf{Q}\mathbf{K}^\top}{\sqrt{d_k}}\right)\mathbf{V},\]
where $\mathbf{Q}$, $\mathbf{K}$, and $\mathbf{V}$ denote the trainable query, key, and value matrices derived from the input sequence, and $d_k$ is the feature dimension. This mechanism enables the transformer to effectively learn context-dependent relationships, making it the backbone of modern LLMs such as GPT.

\smallskip
\centerline{\textbf{Pretraining and finetuning} }
Pretraining and finetuning are a two-stage training paradigm widely used in LLMs such as GPT. In the pretraining stage, the model learns general linguistic patterns and representations from massive unlabeled text corpora using self-supervised objectives.  In quantum applications, this can involve learning measurement outcome distributions across diverse settings.  In the finetuning stage, the pretrained model is further adapted to specific tasks or domains by training on smaller, labeled datasets. This approach enables GPT models to achieve strong performance across a broad range of tasks. For example, fine-tuning a model on entanglement entropy data or specific quantum hardware noise profiles.
\end{tcolorbox}
\end{figure*}

\section{Language model paradigm}
Generative AI~\cite{cao2023comprehensive}, exemplified by \emph{large language models} (LLMs)~\cite{chang2024survey}, has reshaped the landscape of AI research and its societal impact since its rapid rise in the early 2020s. A milestone in this evolution is the development of the GPT framework~\cite{brown2020language}, which established a two-stage training paradigm: pre-training on large-scale unlabeled text corpora followed by task-specific fine-tuning. While neither the transformer architecture~\cite{vaswani2017attention} nor the pre-train and finetune strategy~\cite{radford2018improving,radford2019language,lu2019vilbert,sarzynska2021detecting} was novel on its own, their combination revealed a striking empirical phenomenon---the neural scaling law~\cite{kaplan2020scaling}, where model performance improves predictably with increases in model size, training data, and computational resources. This insight, supported by advances in distributed computing, has enabled LLMs with hundreds of billions of parameters, such as ChatGPT and DeepSeek, to reach and even exceed human-level performance across diverse natural language tasks~\cite{chang2024survey}. The success of LLMs has sparked growing interest in exploring the potential of GPT-like architectures for characterizing and representing scalable quantum systems. In the following, we first outline the general principles underlying this class of approaches, followed by a discussion of their applications and advanced topics.
 
\subsection{General schemes}

GPT-like architectures, when adapted for characterizing quantum systems, typically follow a pre-training-fine-tuning paradigm. In this framework, a foundation model (see Box~\ref{mybox:Terms-LM}) is first pre-trained on a broad corpus of quantum data to capture generalizable patterns and structural features. This pre-trained model is subsequently finetuned for specific downstream tasks, such as predicting key quantum properties of previously unseen states. Through this two-stage process, the resulting foundation model demonstrates strong versatility across diverse tasks and can be readily adapted to emerging tasks with minimal additional training.

\smallskip
\noindent\textbf{Dataset construction}. Applying GPT-like models to learn scalable quantum systems typically relies on two distinct datasets: a large-scale unlabeled dataset $\TLMP$ for pre-training and a smaller labeled dataset $\TLMF$ for fine-tuning with $\TLM=\TLMP\cup \TLMF$. The pre-training dataset $\TLMP=\{\bxi, \bzi, \bsi\}_{i=1}^{n_{P}}$ with $\bzi$ being optional, consisting of many examples with relatively few measurements $T$ per sample, enables the model to learn broad and generalizable patterns across quantum states.

Unlike the pre-training dataset $\TLMP$, the fine-tuning dataset $\TLMF=\{\bxi, \bsi, \bzi, \bhatyi\}_{i=1}^{n_F}$ contains much fewer samples, {\em i.e.}, $n_F\ll n_P$, but provides richer information per sample through a larger number of measurements $T$. This dense supervision enables the model to specialize in the target property prediction tasks. As with prior DL models, the input components $\bxi$, $\bsi$, and $\bzi$ are selectively used during fine-tuning depending on the architecture and the downstream application.

\smallskip
\noindent\textbf{Model implementation and training}. We next describe the implementation of the pre-training and fine-tuning stages employed in GPT models. 

During the pre-training stage, the model undergoes self-supervised training to extract common latent representations of the quantum states under consideration. A standard strategy for this purpose is to solve a generalized version of the implicit quantum state reconstruction problem. The employed transformer, denoted by $h_{\text{LM}}$, operates as an autoregressive model that emulates the measurement outcomes~\cite{melko2024language} associated with any quantum state in $\TLMP$. Mathematically, the model approximates the target distribution $\PP(\bm{s}|\bx)$, in which $\PP(\bx)$ denotes the data distribution over quantum states in $\TLMP$. Consequently, the negative log-likelihood loss function is adapted to the form
\begin{equation}
\mathcal{L}(\btheta) = -\frac{1}{n_P} \sum_{i=1}^{n_P}\frac{1}{T}\sum_{\bm{s}\in \mathcal{T}(\bxi)} \log \Big(\prod_{j=1}^N Q(\bm{s}_j|\bm{s}_{<j})\Big),
\end{equation} 
where $T$ denotes the number of measurement snapshots per state and $\mathcal{T}(\bxi)$ denotes the set of measurement outcomes of $\rho(\bxi)$. Under this formulation, the learning model $h_{\text{LM}}$ approximates the distribution over measurement outcomes as $\QQ(\bs;\bx, \bm{z};\btheta)=\prod_jQ(\bm{s}_j|\bm{s}_{<j};\bx, \bm{z};\btheta)$~\cite{wang2022,PhysRevB.107.075147,PhysRevApplied.21.014037,yao2024shadowgpt,fitzek2024rydberggpt}. 

During the fine-tuning stage, the pre-trained transformer, denoted as $h_{\text{LM}}(\bx, \bs, \bm{z}; \hat{\btheta})$ with $\hat{\btheta}$ being optimized parameters, is further optimized on the fine-tuning dataset $\TLMF$. In practice, additional task-specific layers, such as output layers for entanglement quantification, are often appended to the core transformer architecture. The entire model, or sometimes only the new layers, is then trained using supervised objectives that mirror the discriminative learning procedures of DL models. This approach enables the model to adapt its general knowledge to specialized applications with minimal labeled data.

\smallskip
\noindent\textbf{Model prediction}. The trained GPT-like models can not only be used to predict targeted quantum properties, but can also be readily finetuned to accommodate new prediction tasks.

\subsection{Concrete models and applications}
 
Existing GPT-style foundation models developed for scalable quantum system learning can be categorized into three classes based on their implementation strategies. The first two classes incorporate quantum measurement data: one relies entirely on pre-training, while the other follows a pre-training-then-fine-tuning paradigm. The third class is purely classical and employs a variational approach for quantum state reconstruction, bypassing the need to model the distribution underlying measurement statistics. 
Table~\ref{tab:DL-LM-summary-work} presents several representative results in this regime. 

\smallskip
\noindent\textbf{Foundation models with only pre-training}. The goal of this class of foundation models is to replicate the distribution of random measurement outcomes for previously unseen quantum states $\rho(\bx)$ under a specified informationally complete POVM $\mathcal{M}$. For instance, when $\mathcal{M}$ is set as a randomized Pauli measurement protocol used to collect $\{\bsi\}$, the pre-trained GPT model is designed to emulate the behavior of Pauli-based classical shadows~\cite{huang2020predicting}. That is, the output of the pre-trained transformer can then be employed to estimate the shadow expectation values of many local observables. 

A representative example of this approach is the work by Wang et al.~\cite{wang2022}, who have proposed a foundation model trained solely through pre-training to predict linear properties of two-dimensional random Heisenberg models, using simulations for systems with up to 45 qubits. The adopted measurement protocol was the Pauli-6 POVM. Subsequent studies have extended this framework to different quantum applications and transformer architectures. In particular, a bidirectional transformer architecture, also based on Pauli-6 POVM, has been proposed to predict linear properties of ground states in 
two-dimensional anti-ferromagnetic random Heisenberg models and to address Hamiltonian learning tasks~\cite{PhysRevApplied.21.014037}. In addition, ShadowGPT is developed to predict ground state energy, correlation functions, and entanglement entropy of quantum many-body systems~\cite{yao2024shadowgpt}, while RydbergGPT, which adopts an encoder-decoder transformer architecture, is employed to investigate the properties of ground states in Rydberg atom arrays~\cite{fitzek2024rydberggpt}.

\smallskip
\noindent\textbf{Foundation models with both pre-training and fine-tuning}. A first attempt along this direction 
has been completed by Zhang et al~\cite{PhysRevB.107.075147}. In particular, they have first employed an encoder-only transformer architecture to complete the pre-training towards ground states of a family of Hamiltonians. Then, at the fine-tuning stage, the pre-trained transformer is finetuned on a labeled dataset related to the phase diagram and magnetization strengths for unseen ground states. A subsequent study extended this approach by incorporating the Hamiltonian parameters, {\em i.e.}, the auxiliary information $\bm{z}$, directly into the input of a GPT-style model~\cite{tangtowards}.  
 
\smallskip
\noindent\textbf{Foundation model without quantum data}. The fundamental distinction between this class and the preceding two lies in how implicit quantum state reconstruction is achieved,  as introduced earlier. Rather than relying on measurement data, foundation models employ transformers as variational ans{\"a}tze to approximate ground states for a given class of Hamiltonians. Mathematically, given the input $\bxi$, denote the estimated state represented by the transformer model as $\psi(\btheta;\bxi)$. The objective of the pre-training stage aims to minimize variational energy~\cite{carleo2017solving}, {\em i.e.}, \[\mathcal{L}(\btheta) =\frac{1}{n_P}\sum_{i=1}^{n_P}\frac{\langle\psi(\btheta;\bxi)|\ham(\bxi)|\psi(\btheta;\bxi)\rangle }{ \langle\psi(\btheta;\bxi)|\psi(\btheta;\bxi)\rangle}.\]  This variational optimization can be completed by utilizing advanced techniques such as the minimum-step stochastic reconfiguration algorithm~\cite{chen2024empowering}. A recent representative example is provided by Ref.~\cite{rende2025foundation}, where the proposed foundation model enables efficient simulation of disordered systems, estimation of the quantum geometric tensor in coupling space, and uncovering quantum phase transitions.  

\begin{figure}
    \centering
\includegraphics[width=0.95\linewidth]{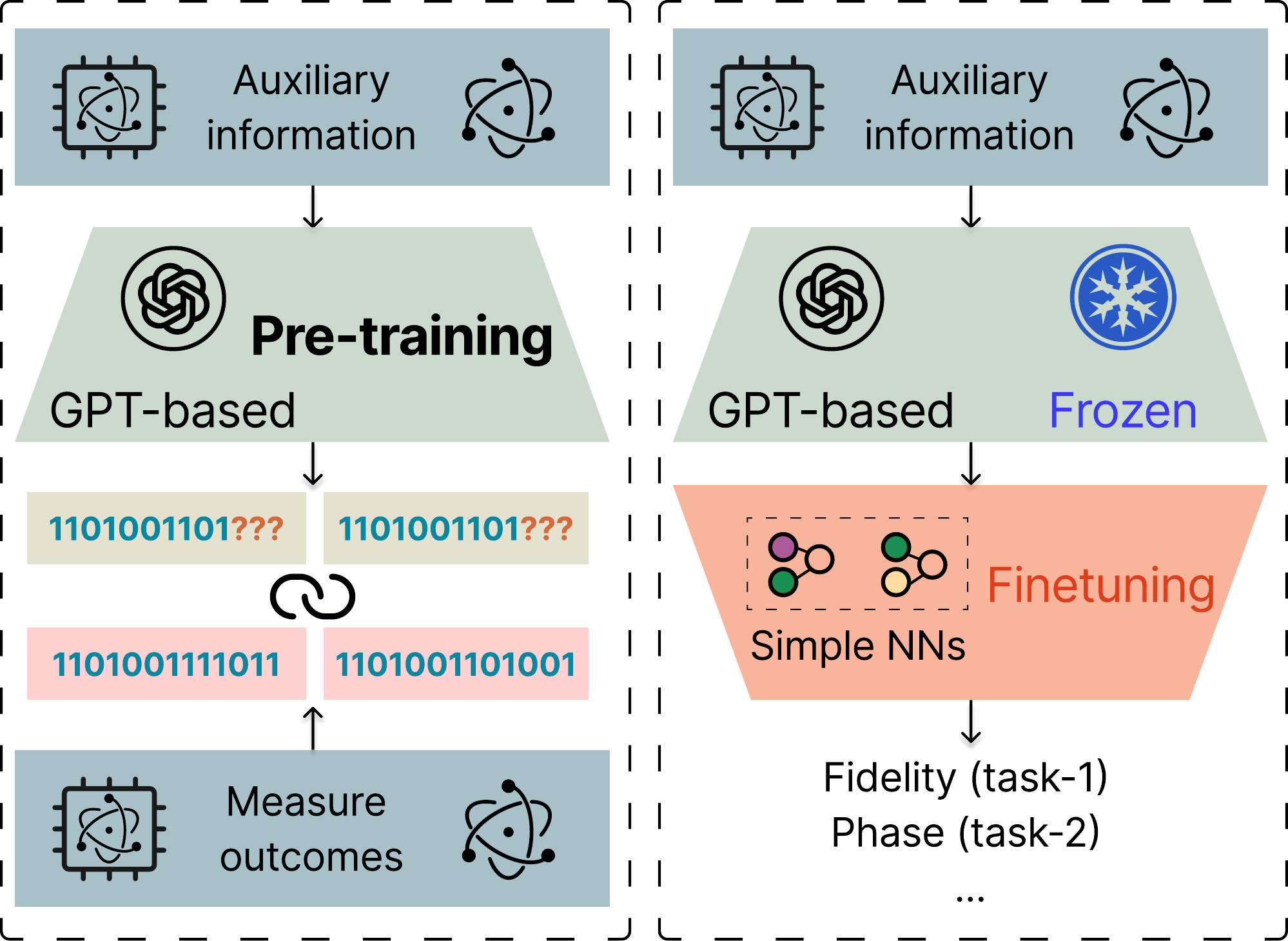}
    \caption{\small{ \textbf{Schematic representation of a pre-training-and-fine-tuning paradigm for quantum property prediction with GPT-based models.} In the pre-training stage (left panel), 
    auxiliary information and measurement outcomes from quantum systems are used to train a generative language model on large, diverse datasets. In the fine-tuning stage (right panel), the pre-trained GPT model is kept fixed (frozen) while lightweight neural networks are trained on task-specific data to predict properties such as fidelity and entropy. This approach leverages the generalization capabilities of large language models and adapts them to specialized quantum tasks through targeted fine-tuning.}}
    \label{fig:fm}
\end{figure}

\subsection{Advanced topics}

Beyond training GPT-like models from scratch to develop foundation models for scalable quantum systems, early efforts have also begun to explore alternative generative AI techniques in this domain.

\smallskip
\noindent\textbf{Generative AI beyond GPT}. Diffusion models represent another prominent class of generative AI, which learn to reverse a gradual noising process and synthesize data by iteratively denoising random inputs~\cite{ho2020denoising,Song2021Score,yang2023diffusion}. Building on this principle, a recent study proposed QuaDiM, a diffusion-based framework for quantum system learning~\cite{tang2025quadim}. By conditioning the noise-to-data generation process on Hamiltonian parameters, QuaDiM enables a single model to approximate ground states across an entire phase diagram, effectively unifying property prediction with state synthesis. In addition, diffusion models, associated with text-conditioning, have been utilized to accomplish entanglement generation and unitary compilation tasks~\cite{furrutter2024quantum}.

\smallskip
\noindent\textbf{LLM for quantum algorithms}. Alongside efforts to train GPT-like models from scratch, a growing direction is to employ pretrained LLMs, such as ChatGPT, to enhance quantum algorithm design~\cite{dupuis2024qiskit, yang2024qcircuitnet,campbell2025enhancing}. Notably, LLMs have been finetuned to specialize in quantum computing programming tasks. Moreover, both open- and closed-source LLMs have also been utilized to generate high-quality ans{\"a}tze, contributing to improved performance in VQAs~\cite{liang2023unleashing,nakaji2024generative,minami2025generative}.

  \begin{table*}[]
     \centering
     \resizebox{0.95\textwidth}{!}{%
     \begin{tabular}{@{}ccccccc@{}}
     \toprule
      &  Task & Learning model &  & & Property & Simulation/experimental scale\\ 
          \midrule
        \multicolumn{1}{c|}{\multirow{12}{*}{DL}} 
        &\multicolumn{1}{c|}{\multirow{6}{*}{ \begin{tabular}[c]{@{}c@{}} Single property\\ prediction\end{tabular}}} 
        & \multicolumn{1}{c|}{FCNN, LSTM~\cite{huang2025direct}} 
        & \multicolumn{1}{c|}{$\mathsf{G.S.}$ \& $\mathsf{D.S.}$} 
        & \multicolumn{1}{c|}{$\mathsf{M.B.}$} 
        & \multicolumn{1}{c|}{$\mathsf{entanglement}$} 
        & \begin{tabular}[c]{@{}c@{}} Simulation \\ XYZ model ($100$ Q)\end{tabular} \\  
        \cmidrule{3-7}
        \multicolumn{1}{c|}{} 
        &  \multicolumn{1}{c|}{}
        & \multicolumn{1}{c|}{Permutation invariant NN+GNN~\cite{qian2023multimodal}} 
         & \multicolumn{1}{c|}{$\mathsf{Q.C.}$}
         & \multicolumn{1}{c|}{$\mathsf{M.B.}$} 
         & \multicolumn{1}{c|}{$\mathsf{similarity}$} 
         & \begin{tabular}[c]{@{}c@{}} Experiment \\ random circuit ($50$ Q)\end{tabular} \\   
         \cmidrule{3-7}
         \multicolumn{1}{c|}{} 
         & \multicolumn{1}{c|}{}
         & \multicolumn{1}{c|}{CNN~\cite{dehghani2023neural}} 
          & \multicolumn{1}{c|}{$\mathsf{Q.C.}$} 
          & \multicolumn{1}{c|}{$\mathsf{M.B.}$} 
          & \multicolumn{1}{c|}{$\mathsf{P.T.}$} 
          & \begin{tabular}[c]{@{}c@{}} Simulation \\ random circuit ($64$ Q) \end{tabular} \\
         \cmidrule{2-7}
        \multicolumn{1}{c|}{} 
        & \multicolumn{1}{c|}{\begin{tabular}[c]{@{}c@{}} Multi-property\\ prediction\end{tabular}}
         & \multicolumn{1}{c|}{GQNN~\cite{zhu2022}, FCNN~\cite{wu2024learning}} 
         & \multicolumn{1}{c|}{$\mathsf{G.S.}$} 
         & \multicolumn{1}{c|}{$\mathsf{M.B.}$} 
         & \multicolumn{1}{c|}{$\mathsf{S.C., E.E., P.T.}$} 
         & \begin{tabular}[c]{@{}c@{}} Simulation \\XXZ model ($50$ Q)\end{tabular} \\ 
        \cmidrule{2-7}
        \multicolumn{1}{c|}{} 
       & \multicolumn{1}{c|}{\multirow{3}{*}{\begin{tabular}[c]{@{}c@{}} State\\ reconstruction\end{tabular}}}
       & \multicolumn{1}{c|}{RNN (data-driven) \cite{carrasquilla2019reconstructing}} 
       & \multicolumn{1}{c|}{$\mathsf{G.S.}$}
       & \multicolumn{1}{c|}{$\mathsf{M.A.}$} 
       & \multicolumn{1}{c|}{-} 
       & \begin{tabular}[c]{@{}c@{}} Simulation \\ GHZ state ($50$ Q)\end{tabular} \\ 
       \cmidrule{3-7}
       \multicolumn{1}{c|}{} 
       &\multicolumn{1}{c|}{} 
       & \multicolumn{1}{c|}{RNN (data-driven+variational) \cite{moss2023enhancing}} 
        & \multicolumn{1}{c|}{$\mathsf{G.S.}$}
        & \multicolumn{1}{c|}{$\mathsf{M.A.}$}
        & \multicolumn{1}{c|}{-} 
        &  \begin{tabular}[c]{@{}c@{}} Experiment \\ Rydberg atoms ($16\times 16$ Q)\end{tabular} \\ 
         \midrule
       \multicolumn{1}{c|}{\multirow{8}{*}{LM}} 
       &  \multicolumn{1}{c|}{\multirow{7}{*}{\begin{tabular}[c]{@{}c@{}} Foundation\\ model\end{tabular}}}
       &  \multicolumn{1}{c|}{GPT (pretrain only)~\cite{wang2022}} 
       &  \multicolumn{1}{c|}{$\mathsf{G.S.}$} 
       & \multicolumn{1}{c|}{$\mathsf{M.A.}$}  
       & \multicolumn{1}{c|}{$\mathsf{S.C., E.E., P.T.}$}  
       & \begin{tabular}[c]{@{}c@{}} Experiment \\ Rydberg atoms ($13\times 13$ Q) \\ Simulation \\ 2D Heisenberg models ($30$ Q) \end{tabular}\\ \cmidrule{3-7} 
        \multicolumn{1}{c|}{} 
         &\multicolumn{1}{c|}{} 
         & \multicolumn{1}{c|}{GPT (pretrain+finetune)~\cite{PhysRevB.107.075147,tangtowards}} 
         & \multicolumn{1}{c|}{$\mathsf{G.S.}$} 
         & \multicolumn{1}{c|}{$\mathsf{M.A.} \& \mathsf{M.B.}$}  
         &  \multicolumn{1}{c|}{\begin{tabular}[c]{@{}c@{}} $\mathsf{energy, magnetization,}$ \\ $\mathsf{Hamiltonian,}$ \\$\mathsf{S.C., E.E., P.T.}$ \end{tabular}  }
         &\begin{tabular}[c]{@{}c@{}}  Simulation \\ TFI model ($40$ Q)~\cite{PhysRevB.107.075147}\\
         Rydberg atoms ($31$)~\cite{tangtowards}
         \end{tabular} \\
         \cmidrule{3-7}
         \multicolumn{1}{c|}{} 
        & \multicolumn{1}{c|}{} 
        & \multicolumn{1}{c|}{GPT (variational)~\cite{rende2025foundation}} 
        & \multicolumn{1}{c|}{$\mathsf{G.S.}$}  
        & \multicolumn{1}{c|}{-} 
        & \multicolumn{1}{c|}{$\mathsf{disorder,\, fidelity\, susceptibility}$}
        &  \begin{tabular}[c]{@{}c@{}} Simulation\\ 2D Heisenberg model ($100$ Q)\end{tabular} \\
         \bottomrule
     \end{tabular}
     }
     \caption{\small{Summary of representative results on representing and characterizing scalable quantum systems with DL and LM-based approaches. The types of the explored quantum systems are denoted by $\mathsf{G.S.}$ for ground states, $\mathsf{D.S.}$ for dynamical states and $\mathsf{Q.C.}$ for states prepared by quantum circuits. The measurement-agnostic and measurement-based protocols are denoted by $\mathsf{M.A.}$ and $\mathsf{M.B.}$, respectively. The properties of spin correlation, entanglement entropy and phase transition are denoted by $\mathsf{S.C}$, $\mathsf{E.E.}$, and $\mathsf{P.T.}$, respectively. The notation `TFI' refers to transverse-field Ising and `Q' denotes qubits. }}
     \label{tab:DL-LM-summary-work}
 \end{table*}
 
\section{Challenges and opportunities}

We next discuss the key challenges of existing AI approaches in representing and characterizing large-scale quantum systems. Any progress in addressing these challenges will improve both the scalability and reliability of AI-driven quantum system learning. We organize these challenges along a spectrum from theoretical foundations to practical implementation and elaborate on each. 

\smallskip
\noindent\textit{Question 1: Are there provably efficient ML models for quantum system characterization that extend beyond linear property prediction?} 

While significant progress has been made in developing ML models for predicting linear properties of scalable quantum systems, the existence of provably efficient ML models for broader learning tasks, particularly nonlinear property prediction, remains an open question. An initial attempt by Huang et al.\ demonstrated an efficient ML model for classifying topological quantum phases~\cite{huang2022provably}, which are indistinguishable by linear properties. The key challenge in this context stems from the intricate dependencies of these properties on both the underlying quantum system and classical inputs.  

Building on insights from prior studies on linear property prediction, a promising approach is to leverage structural constraints, {\em e.g.}, locality, smoothness, or intrinsic symmetry of the relevant quantum states to facilitate efficient learning. Advancing this understanding would not only deepen the theoretical foundation of quantum learning theory but also provide concrete guidelines for designing ML models capable of tackling more complex quantum characterization tasks.

\smallskip
\noindent\textit{Question 2: What are the theoretical foundations of classical learning models in the measurement-based protocol?} 

Previous studies have explored the theoretical performance of classical learners in the measurement-agnostic protocol, as illustrated in Fig.~\ref{fig:pipeline}. Specifically, it has been shown that an ML model can predict linear properties of ground states that classical algorithms without data access cannot achieve~\cite{huang2022provably}. On the other side, ML models perform worse than quantum learners in predicting expectation values of an unknown observable for certain classes of ground states~\cite{molteni2024exponential}. Informally, there exists a hierarchical relationship: purely classical methods are a subset of measurement-agnostic classical learners, which in turn are encompassed by the broader class of quantum learning models that are implemented on quantum computers.  

Despite progress in understanding the theoretical foundations of measurement-agnostic protocols, the performance of measurement-based protocols remains largely unexplored. The definition of a measurement-based protocol can be extended in two directions. One perspective aligns with the protocol illustrated in Fig.~\ref{fig:pipeline}, where the classical learner can perform quantum measurements on the system under consideration during the prediction stage. Alternatively, a measurement-based protocol can be understood as a model that adaptively queries data based on prior outcomes throughout the learning process~\cite{quek2021,lange2023adaptive,huang2025sequence}.

Given the extended definition of measurement-based protocols, several fundamental questions regarding their learnability remain open. Foremost among them are whether a computational separation exists between measurement-agnostic protocols and quantum learning models towards analog quantum simulators and digital quantum computers. Additionally, for any scalable quantum system learning task, as summarized in Fig.~\ref{fig:scheme}, it is crucial to determine whether measurement-based classical learners can achieve both sample and computational efficiency or whether they are sample-efficient but computationally intractable. Furthermore, as with Question 1, how can we design provably efficient measurement-based classical learning models for a given task under mild conditions? Addressing these questions would provide valuable insights into when the measurement-based learning protocols should be employed.

\smallskip
\noindent\textit{Question 3: When do DL and LM-based methods outperform ML models in quantum system characterization?}

Prior studies~\cite{tang2024ssl4q,tangtowards} have claimed that DL models outperform traditional ML models in learning scalable quantum systems. However, many of these comparisons lack rigorous benchmarking, as they involve different measurement strategies, quantum computational costs in dataset construction, or variations in training conditions. A systematic evaluation is necessary to assess the actual advantages of DL and LMs in this domain.
 
A recent study~\cite{zhao2025Rethink} makes an initial attempt to address this issue by conducting extensive simulations on the prediction of both linear and nonlinear properties for various classes of ground states relevant to standard benchmarks. Different from prior works that employ varied amounts of quantum resources, this study ensures a fair comparison by controlling the total number of quantum measurements across all learning models. The results reveal that, in terms of prediction error, DL models do not consistently outperform traditional ML models on commonly studied Hamiltonian systems. Besides, in the regime of QEM, comparison of the performance of various ML and DL models has been conducted across diverse classes of quantum circuits involving up to 100 qubits~\cite{liao2024machine}. The results show that random forest, an ML-based approach, consistently outperforms DL models, {\em e.g.}, FCNN and GNN. 

These findings highlight the need to explore more challenging tasks, where DL models and LMs may demonstrate clearer advantages. Furthermore, it remains important to investigate whether these models offer benefits in other aspects, such as robustness to noise, generalization to out-of-distribution quantum states, and adaptability to dynamically evolving quantum systems. Parallel to benchmarking efforts, another critical research direction is exploring the theoretical foundations of DL models and LMs in quantum system characterization. A fundamental question is whether these models can provide provable advantages over classical ML models for specific tasks. Addressing this question, alongside the insights from Questions 1 and 2, will deepen our understanding of conditions under which these advanced models offer genuine benefits.

\smallskip
\noindent\textit{Question 4: Can advanced DL research topics improve quantum system characterization and representation?} 

Previous studies have focused on evaluating the performance of AI models based on prediction accuracy. However, this single metric is insufficient to comprehensively assess the capabilities of DL models and LMs. The main reason is that these models have demonstrated success across various domains, where advanced techniques, such as transfer learning~\cite{zhuang2020comprehensive}, continual learning~\cite{wang2024comprehensive}, adversarial learning~\cite{kaur2022trustworthy}, and meta-learning~\cite{hospedales2021meta} have improved adaptability, robustness, and efficiency. This raises a fundamental question: can these advanced DL methods offer comparable advantages for scalable quantum systems?

Several studies have shown positive results in applying advanced transfer learning techniques to enhance the performance of DL models in property prediction and quantum state reconstruction~\cite{PhysRevE.101.053301,wang2022, wu2024learning}. A common feature of these models is that training is initially performed on small or classically simulatable quantum systems, while the trained models aided by transfer learning, can be extended to larger and more computationally challenging quantum systems.

Despite these advancements, several important questions remain unanswered. First, the potential of advanced techniques such as continual learning, adversarial learning, and meta-learning in representing and characterizing scalable quantum systems remains largely unexplored. Second, systematic benchmarking studies are desired to compare the performance of different approaches across various evaluation metrics, similar to the benchmark developed for variational calculations of quantum many-body systems~\cite{wu2024variational}. Last, the theoretical foundations of these novel methods are still underdeveloped.  
 
\smallskip
\noindent\textit{Question 5: How can novel task-specific DL models and general-purpose LMs be designed to effectively represent and characterize quantum systems at scale?} 

Symmetries are fundamental in nature and provide powerful guiding principles for designing more efficient DL models and LMs for scalable quantum systems. However, most existing approaches in learning quantum systems are solely data-driven, without explicitly leveraging underlying physical structures. Incorporating intrinsic physical knowledge into model architectures is crucial for enhancing learning performance and improving model interpretability~\cite{PhysRevResearch.2.023358,PhysRevA.104.012401,PRXQuantum.2.040355}.

Few studies have explicitly demonstrated how to incorporate permutation symmetry into learning models to enforce the invariance of measurement outcomes across multiple shots~\cite{qian2023multimodal,kim2024attention}. Beyond permutation invariance, fundamental structures such as gauge symmetries~\cite{chen2023learnability}, locality constraints~\cite{PhysRevX.14.031035}, and Lie-algebraic properties~\cite{zeier2011symmetry} are intrinsic to many quantum systems and can be embedded into network architectures to enhance learning efficiency. Developing effective strategies for integrating these physical prior knowledge has the potential to further improve both the efficiency and adaptability of deep learning models for quantum learning tasks. 
 
Moreover, although progress has been made, a general-purpose foundation model for quantum systems, which is capable of learning from diverse forms of quantum data, such as measurement outcomes, quantum circuits, and Hamiltonians, has yet to be realized. Current research has mainly focused on approaches that work well for specific families of quantum states, enabling some knowledge transfer between related tasks. While these methods are promising, extending them to a broader range of quantum systems will likely require substantial computational resources. Continued advances in this area could ultimately transform our ability to model, simulate, and understand large-scale quantum systems.
 
\smallskip
\noindent\textit{Question 6: How can we employ LLMs to enhance the characterization of scalable quantum systems?} 

Recent advances in LLMs have demonstrated remarkable capabilities in scientific discovery~\cite{wang2023scientific}. For instance, LLMs have been employed to assist in mathematical problem-solving~\cite{romera2024mathematical}. In chemistry and materials science, LLMs have been used to automate research~\cite{boiko2023autonomous}. In quantum computing, LLMs have contributed to hardware fabrication~\cite{cao2024agents} and to quantum error correction~\cite{cao2023qecgpt}. A natural question is whether and how modern LLMs can be employed to enhance the characterization and representation of large-scale quantum systems, particularly in addressing challenges faced by earlier AI models, such as data scarcity, generalization constraints, and limited interpretability.

\section{Outlook}
The coming years will witness a transition in AI-driven quantum system characterization and representation from conceptual development to large-scale implementation. While most current studies have focused on toy models and specific tasks, future research will increasingly target realistic and experimentally relevant quantum systems. These efforts will not only refine existing learning strategies but also push the frontier of quantum technologies. 

To fully unlock the potential of AI in this field, it is essential to broaden and refine the relevant computational toolbox. This includes improving the synergy between graphical processing units and quantum processing units to accelerate hybrid quantum-classical computations and integrating AI frameworks with quantum cloud platforms to enable more scalable and accessible learning models. 

The development of open datasets and standardized benchmarks is equally critical, as these resources will foster collaboration between AI and quantum communities. In particular, these datasets encourage AI researchers to contribute to developing novel AI models with better performance, bridging expertise across disciplines. These initiatives will play a pivotal role in advancing interdisciplinary efforts and accelerating the realization of the practical utility of Megaquop machine~\cite{preskill2025beyond}, with the long-term goal to demonstrate quantum advantage in practical applications ranging from fundamental physics to molecular simulations~\cite{zhang2025fault}.

Looking ahead, AI will continue to play a fundamental role in the fault-tolerant quantum computing era. An ongoing effort to apply AI in fault-tolerant quantum computing focuses on effectively learning the decoder of a surface code in the context of realistic noise models~\cite{bausch2024learning,PhysRevResearch.7.013029,lange2023data}. As quantum hardware matures, the focus will expand beyond quantum system certification, benchmarking, and pretraining VQAs to optimizing large-scale quantum algorithms with provable advantages and automating circuit simplifications across different quantum computing platforms~\cite{ge2024quantum}.

AI-driven approaches will eventually shape quantum computing from the software level to the hardware level, facilitating end-to-end automation in quantum algorithm design, compilation, and execution~\cite{alexeev2024artificial}. Some premature yet important efforts have been made toward these long-term goals~\cite{PRXQuantum.1.010301,lecompte2023machine,furrutter2024quantum,ai2024graph,ruiz2025quantum}. A major milestone in this direction is the development of AI-assisted quantum computing ecosystems, where intelligent models autonomously select optimal quantum circuits and optimize error correction strategies that integrate problem formulation, resource allocation, and real-time feedback mechanisms. 

AI is set to transform quantum many-body physics by uncovering new insights into complex quantum systems. Looking ahead, AI-driven models will play a crucial role in characterizing emergent phenomena, discovering novel quantum phases~\cite{PhysRevLett.125.170603,PhysRevResearch.5.013026}, refining theoretical frameworks for strongly correlated systems~\cite{hou2024unsupervised,venturella2025unified}, and surpassing the capabilities of conventional numerical methods~\cite{huang2022provably,PhysRevB.106.205136}. Ultimately, AI-driven approaches will provide a powerful framework for probing fundamental questions in quantum science~\cite{carleo2019machine,CUP}. By systematically advancing these directions, AI will serve as an indispensable tool in unlocking the full computational power of quantum technologies.

\section{Acknowledgments}
YXD acknowledges the funding from SUG (025257-00001), NTU. YZ and GC acknowledge the funding from the Ministry of Science and Technology of China (MOST2030) under Grant No. 2023200300600. 
YDW acknowledges funding from the National Natural Science Foundation of China through grants no.\ 12405022. 
JE acknowledges funding by the 
German BMFTR, Berlin Quantum, the Munich Quantum Valley, the Quantum Flagship (Millenion and PasQuans2), and the European Research Council.
GC acknowledges support from the Hong Kong Research Grant Council through Grants No. 17307520, No. R7035-21F, and No. T45-406/23-R, by the Ministry of Science and Technology through Grant No. 2023ZD0300600, and by the John Templeton Foundation through Grant 62312, The Quantum Information Structure of Spacetime. The opinions expressed in this publication are those of the authors and do not necessarily reflect the views of the John Templeton Foundation. Research at the Perimeter Institute is supported by the Government of Canada through the Department of Innovation, Science and Economic Development Canada and by the Province of Ontario through the Ministry of Research, Innovation and Science.


\begin{thebibliography}{100}

\bibitem{aaronson2018shadow}
S.~Aaronson.
\newblock Shadow tomography of quantum states.
\newblock In {\em Proceedings of the 50th annual ACM SIGACT symposium on theory
  of computing}, pages 325--338, 2018.

\bibitem{aaronson2004improved}
S.~Aaronson and D.~Gottesman.
\newblock Improved simulation of stabilizer circuits.
\newblock {\em Physical Review A—Atomic, Molecular, and Optical Physics},
  70:052328, 2004.

\bibitem{Giovanni2025}
G.~Acampora et~al.
\newblock Quantum computing and artificial intelligence: status and
  perspectives.
\newblock {\em arXiv:2505.23860}, 2025.

\bibitem{PhysRevX.14.041012}
U.~Agrawal, J.~Lopez-Piqueres, R.~Vasseur, S.~Gopalakrishnan, and A.~C. Potter.
\newblock Observing quantum measurement collapse as a learnability phase
  transition.
\newblock {\em Phys. Rev. X}, 14:041012, Oct 2024.

\bibitem{aharonov2023polynomial}
D.~Aharonov, X.~Gao, Z.~Landau, Y.~Liu, and U.~Vazirani.
\newblock A polynomial-time classical algorithm for noisy random circuit
  sampling.
\newblock In {\em Proceedings of the 55th Annual ACM Symposium on Theory of
  Computing}, pages 945--957, 2023.

\bibitem{ahmed2021quantum}
S.~Ahmed, C.~S. Mu{\~n}oz, F.~Nori, and A.~F. Kockum.
\newblock Quantum state tomography with conditional generative adversarial
  networks.
\newblock {\em Phys. Rev. Lett.}, 127:140502, 2021.

\bibitem{Quantum2025Google}
G.~Q. AI and Collaborators.
\newblock Quantum error correction below the surface code threshold.
\newblock {\em Nature}, 638:920--926, 2025.

\bibitem{ai2024graph}
H.~Ai and Y.-X. Liu.
\newblock Scalable parameter design for superconducting quantum circuits with
  graph neural networks.
\newblock {\em Phys. Rev. Lett.}, 135:040601, 2025.

\bibitem{PhysRevB.109.094209}
A.~A. Akhtar, H.-Y. Hu, and Y.-Z. You.
\newblock Measurement-induced criticality is tomographically optimal.
\newblock {\em Phys. Rev. B}, 109:094209, Mar 2024.

\bibitem{alexeev2024artificial}
Y.~Alexeev et~al.
\newblock Artificial intelligence for quantum computing.
\newblock {\em arXiv:2411.09131}, 2024.

\bibitem{PhysRevLett.134.120202}
Z.~An, J.~Wu, Z.~Lin, X.~Yang, K.~Li, and B.~Zeng.
\newblock Dual-capability machine learning models for quantum hamiltonian
  parameter estimation and dynamics prediction.
\newblock {\em Phys. Rev. Lett.}, 134:120202, 2025.

\bibitem{PhysRevApplied.21.014037}
Z.~An, J.~Wu, M.~Yang, D.~L. Zhou, and B.~Zeng.
\newblock {Unified quantum state tomography and Hamiltonian learning: A
  language-translation-like approach for quantum systems}.
\newblock {\em Phys. Rev. Appl.}, 21:014037, 2024.

\bibitem{andersen2025thermalization}
T.~I. Andersen et~al.
\newblock Thermalization and criticality on an analogue--digital quantum
  simulator.
\newblock {\em Nature}, 638:79--85, 2025.

\bibitem{anshu2024survey}
A.~Anshu and S.~Arunachalam.
\newblock A survey on the complexity of learning quantum states.
\newblock {\em Nature Rev. Phys.}, 6:59--69, 2024.

\bibitem{arnold2022replacing}
J.~Arnold and F.~Sch{\"a}fer.
\newblock Replacing neural networks by optimal analytical predictors for the
  detection of phase transitions.
\newblock {\em Phys. Rev. X}, 12:031044, 2022.

\bibitem{PhysRevResearch.3.033052}
J.~Arnold, F.~Sch\"afer, M.~\ifmmode~\check{Z}\else \v{Z}\fi{}onda, and
  A.~U.~J. Lode.
\newblock Interpretable and unsupervised phase classification.
\newblock {\em Phys. Rev. Res.}, 3:033052, 2021.

\bibitem{arunachalam2017guest}
S.~Arunachalam and R.~de~Wolf.
\newblock Guest column: A survey of quantum learning theory.
\newblock {\em ACM Sigact News}, 48:41--67, 2017.

\bibitem{asif2023entanglement}
N.~Asif, U.~Khalid, A.~Khan, T.~Q. Duong, and H.~Shin.
\newblock Entanglement detection with artificial neural networks.
\newblock {\em Sci. Rep.}, 13:1562, 2023.

\bibitem{baltruvsaitis2018multimodal}
T.~Baltru{\v{s}}aitis, C.~Ahuja, and L.-P. Morency.
\newblock Multimodal machine learning: A survey and taxonomy.
\newblock {\em IEEE Trans. Patt. Ana. Mach. Int.}, 41:423--443, 2018.

\bibitem{bao2025beyond}
T.~Bao, X.~Ye, H.~Ruan, C.~Liu, W.~Wu, and J.~Yan.
\newblock Beyond circuit connections: A non-message passing graph transformer
  approach for quantum error mitigation.
\newblock In {\em The Thirteenth International Conference on Learning
  Representations}.

\bibitem{PhysRevLett.129.200602}
F.~Barratt, U.~Agrawal, A.~C. Potter, S.~Gopalakrishnan, and R.~Vasseur.
\newblock Transitions in the learnability of global charges from local
  measurements.
\newblock {\em Phys. Rev. Lett.}, 129:200602, Nov 2022.

\bibitem{bausch2024learning}
J.~Bausch et~al.
\newblock Learning high-accuracy error decoding for quantum processors.
\newblock {\em Nature}, pages 1--7, 2024.

\bibitem{beguvsic2024fast}
T.~Begu{\v{s}}i{\'c}, J.~Gray, and G.~K.-L. Chan.
\newblock Fast and converged classical simulations of evidence for the utility
  of quantum computing before fault tolerance.
\newblock {\em Sci. Adv.}, 10:eadk4321, 2024.

\bibitem{bengio2013representation}
Y.~Bengio, A.~Courville, and P.~Vincent.
\newblock Representation learning: A review and new perspectives.
\newblock {\em IEEE Trans. Patt. Ana. Mach. Int.}, 35:1798--1828, 2013.

\bibitem{bennewitz2022neural}
E.~R. Bennewitz, F.~Hopfmueller, B.~Kulchytskyy, J.~Carrasquilla, and
  P.~Ronagh.
\newblock Neural error mitigation of near-term quantum simulations.
\newblock {\em Nature Mach. Intell.}, 4:618--624, 2022.

\bibitem{PhysRevLett.133.020602}
C.~Bertoni, J.~Haferkamp, M.~Hinsche, M.~Ioannou, J.~Eisert, and H.~Pashayan.
\newblock Shallow shadows: Expectation estimation using low-depth random
  clifford circuits.
\newblock {\em Phys. Rev. Lett.}, 133:020602, 2024.

\bibitem{bharti2021noisy}
K.~Bharti et~al.
\newblock Noisy intermediate-scale quantum algorithms.
\newblock {\em Rev. Mod. Phys.}, 94:015004, 2022.

\bibitem{bishop2006pattern}
C.~M. Bishop and N.~M. Nasrabadi.
\newblock {\em Pattern recognition and machine learning}, volume~4.
\newblock Springer, 2006.

\bibitem{bluvstein2024logical}
D.~Bluvstein et~al.
\newblock Logical quantum processor based on reconfigurable atom arrays.
\newblock {\em Nature}, 626:58--65, 2024.

\bibitem{PhysRevLett.127.150504}
A.~Bohrdt, S.~Kim, A.~Lukin, M.~Rispoli, R.~Schittko, M.~Knap, M.~Greiner, and
  J.~L\'eonard.
\newblock Analyzing nonequilibrium quantum states through snapshots with
  artificial neural networks.
\newblock {\em Phys. Rev. Lett.}, 127:150504, 2021.

\bibitem{boiko2023autonomous}
D.~A. Boiko, R.~MacKnight, B.~Kline, and G.~Gomes.
\newblock Autonomous chemical research with large language models.
\newblock {\em Nature}, 624:570--578, 2023.

\bibitem{bouland2023public}
A.~Bouland, B.~Fefferman, S.~Ghosh, T.~Metger, U.~Vazirani, C.~Zhang, and
  Z.~Zhou.
\newblock Public-key pseudoentanglement and the hardness of learning ground
  state entanglement structure.
\newblock {\em arXiv:2311.12017}, 2023.

\bibitem{bouland2024hardness}
A.~Bouland, C.~Zhang, and Z.~Zhou.
\newblock {On the hardness of learning ground state entanglement of
  geometrically local Hamiltonians}.
\newblock {\em arXiv:2411.04353}, 2024.

\bibitem{Bourgund2025Formation}
D.~Bourgund et~al.
\newblock {Formation of individual stripes in a mixed-dimensional cold-atom
  Fermi--Hubbard system}.
\newblock {\em Nature}, 637:57--62, 2025.

\bibitem{bravyi2016improved}
S.~Bravyi and D.~Gosset.
\newblock {Improved classical simulation of quantum circuits dominated by
  Clifford gates}.
\newblock {\em Phys. Rev. Lett.}, 116:250501, 2016.

\bibitem{brown2020language}
T.~Brown et~al.
\newblock Language models are few-shot learners.
\newblock {\em Adv. Neur. Inf. Proc. Sys.}, 33:1877--1901, 2020.

\bibitem{RevModPhys.95.045005}
Z.~Cai, R.~Babbush, S.~C. Benjamin, S.~Endo, W.~J. Huggins, Y.~Li, J.~R.
  McClean, and T.~E. O'Brien.
\newblock Quantum error mitigation.
\newblock {\em Rev. Mod. Phys.}, 95:045005, 2023.

\bibitem{campaioli2024quantum}
F.~Campaioli, J.~H. Cole, and H.~Hapuarachchi.
\newblock Quantum master equations: Tips and tricks for quantum optics, quantum
  computing, and beyond.
\newblock {\em PRX Quantum}, 5:020202, 2024.

\bibitem{campbell2025enhancing}
C.~Campbell, H.~M. Chen, W.~Luk, and H.~Fan.
\newblock {Enhancing LLM-based quantum code generation with multi-agent
  optimization and quantum error correction}.
\newblock {\em arXiv:2504.14557}, 2025.

\bibitem{cao2023qecgpt}
H.~Cao, F.~Pan, Y.~Wang, and P.~Zhang.
\newblock {qecGPT: decoding quantum error-correcting codes with generative
  pre-trained transformers}.
\newblock {\em arXiv:2307.09025}, 2023.

\bibitem{cao2024agents}
S.~Cao, Z.~Zhang, M.~Alghadeer, S.~D. Fasciati, M.~Piscitelli, M.~Bakr,
  P.~Leek, and A.~Aspuru-Guzik.
\newblock Agents for self-driving laboratories applied to quantum computing.
\newblock {\em arXiv:2412.07978}, 2024.

\bibitem{cao2023comprehensive}
Y.~Cao, S.~Li, Y.~Liu, Z.~Yan, Y.~Dai, P.~S. Yu, and L.~Sun.
\newblock {A comprehensive survey of AI-generated content (AIGC): A history of
  generative AI from GAN to ChatGPT}.
\newblock {\em arXiv:2303.04226}, 2023.

\bibitem{carleo2019machine}
G.~Carleo, I.~Cirac, K.~Cranmer, L.~Daudet, M.~Schuld, N.~Tishby,
  L.~Vogt-Maranto, and L.~Zdeborov{\'a}.
\newblock Machine learning and the physical sciences.
\newblock {\em Rev. Mod. Phys.}, 91:045002, 2019.

\bibitem{carleo2017solving}
G.~Carleo and M.~Troyer.
\newblock Solving the quantum many-body problem with artificial neural
  networks.
\newblock {\em Science}, 355:602--606, 2017.

\bibitem{PRXQuantum.2.010102}
J.~Carrasco, A.~Elben, C.~Kokail, B.~Kraus, and P.~Zoller.
\newblock Theoretical and experimental perspectives of quantum verification.
\newblock {\em PRX Quantum}, 2:010102, 2021.

\bibitem{carrasquilla2020machine}
J.~Carrasquilla.
\newblock Machine learning for quantum matter.
\newblock {\em Adv. Phys. X}, 5:1797528, 2020.

\bibitem{carrasquilla2017machine}
J.~Carrasquilla and R.~G. Melko.
\newblock Machine learning phases of matter.
\newblock {\em Nature Phys.}, 13:431, 2017.

\bibitem{carrasquilla2021use}
J.~Carrasquilla and G.~Torlai.
\newblock How to use neural networks to investigate quantum many-body physics.
\newblock {\em PRX Quantum}, 2:040201, 2021.

\bibitem{carrasquilla2019reconstructing}
J.~Carrasquilla, G.~Torlai, R.~G. Melko, and L.~Aolita.
\newblock Reconstructing quantum states with generative models.
\newblock {\em Nature Mach. Intell.}, 1:155--161, 2019.

\bibitem{cerezo2021variational}
M.~Cerezo et~al.
\newblock Variational quantum algorithms.
\newblock {\em Nature Rev. Phys.}, 3:625--644, 2021.

\bibitem{cerezo2023does}
M.~Cerezo et~al.
\newblock Does provable absence of barren plateaus imply classical
  simulability? or, why we need to rethink variational quantum computing.
\newblock {\em arXiv:2312.09121}, 2023.

\bibitem{Dequantization2}
M.~Cerezo, M.~Larocca, D.~García-Martín, N.~L. Diaz, P.~Braccia, E.~Fontana,
  M.~S. Rudolph, S.~T. Pablo~Bermejo, Aroosa~Ijaz, E.~R. Anschuetz, and
  Z.~Holmes.
\newblock Does provable absence of barren plateaus imply classical
  simulability.
\newblock {\em Nature Comm.}, 16:7907, 2025.

\bibitem{cervera2021meta}
A.~Cervera-Lierta, J.~S. Kottmann, and A.~Aspuru-Guzik.
\newblock {Meta-variational quantum eigensolver: Learning energy profiles of
  parameterized Hamiltonians for quantum simulation}.
\newblock {\em PRX Quantum}, 2:020329, 2021.

\bibitem{cha2021attention}
P.~Cha, P.~Ginsparg, F.~Wu, J.~Carrasquilla, P.~L. McMahon, and E.-A. Kim.
\newblock Attention-based quantum tomography.
\newblock {\em Mach. Learn. Sci. Technol.}, 3:01LT01, 2021.

\bibitem{chang2024survey}
Y.~Chang et~al.
\newblock A survey on evaluation of large language models.
\newblock {\em ACM transactions on intelligent systems and technology},
  15:1--45, 2024.

\bibitem{PhysRevResearch.3.023246}
L.~Che, C.~Wei, Y.~Huang, D.~Zhao, S.~Xue, X.~Nie, J.~Li, D.~Lu, and T.~Xin.
\newblock {Learning quantum Hamiltonians from single-qubit measurements}.
\newblock {\em Phys. Rev. Res.}, 3:023246, 2021.

\bibitem{che2024exponentially}
Y.~Che, C.~Gneiting, and F.~Nori.
\newblock Exponentially improved efficient machine learning for quantum
  many-body states with provable guarantees.
\newblock {\em Phys. Rev. Res.}, 6:033035, 2024.

\bibitem{che2025quantum}
Y.~Che, C.~Gneiting, X.~Wang, and F.~Nori.
\newblock Quantum circuit complexity and unsupervised machine learning of
  topological order.
\newblock {\em arXiv preprint arXiv:2508.04486}, 2025.

\bibitem{chen2024empowering}
A.~Chen and M.~Heyl.
\newblock Empowering deep neural quantum states through efficient optimization.
\newblock {\em Nature Phys.}, 20:1476--1481, 2024.

\bibitem{chen2021entanglement}
C.~Chen, C.~Ren, H.~Lin, and H.~Lu.
\newblock Entanglement structure detection via machine learning.
\newblock {\em Quantum Sci. Tech.}, 6:035017, 2021.

\bibitem{chen2023learnability}
S.~Chen, Y.~Liu, M.~Otten, A.~Seif, B.~Fefferman, and L.~Jiang.
\newblock {The learnability of Pauli noise}.
\newblock {\em Nature Comm.}, 14:52, 2023.

\bibitem{chen2024predicting}
S.~Chen, J.~d.~D. Pont, J.-T. Hsieh, H.-Y. Huang, J.~Lange, and J.~Li.
\newblock Predicting quantum channels over general product distributions.
\newblock {\em arXiv:2409.03684}, 2024.

\bibitem{chen2021detecting}
Y.~Chen, Y.~Pan, G.~Zhang, and S.~Cheng.
\newblock Detecting quantum entanglement with unsupervised learning.
\newblock {\em Quantum Sci. Tech.}, 7:015005, 2021.

\bibitem{chen2023certifying}
Z.~Chen, X.~Lin, and Z.~Wei.
\newblock Certifying unknown genuine multipartite entanglement by neural
  networks.
\newblock {\em Quantum Sci. Tech.}, 8:035029, 2023.

\bibitem{cho2024machine}
G.~Cho and D.~Kim.
\newblock Machine learning on quantum experimental data toward solving quantum
  many-body problems.
\newblock {\em Nature Comm.}, 15:7552, 2024.

\bibitem{PhysRevA.105.042403}
M.~Choi, D.~Flam-Shepherd, T.~H. Kyaw, and A.~Aspuru-Guzik.
\newblock Learning quantum dynamics with latent neural ordinary differential
  equations.
\newblock {\em Phys. Rev. A}, 105:042403, 2022.

\bibitem{PhysRevLett.125.160504}
V.~Cimini, M.~Barbieri, N.~Treps, M.~Walschaers, and V.~Parigi.
\newblock {Neural networks for detecting multimode Wigner negativity}.
\newblock {\em Phys. Rev. Lett.}, 125:160504, 2020.

\bibitem{RevModPhys.93.045003}
J.~I. Cirac, D.~P\'erez-Garc\'{\i}a, N.~Schuch, and F.~Verstraete.
\newblock Matrix product states and projected entangled pair states: Concepts,
  symmetries, theorems.
\newblock {\em Rev. Mod. Phys.}, 93:045003, 2021.

\bibitem{cirstoiu2024fourier}
C.~Cirstoiu.
\newblock {A Fourier analysis framework for approximate classical simulations
  of quantum circuits}.
\newblock {\em arXiv:2410.13856}, 2024.

\bibitem{Cramer-NatComm-2010}
M.~Cramer, M.~B. Plenio, S.~T. Flammia, R.~Somma, D.~Gross, S.~Bartlett,
  O.~Landon-Cardinal, D.~Poulin, and Y.-K. Liu.
\newblock Efficient quantum state tomography.
\newblock {\em Nature Comm.}, 1:149, 2010.

\bibitem{cybinski2025characterizing}
K.~Cybi{\'n}ski, M.~P{\l}odzie{\'n}, M.~Tomza, M.~Lewenstein, A.~Dauphin, and
  A.~Dawid.
\newblock {Characterizing out-of-distribution generalization of neural
  networks: application to the disordered Su-Schrieffer-Heeger model}.
\newblock {\em Mach. Learn. Sci. Technol.}, 6:015014, 2025.

\bibitem{czarnik2021error}
P.~Czarnik, A.~Arrasmith, P.~J. Coles, and L.~Cincio.
\newblock Error mitigation with clifford quantum-circuit data.
\newblock {\em Quantum}, 5:592, 2021.

\bibitem{czarnik2022improving}
P.~Czarnik, M.~McKerns, A.~T. Sornborger, and L.~Cincio.
\newblock Improving the efficiency of learning-based error mitigation.
\newblock {\em arXiv:2204.07109}, 2022.

\bibitem{PhysRevB.105.205108}
S.~Czischek, M.~S. Moss, M.~Radzihovsky, E.~Merali, and R.~G. Melko.
\newblock {Data-enhanced variational Monte Carlo simulations for Rydberg atom
  arrays}.
\newblock {\em Phys. Rev. B}, 105:205108, 2022.

\bibitem{CUP}
A.~Dawid, J.~Arnold, B.~Requena, A.~Gresch, M.~Płodzień, K.~Donatella, K.~A.
  Nicoli, P.~Stornati, R.~Koch, M.~Büttner, R.~Okuła, G.~Muñoz-Gil, R.~A.
  Vargas-Hernández, A.~Cervera-Lierta, J.~Carrasquilla, V.~Dunjko, M.~Gabrié,
  P.~H.~E. van Nieuwenburg, F.~Vicentini, L.~Wang, S.~J. Wetzel, G.~Carleo,
  E.~Greplová, R.~Krems, F.~Marquardt, M.~Tomza, M.~Lewenstein, and
  A.~Dauphin.
\newblock {\em Machine learning in quantum sciences}.
\newblock Cambridge University Press, Cambridge, 2025.

\bibitem{de2025interpretable}
P.~de~Schoulepnikoff, G.~Mu{\~n}oz-Gil, H.~P. Nautrup, and H.~J. Briegel.
\newblock Interpretable representation learning of quantum data enabled by
  probabilistic variational autoencoders.
\newblock {\em arXiv:2506.11982}, 2025.

\bibitem{decross2024computational}
M.~DeCross et~al.
\newblock The computational power of random quantum circuits in arbitrary
  geometries.
\newblock {\em arXiv:2406.02501}, 2024.

\bibitem{dehghani2023neural}
H.~Dehghani, A.~Lavasani, M.~Hafezi, and M.~J. Gullans.
\newblock Neural-network decoders for measurement induced phase transitions.
\newblock {\em Nature Comm.}, 14:2918, 2023.

\bibitem{denis2023estimation}
J.~Denis, F.~Damanet, and J.~Martin.
\newblock Estimation of the geometric measure of entanglement with wehrl
  moments through artificial neural networks.
\newblock {\em SciPost Physics}, 15:208, 2023.

\bibitem{du2025quantum}
Y.~Du et~al.
\newblock Quantum machine learning: A hands-on tutorial for machine learning
  practitioners and researchers.
\newblock {\em arXiv:2502.01146}, 2025.

\bibitem{du2025efficient}
Y.~Du, M.-H. Hsieh, and D.~Tao.
\newblock Efficient learning for linear properties of bounded-gate quantum
  circuits.
\newblock {\em Nature Comm.}, 16:3790, 2025.

\bibitem{du2023shadownet}
Y.~Du, Y.~Yang, T.~Liu, Z.~Lin, B.~Ghanem, and D.~Tao.
\newblock Shadownet for data-centric quantum system learning.
\newblock {\em arXiv:2308.11290}, 2023.

\bibitem{du2023problem}
Y.~Du, Y.~Yang, D.~Tao, and M.-H. Hsieh.
\newblock Problem-dependent power of quantum neural networks on multiclass
  classification.
\newblock {\em Phys. Rev. Lett.}, 131:140601, 2023.

\bibitem{dupuis2024qiskit}
N.~Dupuis, L.~Buratti, S.~Vishwakarma, A.~V. Forrat, D.~Kremer, I.~Faro,
  R.~Puri, and J.~Cruz-Benito.
\newblock {Qiskit code assistant: Training LLMs for generating quantum
  computing code}.
\newblock In {\em 2024 IEEE LLM Aided Design Workshop (LAD)}, pages 1--4. IEEE,
  2024.

\bibitem{eisert2010colloquium}
J.~Eisert, M.~Cramer, and M.~B. Plenio.
\newblock Colloquium: Area laws for the entanglement entropy.
\newblock {\em Rev. Mod. Phys.}, 82:277, 2010.

\bibitem{eisert2020quantum}
J.~Eisert, D.~Hangleiter, N.~Walk, I.~Roth, D.~Markham, R.~Parekh, U.~Chabaud,
  and E.~Kashefi.
\newblock Quantum certification and benchmarking.
\newblock {\em Nature Rev. Phys.}, 2:382--390, 2020.

\bibitem{elben2020cross}
A.~Elben et~al.
\newblock Cross-platform verification of intermediate scale quantum devices.
\newblock {\em Phys. Rev. Lett.}, 124:010504, 2020.

\bibitem{elben2023randomized}
A.~Elben, S.~T. Flammia, H.-Y. Huang, R.~Kueng, J.~Preskill, B.~Vermersch, and
  P.~Zoller.
\newblock The randomized measurement toolbox.
\newblock {\em Nature Rev. Phys.}, 5:9--24, 2023.

\bibitem{fitzek2024rydberggpt}
D.~Fitzek et~al.
\newblock {RydbergGPT}.
\newblock {\em arXiv:2405.21052}, 2024.

\bibitem{flam2022learning}
D.~Flam-Shepherd, T.~C. Wu, X.~Gu, A.~Cervera-Lierta, M.~Krenn, and
  A.~Aspuru-Guzik.
\newblock Learning interpretable representations of entanglement in quantum
  optics experiments using deep generative models.
\newblock {\em Nature Mach. Intell.}, 4:544--554, 2022.

\bibitem{PhysRevLett.106.230501}
S.~T. Flammia and Y.-K. Liu.
\newblock Direct fidelity estimation from few pauli measurements.
\newblock {\em Phys. Rev. Lett.}, 106:230501, 2011.

\bibitem{fontana2023classical}
E.~Fontana, M.~S. Rudolph, R.~Duncan, I.~Rungger, and C.~C{\^\i}rstoiu.
\newblock Classical simulations of noisy variational quantum circuits.
\newblock {\em arXiv:2306.05400}, 2023.

\bibitem{fosel2021quantum}
T.~F{\"o}sel, M.~Y. Niu, F.~Marquardt, and L.~Li.
\newblock Quantum circuit optimization with deep reinforcement learning.
\newblock {\em arXiv:2103.07585}, 2021.

\bibitem{friedrich2022avoiding}
L.~Friedrich and J.~Maziero.
\newblock Avoiding barren plateaus with classical deep neural networks.
\newblock {\em Phys. Rev. A}, 106:042433, 2022.

\bibitem{frohnert2024explainable}
F.~Frohnert and E.~van Nieuwenburg.
\newblock Explainable representation learning of small quantum states.
\newblock {\em Mach. Learn. Sci. Technol.}, 5:015001, 2024.

\bibitem{furrutter2024quantum}
F.~F{\"u}rrutter, G.~Mu{\~n}oz-Gil, and H.~J. Briegel.
\newblock Quantum circuit synthesis with diffusion models.
\newblock {\em Nature Mach. Intell.}, 6:515--524, 2024.

\bibitem{gan2024concept}
B.~Y. Gan, P.-W. Huang, E.~Gil-Fuster, and P.~Rebentrost.
\newblock Concept learning of parameterized quantum models from limited
  measurements.
\newblock {\em arXiv:2408.05116}, 2024.

\bibitem{gao2025establishing}
D.~Gao et~al.
\newblock Establishing a new benchmark in quantum computational advantage with
  105-qubit zuchongzhi 3.0 processor.
\newblock {\em Phys. Rev. Lett.}, 134:090601, 2025.

\bibitem{PhysRevLett.120.240501}
J.~Gao et~al.
\newblock Experimental machine learning of quantum states.
\newblock {\em Phys. Rev. Lett.}, 120:240501, 2018.

\bibitem{gao2017efficient1}
X.~Gao and L.-M. Duan.
\newblock Efficient representation of quantum many-body states with deep neural
  networks.
\newblock {\em Nature Comm.}, 8:662, 2017.

\bibitem{PhysRevLett.132.220202}
X.~Gao, M.~Isoard, F.~Sun, C.~E. Lopetegui, Y.~Xiang, V.~Parigi, Q.~He, and
  M.~Walschaers.
\newblock Correlation-pattern-based continuous variable entanglement detection
  through neural networks.
\newblock {\em Phys. Rev. Lett.}, 132:220202, 2024.

\bibitem{ge2024quantum}
Y.~Ge et~al.
\newblock Quantum circuit synthesis and compilation optimization: Overview and
  prospects.
\newblock {\em arXiv:2407.00736}, 2024.

\bibitem{gebhart2023learning}
V.~Gebhart et~al.
\newblock Learning quantum systems.
\newblock {\em Nature Rev. Phys.}, pages 1--16, 2023.

\bibitem{PRXQuantum.2.040355}
E.~Genois, J.~A. Gross, A.~Di~Paolo, N.~J. Stevenson, G.~Koolstra, A.~Hashim,
  I.~Siddiqi, and A.~Blais.
\newblock Quantum-tailored machine-learning characterization of a
  superconducting qubit.
\newblock {\em PRX Quantum}, 2:040355, 2021.

\bibitem{Dequantization}
E.~Gil-Fuster, C.~Gyurik, A.~Perez-Salinas, and V.~Dunjko.
\newblock On the relation between trainability and dequantization of
  variational quantum learning models.
\newblock In {\em The Thirteenth International Conference on Learning
  Representations}, 2025.

\bibitem{goh2023lie}
M.~L. Goh, M.~Larocca, L.~Cincio, M.~Cerezo, and F.~Sauvage.
\newblock Lie-algebraic classical simulations for variational quantum
  computing.
\newblock {\em arXiv:2308.01432}, 2023.

\bibitem{goodfellow2016deep}
I.~Goodfellow, Y.~Bengio, and A.~Courville.
\newblock {\em Deep learning}.
\newblock MIT press, 2016.

\bibitem{greplova2020unsupervised}
E.~Greplova, A.~Valenti, G.~Boschung, F.~Sch{\"a}fer, N.~L{\"o}rch, and S.~D.
  Huber.
\newblock Unsupervised identification of topological phase transitions using
  predictive models.
\newblock {\em New J. Phys.}, 22:045003, 2020.

\bibitem{grewal2023efficient}
S.~Grewal, V.~Iyer, W.~Kretschmer, and D.~Liang.
\newblock {Efficient learning of quantum states prepared with few non-Clifford
  gates}.
\newblock {\em arXiv:2305.13409}, 2023.

\bibitem{RigorousTensorNetworkTomography2}
Y.~Guo and S.~Yang.
\newblock Quantum state tomography with locally purified density operators and
  local measurements.
\newblock {\em Commun. Phys.}, 7:322, 2024.

\bibitem{gurvits2003classical}
L.~Gurvits.
\newblock Classical deterministic complexity of edmonds' problem and quantum
  entanglement.
\newblock In {\em Proceedings of the thirty-fifth annual ACM symposium on
  Theory of computing}, pages 10--19, 2003.

\bibitem{gyurik2022establishing}
C.~Gyurik and V.~Dunjko.
\newblock On establishing learning separations between classical and quantum
  machine learning with classical data.
\newblock {\em arXiv:2208.06339}, 2022.

\bibitem{gyurik2023exponential}
C.~Gyurik and V.~Dunjko.
\newblock Exponential separations between classical and quantum learners.
\newblock {\em arXiv:2306.16028}, 2023.

\bibitem{HamiltonianLearningGoogle}
D.~Hangleiter, I.~Roth, J.~Fuksa, J.~Eisert, and P.~Roushan.
\newblock {Robustly learning the Hamiltonian dynamics of a superconducting
  quantum processor}.
\newblock {\em Nature Comm.}, 15:9595, 2024.

\bibitem{harney2020entanglement}
C.~Harney, S.~Pirandola, A.~Ferraro, and M.~Paternostro.
\newblock Entanglement classification via neural network quantum states.
\newblock {\em New J. Phys.}, 22:045001, 2020.

\bibitem{harrington2022engineered}
P.~M. Harrington, E.~J. Mueller, and K.~W. Murch.
\newblock Engineered dissipation for quantum information science.
\newblock {\em Nature Rev. Phys.}, 4:660--671, 2022.

\bibitem{hartmann2019neural}
M.~J. Hartmann and G.~Carleo.
\newblock Neural-network approach to dissipative quantum many-body dynamics.
\newblock {\em Phys. Rev. Lett.}, 122:250502, 2019.

\bibitem{hartnett2024learning}
G.~S. Hartnett, A.~Barbosa, P.~S. Mundada, M.~Hush, M.~J. Biercuk, and Y.~Baum.
\newblock Learning to rank quantum circuits for hardware-optimized performance
  enhancement.
\newblock {\em Quantum}, 8:1542, 2024.

\bibitem{he2023gnn}
Z.~He, X.~Zhang, C.~Chen, Z.~Huang, Y.~Zhou, and H.~Situ.
\newblock A gnn-based predictor for quantum architecture search.
\newblock {\em Quantum Information Processing}, 22:128, 2023.

\bibitem{PhysRevResearch.2.023358}
M.~Hibat-Allah, M.~Ganahl, L.~E. Hayward, R.~G. Melko, and J.~Carrasquilla.
\newblock Recurrent neural network wave functions.
\newblock {\em Phys. Rev. Res.}, 2:023358, 2020.

\bibitem{hinton2012practical}
G.~E. Hinton.
\newblock A practical guide to training restricted boltzmann machines.
\newblock In {\em Neural networks: Tricks of the trade}, pages 599--619.
  Springer, 2012.

\bibitem{ho2020denoising}
J.~Ho, A.~Jain, and P.~Abbeel.
\newblock Denoising diffusion probabilistic models.
\newblock {\em Adv. Neur. Inf. Proc. Sys.}, 33:6840--6851, 2020.

\bibitem{horodecki2009quantum}
R.~Horodecki, P.~Horodecki, M.~Horodecki, and K.~Horodecki.
\newblock Quantum entanglement.
\newblock {\em Rev. Mod. Phys.}, 81:865--942, 2009.

\bibitem{hospedales2021meta}
T.~Hospedales, A.~Antoniou, P.~Micaelli, and A.~Storkey.
\newblock Meta-learning in neural networks: A survey.
\newblock {\em IEEE Trans. Patt. Ana. Mach. Int.}, 44:5149--5169, 2021.

\bibitem{hothem2024my}
D.~Hothem, A.~Miller, and T.~Proctor.
\newblock What is my quantum computer good for? quantum capability learning
  with physics-aware neural networks.
\newblock {\em arXiv:2406.05636}, 2024.

\bibitem{hothem2024learning}
D.~Hothem, K.~Young, T.~Catanach, and T.~Proctor.
\newblock Learning a quantum computer's capability.
\newblock {\em IEEE Trans. Quant. Eng.}, 2024.

\bibitem{hou2024unsupervised}
B.~Hou, J.~Wu, and D.~Y. Qiu.
\newblock Unsupervised representation learning of kohn--sham states and
  consequences for downstream predictions of many-body effects.
\newblock {\em Nature Comm.}, 15:9481, 2024.

\bibitem{hu2025demonstration}
H.-Y. Hu, A.~Gu, S.~Majumder, H.~Ren, Y.~Zhang, D.~S. Wang, Y.-Z. You,
  Z.~Minev, S.~F. Yelin, and A.~Seif.
\newblock Demonstration of robust and efficient quantum property learning with
  shallow shadows.
\newblock {\em Nature Communications}, 16(1):2943, 2025.

\bibitem{huang2022learning}
H.-Y. Huang.
\newblock Learning quantum states from their classical shadows.
\newblock {\em Nature Rev. Phys.}, 4:81--81, 2022.

\bibitem{huang2023learning}
H.-Y. Huang, S.~Chen, and J.~Preskill.
\newblock Learning to predict arbitrary quantum processes.
\newblock {\em PRX Quantum}, 4:040337, 2023.

\bibitem{huang2020predicting}
H.-Y. Huang, R.~Kueng, and J.~Preskill.
\newblock Predicting many properties of a quantum system from very few
  measurements.
\newblock {\em Nature Phys.}, 16:1050--1057, 2020.

\bibitem{huang2022provably}
H.-Y. Huang, R.~Kueng, G.~Torlai, V.~V. Albert, and J.~Preskill.
\newblock Provably efficient machine learning for quantum many-body problems.
\newblock {\em Science}, 377:eabk3333, 2022.

\bibitem{huang2024learning}
H.-Y. Huang, Y.~Liu, M.~Broughton, I.~Kim, A.~Anshu, Z.~Landau, and J.~R.
  McClean.
\newblock Learning shallow quantum circuits.
\newblock In {\em Proceedings of the 56th Annual ACM Symposium on Theory of
  Computing}, pages 1343--1351, 2024.

\bibitem{huang2025sequence}
J.~Huang, Y.~Zhu, G.~Chiribella, and Y.-D. Wu.
\newblock Sequence-model-guided measurement selection for quantum state
  learning.
\newblock {\em arXiv preprint arXiv:2507.09891}, 2025.

\bibitem{huang2022learningto}
R.~Huang, X.~Tan, and Q.~Xu.
\newblock Learning to learn variational quantum algorithm.
\newblock {\em IEEE Trans. Neur. Net. Learn. Sys.}, 34:8430--8440, 2022.

\bibitem{huang2025direct}
Y.~Huang, L.~Che, C.~Wei, F.~Xu, X.~Nie, J.~Li, D.~Lu, and T.~Xin.
\newblock Direct entanglement detection of quantum systems using machine
  learning.
\newblock {\em npjqi}, 11:29, 2025.

\bibitem{huggins2021virtual}
W.~J. Huggins et~al.
\newblock Virtual distillation for quantum error mitigation.
\newblock {\em Phys. Rev. X}, 11:041036, 2021.

\bibitem{PRXQuantum.5.020304}
M.~Ippoliti and V.~Khemani.
\newblock Learnability transitions in monitored quantum dynamics via
  eavesdropper's classical shadows.
\newblock {\em PRX Quantum}, 5:020304, 2024.

\bibitem{iten2020}
R.~Iten, T.~Metger, H.~Wilming, L.~del Rio, and R.~Renner.
\newblock Discovering physical concepts with neural networks.
\newblock {\em Phys. Rev. Lett.}, 124:010508, 2020.

\bibitem{jain2022graph}
N.~Jain, B.~Coyle, E.~Kashefi, and N.~Kumar.
\newblock Graph neural network initialisation of quantum approximate
  optimisation.
\newblock {\em Quantum}, 6:861, 2022.

\bibitem{RigorousTensorNetworkTomography1}
C.~Jameson, Z.~Qin, A.~Goldar, M.~B. Wakin, Z.~Zhu, and Z.~Gong.
\newblock Optimal quantum state tomography with local informationally complete
  measurements.
\newblock {\em arXiv:2408.07115}.

\bibitem{jiang2023adversarial}
S.~Jiang, S.~Lu, and D.-L. Deng.
\newblock Adversarial machine learning phases of matter.
\newblock {\em Quantum Frontiers}, 2:15, 2023.

\bibitem{kaplan2020scaling}
J.~Kaplan et~al.
\newblock Scaling laws for neural language models.
\newblock {\em arXiv:2001.08361}, 2020.

\bibitem{kaur2022trustworthy}
D.~Kaur, S.~Uslu, K.~J. Rittichier, and A.~Durresi.
\newblock Trustworthy artificial intelligence: a review.
\newblock {\em ACM Comp. Surv. (CSUR)}, 55:1--38, 2022.

\bibitem{khairy2020learning}
S.~Khairy, R.~Shaydulin, L.~Cincio, Y.~Alexeev, and P.~Balaprakash.
\newblock Learning to optimize variational quantum circuits to solve
  combinatorial problems.
\newblock In {\em Proceedings of the AAAI conference on artificial
  intelligence}, volume~34, pages 2367--2375, 2020.

\bibitem{kim2020quantum}
C.~Kim, K.~D. Park, and J.-K. Rhee.
\newblock Quantum error mitigation with artificial neural network.
\newblock {\em IEEE Access}, 8:188853--188860, 2020.

\bibitem{kim2024attention}
H.~Kim et~al.
\newblock Attention to quantum complexity.
\newblock {\em arXiv:2405.11632}, 2024.

\bibitem{kim2025learning}
H.~Kim, A.~Kumar, Y.~Zhou, Y.~Xu, R.~Vasseur, and E.-A. Kim.
\newblock Learning measurement-induced phase transitions using attention.
\newblock {\em arXiv preprint arXiv:2508.15895}, 2025.

\bibitem{king2025beyond}
A.~D. King et~al.
\newblock Beyond-classical computation in quantum simulation.
\newblock {\em Science}, page eado6285, 2025.

\bibitem{kookani2024xpookynet}
A.~Kookani, Y.~Mafi, P.~Kazemikhah, H.~Aghababa, K.~Fouladi, and M.~Barati.
\newblock Xpookynet: advancement in quantum system analysis through
  convolutional neural networks for detection of entanglement.
\newblock {\em Quantum Mach. Intell.}, 6:50, 2024.

\bibitem{PhysRevLett.125.170603}
K.~Kottmann, P.~Huembeli, M.~Lewenstein, and A.~Ac\'{\i}n.
\newblock Unsupervised phase discovery with deep anomaly detection.
\newblock {\em Phys. Rev. Lett.}, 125:170603, 2020.

\bibitem{koutny2023deep}
D.~Koutn{\`y} et~al.
\newblock Deep learning of quantum entanglement from incomplete measurements.
\newblock {\em Sci. Adv.}, 9:eadd7131, 2023.

\bibitem{krawczyk2024data}
M.~Krawczyk, J.~Paw{\l}owski, M.~M. Ma{\'s}ka, and K.~Roszak.
\newblock Data-driven criteria for quantum correlations.
\newblock {\em Phys. Rev. A}, 109:022405, 2024.

\bibitem{PhysRevA.107.010101}
M.~Krenn, J.~Landgraf, T.~Foesel, and F.~Marquardt.
\newblock Artificial intelligence and machine learning for quantum
  technologies.
\newblock {\em Phys. Rev. A}, 107:010101, 2023.

\bibitem{landau2024learning}
Z.~Landau and Y.~Liu.
\newblock Learning quantum states prepared by shallow circuits in polynomial
  time.
\newblock {\em arXiv:2410.23618}, 2024.

\bibitem{landman2022classically}
J.~Landman, S.~Thabet, C.~Dalyac, H.~Mhiri, and E.~Kashefi.
\newblock {C}lassically {a}pproximating {v}ariational {q}uantum {m}achine
  {l}earning with {r}andom {F}ourier {f}eatures, 2022.
\newblock arXiv:2210.13200v1.

\bibitem{lange2025transformer}
H.~Lange et~al.
\newblock Transformer neural networks and quantum simulators: a hybrid approach
  for simulating strongly correlated systems.
\newblock {\em Quantum}, 9:1675, 2025.

\bibitem{lange2023adaptive}
H.~Lange, M.~Kebri{\v{c}}, M.~Buser, U.~Schollw{\"o}ck, F.~Grusdt, and
  A.~Bohrdt.
\newblock Adaptive quantum state tomography with active learning.
\newblock {\em Quantum}, 7:1129, 2023.

\bibitem{lange2024architectures}
H.~Lange, A.~Van~de Walle, A.~Abedinnia, and A.~Bohrdt.
\newblock From architectures to applications: A review of neural quantum
  states.
\newblock {\em Quantum Sci. Tech.}, 2024.

\bibitem{lange2023data}
M.~Lange, P.~Havstr{\"o}m, B.~Srivastava, V.~Bergentall, K.~Hammar, O.~Heuts,
  E.~van Nieuwenburg, and M.~Granath.
\newblock Data-driven decoding of quantum error correcting codes using graph
  neural networks.
\newblock {\em arXiv:2307.01241}, 2023.

\bibitem{lecompte2023machine}
T.~LeCompte, F.~Qi, X.~Yuan, N.-F. Tzeng, M.~H. Najafi, and L.~Peng.
\newblock Machine-learning-based qubit allocation for error reduction in
  quantum circuits.
\newblock {\em IEEE Trans. Quant. Eng.}, 4:1--14, 2023.

\bibitem{lee2025q}
J.~Lee, J.~Cho, and S.~Kim.
\newblock Q-maml: Quantum model-agnostic meta-learning for variational quantum
  algorithms.
\newblock {\em arXiv:2501.05906}, 2025.

\bibitem{leone2024learning}
L.~Leone, S.~F. Oliviero, and A.~Hamma.
\newblock Learning t-doped stabilizer states.
\newblock {\em Quantum}, 8:1361, 2024.

\bibitem{lewis2024improved}
L.~Lewis, H.-Y. Huang, V.~T. Tran, S.~Lehner, R.~Kueng, and J.~Preskill.
\newblock Improved machine learning algorithm for predicting ground state
  properties.
\newblock {\em Nature Comm.}, 15:895, 2024.

\bibitem{liang2023unleashing}
Z.~Liang, J.~Cheng, R.~Yang, H.~Ren, Z.~Song, D.~Wu, X.~Qian, T.~Li, and
  Y.~Shi.
\newblock {Unleashing the potential of LLMs for quantum computing: A study in
  quantum architecture design}.
\newblock {\em arXiv:2307.08191}, 2023.

\bibitem{liao2024machine}
H.~Liao, D.~S. Wang, I.~Sitdikov, C.~Salcedo, A.~Seif, and Z.~K. Minev.
\newblock Machine learning for practical quantum error mitigation.
\newblock {\em Nature Mach. Intell.}, pages 1--9, 2024.

\bibitem{liao2025noise}
M.~Liao, Y.~Zhu, G.~Chiribella, and Y.~Yang.
\newblock Noise-agnostic quantum error mitigation with data augmented neural
  models.
\newblock {\em npjqi}, 11:8, 2025.

\bibitem{lin2023quantifying}
X.~Lin, Z.~Chen, and Z.~Wei.
\newblock Quantifying quantum entanglement via a hybrid quantum-classical
  machine learning framework.
\newblock {\em Phys. Rev. A}, 107:062409, 2023.

\bibitem{liu2018}
Y.-H. Liu and E.~P.~L. van Nieuwenburg.
\newblock Discriminative cooperative networks for detecting phase transitions.
\newblock {\em Phys. Rev. Lett.}, 120:176401, 2018.

\bibitem{lu2019vilbert}
J.~Lu, D.~Batra, D.~Parikh, and S.~Lee.
\newblock Vilbert: Pretraining task-agnostic visiolinguistic representations
  for vision-and-language tasks.
\newblock {\em Adv. Neur. Inf. Proc. Sys.}, 32, 2019.

\bibitem{luo2023quack}
D.~Luo, J.~Shen, R.~Dangovski, and M.~Soljacic.
\newblock Quack: accelerating gradient-based quantum optimization with koopman
  operator learning.
\newblock {\em Adv. Neur. Inf. Proc. Sys.}, 36:25662--25692, 2023.

\bibitem{luo2023detecting}
Y.-J. Luo, J.-M. Liu, and C.~Zhang.
\newblock Detecting genuine multipartite entanglement via machine learning.
\newblock {\em Phys. Rev. A}, 108:052424, 2023.

\bibitem{manetsch2024tweezer}
H.~J. Manetsch, G.~Nomura, E.~Bataille, K.~H. Leung, X.~Lv, and M.~Endres.
\newblock A tweezer array with 6100 highly coherent atomic qubits.
\newblock {\em arXiv:2403.12021}, 2024.

\bibitem{manovitz2025quantum}
T.~Manovitz et~al.
\newblock Quantum coarsening and collective dynamics on a programmable
  simulator.
\newblock {\em Nature}, 638:86--92, 2025.

\bibitem{melko2024language}
R.~G. Melko and J.~Carrasquilla.
\newblock Language models for quantum simulation.
\newblock {\em Nature Comp. Sci.}, 4:11--18, 2024.

\bibitem{mello2024retrieving}
A.~F. Mello, G.~Lami, and M.~Collura.
\newblock Retrieving nonstabilizerness with neural networks.
\newblock {\em Phys. Rev. A}, 111:012440, 2025.

\bibitem{miles2021correlator}
C.~Miles et~al.
\newblock Correlator convolutional neural networks as an interpretable
  architecture for image-like quantum matter data.
\newblock {\em Nature Comm.}, 12:3905, 2021.

\bibitem{PhysRevResearch.5.013026}
C.~Miles et~al.
\newblock Machine learning discovery of new phases in programmable quantum
  simulator snapshots.
\newblock {\em Phys. Rev. Res.}, 5:013026, 2023.

\bibitem{minami2025generative}
S.~Minami, K.~Nakaji, Y.~Suzuki, A.~Aspuru-Guzik, and T.~Kadowaki.
\newblock Generative quantum combinatorial optimization by means of a novel
  conditional generative quantum eigensolver.
\newblock {\em arXiv:2501.16986}, 2025.

\bibitem{mohri2018foundations}
M.~Mohri, A.~Rostamizadeh, and A.~Talwalkar.
\newblock {\em Foundations of machine learning}.
\newblock MIT press, 2018.

\bibitem{mohseni2022deep}
N.~Mohseni, T.~F{\"o}sel, L.~Guo, C.~Navarrete-Benlloch, and F.~Marquardt.
\newblock Deep learning of quantum many-body dynamics via random driving.
\newblock {\em Quantum}, 6:714, 2022.

\bibitem{mohseni2025transfer}
N.~Mohseni, F.~Marquardt, and P.~Schmidt.
\newblock Transfer learning in predicting quantum many-body dynamics: from
  physical observables to entanglement entropy.
\newblock {\em Quantum Sci. Tech.}, 2025.

\bibitem{mohseni2024deep}
N.~Mohseni, J.~Shi, T.~Byrnes, and M.~J. Hartmann.
\newblock Deep learning of many-body observables and quantum information
  scrambling.
\newblock {\em Quantum}, 8:1417, 2024.

\bibitem{molteni2024exponential}
R.~Molteni, C.~Gyurik, and V.~Dunjko.
\newblock Exponential quantum advantages in learning quantum observables from
  classical data.
\newblock {\em arXiv:2405.02027}, 2024.

\bibitem{PhysRevA.104.012401}
S.~Morawetz, I.~J.~S. De~Vlugt, J.~Carrasquilla, and R.~G. Melko.
\newblock U(1)-symmetric recurrent neural networks for quantum state
  reconstruction.
\newblock {\em Phys. Rev. A}, 104:012401, 2021.

\bibitem{moss2023enhancing}
M.~S. Moss, S.~Ebadi, T.~T. Wang, G.~Semeghini, A.~Bohrdt, M.~D. Lukin, and
  R.~G. Melko.
\newblock {Enhancing variational Monte Carlo simulations using a programmable
  quantum simulator}.
\newblock {\em Phys. Rev. A}, 109:032410, 2024.

\bibitem{nakaji2024generative}
K.~Nakaji et~al.
\newblock {The generative quantum eigensolver (GQE) and its application for
  ground state search}.
\newblock {\em arXiv:2401.09253}, 2024.

\bibitem{PhysRevResearch.6.023160}
S.~Nandy, M.~Schmitt, M.~Bukov, and Z.~Lenar\ifmmode \check{c}\else
  \v{c}\fi{}i\ifmmode~\check{c}\else \v{c}\fi{}.
\newblock {Reconstructing effective Hamiltonians from nonequilibrium thermal
  and prethermal steady states}.
\newblock {\em Phys. Rev. Res.}, 6:023160, 2024.

\bibitem{Nemkov2023Fourier}
N.~A. Nemkov, E.~O. Kiktenko, and A.~K. Fedorov.
\newblock Fourier expansion in variational quantum algorithms.
\newblock {\em Phys. Rev. A}, 108, 2023.

\bibitem{nielsen2010quantum}
M.~A. Nielsen and I.~L. Chuang.
\newblock {\em Quantum computation and quantum information}.
\newblock Cambridge University Press, 2010.

\bibitem{niu2020learnability}
M.~Y. Niu, A.~M. Dai, L.~Li, A.~Odena, Z.~Zhao, V.~Smelyanskyi, H.~Neven, and
  S.~Boixo.
\newblock Learnability and complexity of quantum samples.
\newblock {\em arXiv:2010.11983}, 2020.

\bibitem{Efficient}
M.~Ohliger, V.~Nesme, and J.~Eisert.
\newblock Efficient and feasible state tomography of quantum many-body systems.
\newblock {\em New J. Phys.}, 15:015024, 2013.

\bibitem{onorati2023provably}
E.~Onorati, C.~Rouz{\'e}, D.~S. Fran{\c{c}}a, and J.~D. Watson.
\newblock Provably efficient learning of phases of matter via dissipative
  evolutions.
\newblock {\em arXiv:2311.07506}, 2023.

\bibitem{orus2019tensor}
R.~Or{\'u}s.
\newblock Tensor networks for complex quantum systems.
\newblock {\em Nature Rev. Phys.}, 1:538--550, 2019.

\bibitem{pawlowski2024identification}
J.~Paw{\l}owski and M.~Krawczyk.
\newblock Identification of quantum entanglement with siamese convolutional
  neural networks and semisupervised learning.
\newblock {\em Phys. Rev. Applied}, 22:014068, 2024.

\bibitem{pouyanfar2018survey}
S.~Pouyanfar et~al.
\newblock A survey on deep learning: Algorithms, techniques, and applications.
\newblock {\em ACM Comp. Surv. (CSUR)}, 51:1--36, 2018.

\bibitem{preskill2025beyond}
J.~Preskill.
\newblock Beyond nisq: The megaquop machine.
\newblock {\em arXiv:2502.17368}, 2025.

\bibitem{proctor2025benchmarking}
T.~Proctor, K.~Young, A.~D. Baczewski, and R.~Blume-Kohout.
\newblock Benchmarking quantum computers.
\newblock {\em Nature Rev. Phys.}, 7:1--14, 2025.

\bibitem{qian2023multimodal}
Y.~Qian, Y.~Du, Z.~He, M.-H. Hsieh, and D.~Tao.
\newblock Multimodal deep representation learning for quantum cross-platform
  verification.
\newblock {\em Phys. Rev. Lett.}, 133:130601, 2024.

\bibitem{qian2024mg}
Y.~Qian, X.~Wang, Y.~Du, Y.~Luo, and D.~Tao.
\newblock Mg-net: Learn to customize qaoa with circuit depth awareness.
\newblock {\em arXiv:2409.18692}, 2024.

\bibitem{PhysRevLett.132.190801}
H.~Qin, L.~Che, C.~Wei, F.~Xu, Y.~Huang, and T.~Xin.
\newblock Experimental direct quantum fidelity learning via a data-driven
  approach.
\newblock {\em Phys. Rev. Lett.}, 132:190801, 2024.

\bibitem{quek2021}
Y.~Quek, S.~Fort, and H.~K. Ng.
\newblock Adaptive quantum state tomography with neural networks.
\newblock {\em npjqi}, 7:105, 2021.

\bibitem{ErrorMitigationObstructions}
Y.~Quek, D.~S. Franca, S.~Khatri, J.~J. Meyer, and J.~Eisert.
\newblock Exponentially tighter bounds on limitations of quantum error
  mitigation.
\newblock {\em Nature Phys.}, 20:1648--1658, 2024.

\bibitem{radford2018improving}
A.~Radford, K.~Narasimhan, T.~Salimans, I.~Sutskever, et~al.
\newblock Improving language understanding by generative pre-training.
\newblock 2018.

\bibitem{radford2019language}
A.~Radford, J.~Wu, R.~Child, D.~Luan, D.~Amodei, I.~Sutskever, et~al.
\newblock Language models are unsupervised multitask learners.
\newblock {\em OpenAI blog}, 1:9, 2019.

\bibitem{rahimi2007random}
A.~Rahimi and B.~Recht.
\newblock Random features for large-scale kernel machines.
\newblock {\em Adv. Neur. Inf. Proc. Sys.}, 20, 2007.

\bibitem{rende2025foundation}
R.~Rende, L.~L. Viteritti, F.~Becca, A.~Scardicchio, A.~Laio, and G.~Carleo.
\newblock {Foundation neural-networks quantum states as a unified ansatz for
  multiple Hamiltonians}.
\newblock {\em Nature Communications}, 16:7213, 2025.

\bibitem{rieger2024sample}
M.~Rieger, M.~Reh, and M.~G{\"a}rttner.
\newblock Sample-efficient estimation of entanglement entropy through
  supervised learning.
\newblock {\em Phys. Rev. A}, 109:012403, 2024.

\bibitem{rocchetto2017stabiliser}
A.~Rocchetto.
\newblock {Stabiliser states are efficiently PAC-learnable}.
\newblock {\em arXiv:1705.00345}, 2017.

\bibitem{roik2022entanglement}
J.~Roik, K.~Bartkiewicz, A.~{\v{C}}ernoch, and K.~Lemr.
\newblock Entanglement quantification from collective measurements processed by
  machine learning.
\newblock {\em Phys. Lett. A}, 446:128270, 2022.

\bibitem{roik2021accuracy}
J.~Roik, K.~Bartkiewicz, A.~\ifmmode~\check{C}\else \v{C}\fi{}ernoch, and
  K.~Lemr.
\newblock Accuracy of entanglement detection via artificial neural networks and
  human-designed entanglement witnesses.
\newblock {\em Phys. Rev. Appl.}, 15:054006, 2021.

\bibitem{romera2024mathematical}
B.~Romera-Paredes et~al.
\newblock Mathematical discoveries from program search with large language
  models.
\newblock {\em Nature}, 625:468--475, 2024.

\bibitem{rouze2024efficient}
C.~Rouz{\'e}, D.~Stilck~Fran{\c{c}}a, E.~Onorati, and J.~D. Watson.
\newblock Efficient learning of ground and thermal states within phases of
  matter.
\newblock {\em Nature Comm.}, 15:7755, 2024.

\bibitem{rudolph2023classical}
M.~S. Rudolph, E.~Fontana, Z.~Holmes, and L.~Cincio.
\newblock Classical surrogate simulation of quantum systems with lowesa.
\newblock {\em arXiv:2308.09109}, 2023.

\bibitem{ruiz2025quantum}
F.~J. Ruiz et~al.
\newblock Quantum circuit optimization with alphatensor.
\newblock {\em Nature Mach. Intell.}, pages 1--12, 2025.

\bibitem{PhysRevResearch.6.013223}
S.~H. Sack and D.~J. Egger.
\newblock Large-scale quantum approximate optimization on nonplanar graphs with
  machine learning noise mitigation.
\newblock {\em Phys. Rev. Res.}, 6:013223, 2024.

\bibitem{sadoune2023unsupervised}
N.~Sadoune, G.~Giudici, K.~Liu, and L.~Pollet.
\newblock Unsupervised interpretable learning of phases from many-qubit
  systems.
\newblock {\em Phys. Rev. Res.}, 5:013082, 2023.

\bibitem{ExplainableAI}
W.~Samek, T.~Wiegand, and K.-R. Müller.
\newblock Explainable artificial intelligence: Understanding, visualizing and
  interpreting deep learning models.
\newblock {\em arXiv:1708.08296}, 2017.

\bibitem{sarzynska2021detecting}
J.~Sarzynska-Wawer, A.~Wawer, A.~Pawlak, J.~Szymanowska, I.~Stefaniak,
  M.~Jarkiewicz, and L.~Okruszek.
\newblock Detecting formal thought disorder by deep contextualized word
  representations.
\newblock {\em Psychiatry Research}, 304:114135, 2021.

\bibitem{sauvage2021flip}
F.~Sauvage, S.~Sim, A.~A. Kunitsa, W.~A. Simon, M.~Mauri, and A.~Perdomo-Ortiz.
\newblock Flip: A flexible initializer for arbitrarily-sized parametrized
  quantum circuits.
\newblock {\em arXiv:2103.08572}, 2021.

\bibitem{PhysRevE.99.062107}
F.~Sch\"afer and N.~L\"orch.
\newblock Vector field divergence of predictive model output as indication of
  phase transitions.
\newblock {\em Phys. Rev. E}, 99:062107, 2019.

\bibitem{PhysRevB.95.245134}
F.~Schindler, N.~Regnault, and T.~Neupert.
\newblock Probing many-body localization with neural networks.
\newblock {\em Phys. Rev. B}, 95:245134, 2017.

\bibitem{schmale2022efficient}
T.~Schmale, M.~Reh, and M.~G{\"a}rttner.
\newblock Efficient quantum state tomography with convolutional neural
  networks.
\newblock {\em npjqi}, 8:115, 2022.

\bibitem{PhysRevB.106.L041110}
M.~Schmitt and Z.~Lenar\ifmmode \check{c}\else
  \v{c}\fi{}i\ifmmode~\check{c}\else \v{c}\fi{}.
\newblock From observations to complexity of quantum states via unsupervised
  learning.
\newblock {\em Phys. Rev. B}, 106:L041110, 2022.

\bibitem{scholkopf2002learning}
B.~Sch{\"o}lkopf and A.~J. Smola.
\newblock {\em Learning with kernels: support vector machines, regularization,
  optimization, and beyond}.
\newblock MIT press, 2002.

\bibitem{schreiber2022classical}
F.~J. Schreiber, J.~Eisert, and J.~J. Meyer.
\newblock Classical surrogates for quantum learning models.
\newblock {\em Phys. Rev. Lett.}, 131:100803, 2023.

\bibitem{Schuld2021Effect}
M.~Schuld, R.~Sweke, and J.~J. Meyer.
\newblock Effect of data encoding on the expressive power of variational
  quantum-machine-learning models.
\newblock {\em Phys. Rev. A}, 103, 2021.

\bibitem{ExtremelyShallow}
T.~Schuster, J.~Haferkamp, and H.-Y. Huang.
\newblock Random unitaries in extremely low depth.
\newblock {\em Science}, 389(6755):92--96, 2025.

\bibitem{servedio2004equivalences}
R.~A. Servedio and S.~J. Gortler.
\newblock Equivalences and separations between quantum and classical
  learnability.
\newblock {\em SIAM Journal on Computing}, 33:1067--1092, 2004.

\bibitem{PhysRevLett.124.020503}
O.~Sharir, Y.~Levine, N.~Wies, G.~Carleo, and A.~Shashua.
\newblock Deep autoregressive models for the efficient variational simulation
  of many-body quantum systems.
\newblock {\em Phys. Rev. Lett.}, 124:020503, 2020.

\bibitem{PhysRevB.106.205136}
O.~Sharir, A.~Shashua, and G.~Carleo.
\newblock Neural tensor contractions and the expressive power of deep neural
  quantum states.
\newblock {\em Phys. Rev. B}, 106:205136, 2022.

\bibitem{shaw2024benchmarking}
A.~L. Shaw et~al.
\newblock Benchmarking highly entangled states on a 60-atom analogue quantum
  simulator.
\newblock {\em Nature}, 628:71--77, 2024.

\bibitem{sinibaldi2025non}
A.~Sinibaldi, A.~F. Mello, M.~Collura, and G.~Carleo.
\newblock Non-stabilizerness of neural quantum states.
\newblock {\em arXiv:2502.09725}, 2025.

\bibitem{PhysRevX.9.031009}
B.~Skinner, J.~Ruhman, and A.~Nahum.
\newblock Measurement-induced phase transitions in the dynamics of
  entanglement.
\newblock {\em Phys. Rev. X}, 9:031009, 2019.

\bibitem{vsmid2024accurate}
{\v{S}}.~{\v{S}}m{\'\i}d and R.~Bondesan.
\newblock Accurate learning of equivariant quantum systems from a single ground
  state.
\newblock {\em arXiv:2405.12309}, 2024.

\bibitem{vsmid2025efficient}
{\v{S}}.~{\v{S}}m{\'\i}d and R.~Bondesan.
\newblock Efficient learning of long-range and equivariant quantum systems.
\newblock {\em Quantum}, 9:1597, 2025.

\bibitem{smith2021}
A.~W.~R. Smith, J.~Gray, and M.~S. Kim.
\newblock Efficient quantum state sample tomography with basis-dependent neural
  networks.
\newblock {\em PRX Quantum}, 2:020348, 2021.

\bibitem{Song2021Score}
Y.~Song, J.~Sohl{-}Dickstein, D.~P. Kingma, A.~Kumar, S.~Ermon, and B.~Poole.
\newblock Score-based generative modeling through stochastic differential
  equations.
\newblock In {\em 9th International Conference on Learning Representations,
  {ICLR} 2021, Virtual Event, Austria, May 3-7, 2021}. OpenReview.net, 2021.

\bibitem{strikis2021learning}
A.~Strikis, D.~Qin, Y.~Chen, S.~C. Benjamin, and Y.~Li.
\newblock Learning-based quantum error mitigation.
\newblock {\em PRX Quantum}, 2:040330, 2021.

\bibitem{Reinforcement}
R.~Sweke, M.~S. Kesselring, E.~P.~L. van Nieuwenburg, and J.~Eisert.
\newblock Reinforcement learning decoders for fault-tolerant quantum
  computation.
\newblock {\em Mach. Learn. Sci. Technol.}, 2:025005, 2021.

\bibitem{sweke2025potential}
R.~Sweke, E.~Recio-Armengol, S.~Jerbi, E.~Gil-Fuster, B.~Fuller, J.~Eisert, and
  J.~J. Meyer.
\newblock {Potential and limitations of random Fourier features for
  dequantizing quantum machine learning}.
\newblock {\em Quantum}, 9:1640, 2025.

\bibitem{taghadomi2024effective}
N.~Taghadomi, A.~Mani, A.~Fahim, A.~Bakoui, and M.~S. Salami.
\newblock Effective detection of quantum discord by using convolutional neural
  networks.
\newblock {\em arXiv:2401.07405}, 2024.

\bibitem{ErrorMitigationObstructionsOld}
R.~Takagi, S.~Endo, S.~Minagawa, and M.~Gu.
\newblock Fundamental limits of quantum error mitigation.
\newblock {\em npj Quant. Inf.}, 8:114, 2022.
\newblock arXiv:2210.11505.

\bibitem{tang2025quadim}
Y.~Tang, M.~Long, and J.~Yan.
\newblock Quadim: A conditional diffusion model for quantum state property
  estimation.
\newblock In {\em The Thirteenth International Conference on Learning
  Representations}, 2025.

\bibitem{tangtowards}
Y.~Tang, H.~Xiong, N.~Yang, T.~Xiao, and J.~Yan.
\newblock Towards llm4qpe: Unsupervised pretraining of quantum property
  estimation and a benchmark.
\newblock In {\em The Twelfth International Conference on Learning
  Representations, {ICLR} 2024, Vienna, Austria, May 7-11, 2024}.
  OpenReview.net, 2024.

\bibitem{tang2024ssl4q}
Y.~Tang, N.~Yang, M.~Long, and J.~Yan.
\newblock Ssl4q: semi-supervised learning of quantum data with application to
  quantum state classification.
\newblock In {\em Forty-first International Conference on Machine Learning},
  ICML'24, 2024.

\bibitem{temme2017error}
K.~Temme, S.~Bravyi, and J.~M. Gambetta.
\newblock Error mitigation for short-depth quantum circuits.
\newblock {\em Phys. Rev. Lett.}, 119:180509, 2017.

\bibitem{thabet2024quantum}
S.~Thabet, L.~Monbroussou, E.~Z. Mamon, and J.~Landman.
\newblock When quantum and classical models disagree: Learning beyond minimum
  norm least square.
\newblock {\em arXiv:2411.04940}, 2024.

\bibitem{tian2022recent}
J.~Tian et~al.
\newblock Recent advances for quantum neural networks in generative learning.
\newblock {\em arXiv:2206.03066}, 2022.

\bibitem{tilly2022variational}
J.~Tilly et~al.
\newblock The variational quantum eigensolver: a review of methods and best
  practices.
\newblock {\em Phys. Rep.}, 986:1--128, 2022.

\bibitem{tiunov2020experimental}
E.~S. Tiunov, V.~Tiunova, A.~E. Ulanov, A.~Lvovsky, and A.~K. Fedorov.
\newblock Experimental quantum homodyne tomography via machine learning.
\newblock {\em Optica}, 7:448--454, 2020.

\bibitem{torlai2018}
G.~Torlai, G.~Mazzola, J.~Carrasquilla, M.~Troyer, R.~Melko, and G.~Carleo.
\newblock Neural-network quantum state tomography.
\newblock {\em Nature Phys.}, 14:447--450, 2018.

\bibitem{vadali2024quantum}
A.~Vadali, R.~Kshirsagar, P.~Shyamsundar, and G.~N. Perdue.
\newblock Quantum circuit fidelity estimation using machine learning.
\newblock {\em Quantum Mach. Intell.}, 6:1, 2024.

\bibitem{PhysRevA.105.023302}
A.~Valenti, G.~Jin, J.~L\'eonard, S.~D. Huber, and E.~Greplova.
\newblock Scalable hamiltonian learning for large-scale out-of-equilibrium
  quantum dynamics.
\newblock {\em Phys. Rev. A}, 105:023302, 2022.

\bibitem{PhysRevResearch.1.033092}
A.~Valenti, E.~van Nieuwenburg, S.~Huber, and E.~Greplova.
\newblock Hamiltonian learning for quantum error correction.
\newblock {\em Phys. Rev. Res.}, 1:033092, 2019.

\bibitem{van2024advanced}
J.~Van~Damme et~al.
\newblock Advanced cmos manufacturing of superconducting qubits on 300 mm
  wafers.
\newblock {\em Nature}, 634:74--79, 2024.

\bibitem{van2008visualizing}
L.~Van~der Maaten and G.~Hinton.
\newblock Visualizing data using t-sne.
\newblock {\em J. Mach. Learn. Res.}, 9, 2008.

\bibitem{van2020survey}
J.~E. Van~Engelen and H.~H. Hoos.
\newblock A survey on semi-supervised learning.
\newblock {\em Machine learning}, 109:373--440, 2020.

\bibitem{van2017learning}
E.~P. Van~Nieuwenburg, Y.-H. Liu, and S.~D. Huber.
\newblock Learning phase transitions by confusion.
\newblock {\em Nature Phys.}, 13:435, 2017.

\bibitem{PhysRevResearch.7.013029}
B.~M. Varbanov, M.~Serra-Peralta, D.~Byfield, and B.~M. Terhal.
\newblock Neural network decoder for near-term surface-code experiments.
\newblock {\em Phys. Rev. Res.}, 7:013029, 2025.

\bibitem{vaswani2017attention}
A.~Vaswani, N.~Shazeer, N.~Parmar, J.~Uszkoreit, L.~Jones, A.~N. Gomez,
  {\L}.~Kaiser, and I.~Polosukhin.
\newblock Attention is all you need.
\newblock {\em Adv. Neur. Inf. Proc. Sys.}, 30, 2017.

\bibitem{venturella2025unified}
C.~Venturella, J.~Li, C.~Hillenbrand, X.~Leyva~Peralta, J.~Liu, and T.~Zhu.
\newblock Unified deep learning framework for many-body quantum chemistry via
  green's functions.
\newblock {\em Nature Comp. Sci.}, pages 1--12, 2025.

\bibitem{verdon2019learning}
G.~Verdon, M.~Broughton, J.~R. McClean, K.~J. Sung, R.~Babbush, Z.~Jiang,
  H.~Neven, and M.~Mohseni.
\newblock Learning to learn with quantum neural networks via classical neural
  networks.
\newblock {\em arXiv:1907.05415}, 2019.

\bibitem{PhysRevX.14.031035}
B.~Vermersch, M.~Ljubotina, J.~I. Cirac, P.~Zoller, M.~Serbyn, and L.~Piroli.
\newblock Many-body entropies and entanglement from polynomially many local
  measurements.
\newblock {\em Phys. Rev. X}, 14:031035, 2024.

\bibitem{vidal2020input}
J.~G. Vidal and D.~O. Theis.
\newblock Input redundancy for parameterized quantum circuits, 2020.

\bibitem{PRXQuantum.1.010301}
J.~Walln\"ofer, A.~A. Melnikov, W.~D\"ur, and H.~J. Briegel.
\newblock Machine learning for long-distance quantum communication.
\newblock {\em PRX Quantum}, 1:010301, 2020.

\bibitem{wang2019emergent}
C.~Wang, H.~Zhai, and Y.-Z. You.
\newblock Emergent schr{\"o}dinger equation in an introspective machine
  learning architecture.
\newblock {\em Science Bulletin}, 64(17):1228--1233, 2019.

\bibitem{wang2022quest}
H.~Wang et~al.
\newblock Quest: Graph transformer for quantum circuit reliability estimation.
\newblock {\em arXiv:2210.16724}, 2022.

\bibitem{wang2023scientific}
H.~Wang et~al.
\newblock Scientific discovery in the age of artificial intelligence.
\newblock {\em Nature}, 620:47--60, 2023.

\bibitem{wang2023transformer}
H.~Wang, P.~Liu, K.~Shao, D.~Li, J.~Gu, D.~Z. Pan, Y.~Ding, and S.~Han.
\newblock {Transformer-QEC: quantum error correction code decoding with
  transferable transformers}.
\newblock {\em arXiv:2311.16082}, 2023.

\bibitem{wang2022}
H.~Wang, M.~Weber, J.~Izaac, and C.~Y.-Y. Lin.
\newblock Predicting properties of quantum systems with conditional generative
  models.
\newblock {\em arXiv:2211.16943}, 2022.

\bibitem{wang2017experimental}
J.~Wang et~al.
\newblock Experimental quantum hamiltonian learning.
\newblock {\em Nature Phys.}, 13:551--555, 2017.

\bibitem{PhysRevB.94.195105}
L.~Wang.
\newblock Discovering phase transitions with unsupervised learning.
\newblock {\em Phys. Rev. B}, 94:195105, 2016.

\bibitem{wang2024comprehensive}
L.~Wang, X.~Zhang, H.~Su, and J.~Zhu.
\newblock A comprehensive survey of continual learning: Theory, method and
  application.
\newblock {\em IEEE Trans. Patt. Ana. Mach. Int.}, 2024.

\bibitem{wanner2024predicting}
M.~Wanner, L.~Lewis, C.~Bhattacharyya, D.~Dubhashi, and A.~Gheorghiu.
\newblock Predicting ground state properties: Constant sample complexity and
  deep learning algorithms.
\newblock {\em arXiv:2405.18489}, 2024.

\bibitem{PhysRevE.96.022140}
S.~J. Wetzel.
\newblock Unsupervised learning of phase transitions: From principal component
  analysis to variational autoencoders.
\newblock {\em Phys. Rev. E}, 96:022140, 2017.

\bibitem{wetzel2025interpretable}
S.~J. Wetzel, S.~Ha, R.~Iten, M.~Klopotek, and Z.~Liu.
\newblock Interpretable machine learning in physics: A review.
\newblock {\em arXiv:2503.23616}, 2025.

\bibitem{wiebe2014hamiltonian}
N.~Wiebe, C.~Granade, C.~Ferrie, and D.~G. Cory.
\newblock Hamiltonian learning and certification using quantum resources.
\newblock {\em Phys. Rev. Lett.}, 112:190501, 2014.

\bibitem{PRXQuantum.2.010316}
D.~F. Wise, J.~J. Morton, and S.~Dhomkar.
\newblock Using deep learning to understand and mitigate the qubit noise
  environment.
\newblock {\em PRX Quantum}, 2:010316, 2021.

\bibitem{wu2024variational}
D.~Wu et~al.
\newblock Variational benchmarks for quantum many-body problems.
\newblock {\em Science}, 386:296--301, 2024.

\bibitem{PhysRevLett.122.080602}
D.~Wu, L.~Wang, and P.~Zhang.
\newblock Solving statistical mechanics using variational autoregressive
  networks.
\newblock {\em Phys. Rev. Lett.}, 122:080602, 2019.

\bibitem{wu2023}
Y.-D. Wu, Y.~Zhu, G.~Bai, Y.~Wang, and G.~Chiribella.
\newblock Quantum similarity testing with convolutional neural networks.
\newblock {\em Phys. Rev. Lett.}, 130:210601, 2023.

\bibitem{wu2024learning}
Y.-D. Wu, Y.~Zhu, Y.~Wang, and G.~Chiribella.
\newblock Learning quantum properties from short-range correlations using
  multi-task networks.
\newblock {\em Nature Comm.}, 15, 2024.

\bibitem{xiao2022}
T.~Xiao, J.~Huang, H.~Li, J.~Fan, and G.~Zeng.
\newblock Intelligent certification for quantum simulators via machine
  learning.
\newblock {\em npjqi}, 8:138, 2022.

\bibitem{xu2024non}
S.~Xu et~al.
\newblock Non-abelian braiding of fibonacci anyons with a superconducting
  processor.
\newblock {\em Nature Phys.}, 20:1469--1475, 2024.

\bibitem{yang2023diffusion}
L.~Yang, Z.~Zhang, Y.~Song, S.~Hong, R.~Xu, Y.~Zhao, W.~Zhang, B.~Cui, and
  M.-H. Yang.
\newblock Diffusion models: A comprehensive survey of methods and applications.
\newblock {\em ACM Computing Surveys}, 56:1--39, 2023.

\bibitem{yang2024qcircuitnet}
R.~Yang, Y.~Gu, Z.~Wang, Y.~Liang, and T.~Li.
\newblock {QCircuitNet: A large-scale hierarchical dataset for quantum
  algorithm design}.
\newblock {\em arXiv:2410.07961}, 2024.

\bibitem{yang2024can}
T.-H. Yang, M.~Soleimanifar, T.~Bergamaschi, and J.~Preskill.
\newblock When can classical neural networks represent quantum states?
\newblock {\em arXiv:2410.23152}, 2024.

\bibitem{yao2024shadowgpt}
J.~Yao and Y.-Z. You.
\newblock Shadowgpt: Learning to solve quantum many-body problems from
  randomized measurements.
\newblock {\em arXiv:2411.03285}, 2024.

\bibitem{zeier2011symmetry}
R.~Zeier and T.~Schulte-Herbr{\"u}ggen.
\newblock Symmetry principles in quantum systems theory.
\newblock {\em J. Math. Phys.}, 52, 2011.

\bibitem{PhysRevE.101.053301}
R.~Zen, L.~My, R.~Tan, F.~H\'ebert, M.~Gattobigio, C.~Miniatura, D.~Poletti,
  and S.~Bressan.
\newblock Transfer learning for scalability of neural-network quantum states.
\newblock {\em Phys. Rev. E}, 101:053301, 2020.

\bibitem{zhang2022}
H.~Zhang et~al.
\newblock Experimental demonstration of adversarial examples in learning
  topological phases.
\newblock {\em Nature Comm.}, 13:4993, 2022.

\bibitem{zhang2021neural}
S.-X. Zhang, C.-Y. Hsieh, S.~Zhang, and H.~Yao.
\newblock Neural predictor based quantum architecture search.
\newblock {\em Mach. Learn. Sci. Technol.}, 2:045027, 2021.

\bibitem{zhang2021}
X.~Zhang, M.~Luo, Z.~Wen, Q.~Feng, S.~Pang, W.~Luo, and X.~Zhou.
\newblock Direct fidelity estimation of quantum states using machine learning.
\newblock {\em Phys. Rev. Lett.}, 127:130503, 2021.

\bibitem{zhang2025fault}
Y.~Zhang, X.~Zhang, J.~Sun, H.~Lin, Y.~Huang, D.~Lv, and X.~Yuan.
\newblock Fault-tolerant quantum algorithms for quantum molecular systems: A
  survey.
\newblock {\em arXiv:2502.02139}, 2025.

\bibitem{PhysRevB.107.075147}
Y.-H. Zhang and M.~Di~Ventra.
\newblock Transformer quantum state: A multipurpose model for quantum many-body
  problems.
\newblock {\em Phys. Rev. B}, 107:075147, 2023.

\bibitem{zhang2024observing}
Z.~Zhang and Y.-Z. You.
\newblock {Observing Schr{\"o}dinger's cat with artificial intelligence:
  emergent classicality from information bottleneck}.
\newblock {\em Mach. Learn. Sci. Technol.}, 5:015051, 2024.

\bibitem{zhao2023provable}
L.~Zhao, N.~Guo, M.-X. Luo, and P.~Rebentrost.
\newblock Provable learning of quantum states with graphical models.
\newblock {\em arXiv:2309.09235}, 2023.

\bibitem{zhao2025Rethink}
Y.~Zhao, C.~Zhang, and Y.~Du.
\newblock {R}ethink the {r}ole of {d}eep {l}earning towards {l}arge-scale
  {q}uantum {s}ystems, 2025.
\newblock arXiv:2505.13852v1.

\bibitem{zhong2022quantum}
L.~Zhong, C.~Guo, and X.~Wang.
\newblock Quantum state tomography inspired by language modeling.
\newblock {\em arXiv:2212.04940}, 2022.

\bibitem{zhou2025learning}
Y.~Zhou, C.~Wan, Y.~Xu, J.~P. Zhou, K.~Q. Weinberger, and E.-A. Kim.
\newblock Learning to decode logical circuits.
\newblock {\em arXiv:2504.16999}, 2025.

\bibitem{zhu2022}
Y.~Zhu, Y.-D. Wu, G.~Bai, D.-S. Wang, Y.~Wang, and G.~Chiribella.
\newblock Flexible learning of quantum states with generative query neural
  networks.
\newblock {\em Nature Comm.}, 13:6222, 2022.

\bibitem{zhu2023predictive}
Y.~Zhu, Y.-D. Wu, Q.~Liu, Y.~Wang, and G.~Chiribella.
\newblock Predictive modelling of quantum process with neural networks.
\newblock {\em arXiv:2308.08815}, 2023.

\bibitem{zhuang2020comprehensive}
F.~Zhuang, Z.~Qi, K.~Duan, D.~Xi, Y.~Zhu, H.~Zhu, H.~Xiong, and Q.~He.
\newblock A comprehensive survey on transfer learning.
\newblock {\em Proc. IEEE}, 109:43--76, 2020.

\bibitem{zhukov2022quantum}
A.~Zhukov and W.~Pogosov.
\newblock Quantum error reduction with deep neural network applied at the
  post-processing stage.
\newblock {\em Quantum Information Processing}, 21:93, 2022.

\bibitem{zou2005regularization}
H.~Zou and T.~Hastie.
\newblock Regularization and variable selection via the elastic net.
\newblock {\em Journal of the Royal Statistical Society Series B: Statistical
  Methodology}, 67:301--320, 2005.

\end{thebibliography}
\end{document}